\documentclass[onecolumn, numberedappendix]{aastex6}
\usepackage{graphicx}
\usepackage{epsfig}
\usepackage{natbib}
\usepackage{amssymb}
\usepackage{amsbsy}
\usepackage{natbib}
\usepackage{url}
\usepackage[mathcal]{euscript}
\bibpunct{(}{)}{,}{a}{}{,}

\newcommand{\rsun}{R$_{\sun}$}
\newcommand{\msun}{M$_{\sun}$}

\newcommand{\mum}{$\mu$m}

\newcommand{\etal}{et~al.}

\begin{document}
 
\def\simlt{\vcenter{\hbox{$<$}\offinterlineskip\hbox{$\sim$}}}
\def\simgt{\vcenter{\hbox{$>$}\offinterlineskip\hbox{$\sim$}}}
\def\etal{et al.\ }
\def\kms{km s$^{-1}$}

\title{Orbiting Clouds of Material at the Keplerian Co-Rotation Radius of Rapidly Rotating Low Mass WTTs in Upper Sco}
\author{John Stauffer\altaffilmark{1}, 
Andrew Collier Cameron\altaffilmark{5},
Moira Jardine\altaffilmark{5},
Trevor J. David\altaffilmark{2},
Luisa Rebull\altaffilmark{1,3}, 
Ann Marie Cody\altaffilmark{4},
Lynne A. Hillenbrand\altaffilmark{2},
David Barrado\altaffilmark{6},
Scott Wolk\altaffilmark{7},
James Davenport\altaffilmark{8},
Marc Pinsonneault\altaffilmark{9}
}
\altaffiltext{1}{Spitzer Science Center (SSC), California Institute of
Technology, Pasadena, CA 91125, USA}
\altaffiltext{2}{Astronomy Department, 
California Institute of Technology, Pasadena, CA 91125 USA}
\altaffiltext{3}{Infrared Science Archive (IRSA), 1200 E. California Blvd, 
MS 314-6, California Institute of Technology, Pasadena, CA 91125 USA}
\altaffiltext{4}{NASA Ames Research Center, Space Sciences and
Astrobiology Division, MS245-3, Moffett Field, CA 94035 USA}
\altaffiltext{5}{School of Physics and Astronomy, University of St. Andrews, North
Haugh, St Andrews KY16922, UK}
\altaffiltext{6}{Centro de Astrobiolog\'ia, Dpto. de
Astrof\'isica, INTA-CSIC, E-28692, ESAC Campus, Villanueva de
la Ca\~nada, Madrid, Spain}
\altaffiltext{7}{Harvard-Smithsonian Center for Astrophysics, 60
Garden Street, Cambridge, MA 02138, USA}
\altaffiltext{8}{Department of Physics \& Astronomy, Western Washington University, 
  516 High Street, Bellingham, WA 98225, USA} 
\altaffiltext{9}{Department of Astronomy, The Ohio State University, Columbus, OH 43210, USA}

\email{stauffer@ipac.caltech.edu}

\begin{abstract}

Using {\em K2} data, we have identified 23 very low mass members of
the $\rho$ Oph and Upper Scorpius star-forming region as having
periodic photometric variability not easily explained by
well-established physical mechanisms such as star spots, eclipsing
binaries, or pulsation.  All of these unusual stars are mid-to-late M
dwarfs without evidence of active accretion, and with photometric periods
generally $<$1 day.  Often the unusual light curve signature
takes the form of narrow flux dips; when we also have rotation periods
from star spots, the two periods agree, suggesting that
the flux dips are due to material orbiting the star at the Keplerian
co-rotation radius.  We sometimes see
``state-changes" in the phased light curve morphologies where
$\sim$25\% of the waveform changes shape on timescales less than a
day; often, the ``state-change" takes place immediately after a
strong flare.   For the group of stars with these sudden light curve
morphology shifts,  we attribute their flux dips as most probably
arising from eclipses of warm coronal gas clouds, analagous to the
sling-shot prominences postulated to explain transient H$\alpha$\
absorption features in AB Doradus itself and other rapidly rotating late type
stars.  For another group of stars with somewhat longer periods, we
find the  short duration flux dips to be highly variable on both short
and long timescales, with generally asymmetric flux dip profiles.  We
believe that these flux dips are due to particulate clouds possibly
associated with a close-in planet or resulting from a recent
collisional event.

\end{abstract}

\section{Introduction}

We have entered into a new era, where synoptic photometry for millions
of stars is currently being collected and analysed (e.g., OGLE, Gaia,
WASP).  
In the relatively near future, LSST will obtain synoptic
photometry for billions of stars.  In order to make the most of those
data, we need to develop automated routines to sift through enormous
datasets in order to identify light curves of interest.   A precursor
to that activity is to develop a better understanding of all the types
of photometric variability present in real stars, and where possible
to assign physical mechanisms to those variability classes.

The {\em Kepler} mission provided one very well characterized training set
of light curves for $>$100,000 stars.   However, because it was tied
to a single position in the sky and the target stars were primarily
chosen for one science goal, the {\em Kepler} data only provide templates
for a limited range of stellar variability types. The {\em K2} mission
(Howell \etal\ 2014) is providing a greatly augmented set of data by
observing a new group of $\sim$20000 stars every three months,  with
those stars having a much wider range of characteristics. One specific
variability type for which {\em K2} is providing a wealth of new data is
young stellar objects (YSOs).   

During Campaign 2 of the {\em K2} mission, nearly 1500 candidate members of
the star-forming regions comprised of the $\rho$ Oph dark-cloud complex
and the Upper Scorpius association were targeted.   The data from that
78 day campaign form the largest set of precision YSO optical light curves
that have ever been obtained.    Those light curves provide a hunting
ground for the search for newly formed planets via their transits
(David \etal\ 2016a; Mann \etal\ 2016),  for the identification of YSO
eclipsing binaries (Alonso \etal\ 2015; David \etal\ 2016b; Kraus
\etal\ 2015; Lodieu \etal\ 2015),  and for the determination of the
rotational periods for a large sample of low mass stars of 5-10 Myr
age (Rebull \etal, in preparation).    A small fraction of the Upper
Sco YSOs have IR excesses and are likely classical T~Tauri stars
(CTTs); a sub-group of ``burster" lightcurves associated with some of
these accreting systems is discussed in Cody \etal\ (2017a).

The discussion of our process for identifying a best set of probable 
Upper Sco members,  and the categorization of light
curve morphologies for the disked stars is provided in 
Cody et al. (2017b, in preparation).
Our sorting of those members into stars with primordial disks and those 
without disks (weak-lined T Tauri stars, WTTs) is
described briefly below, and more fully in Cody etal (2017b).

About 1100 of the Upper Sco and $\rho$ Oph stars appear not to have
infrared (IR) excesses and presumably have shed their primordial
circumstellar disks.  In the course of our processing and  analysis of
those light curves, the vast majority were identified as showing
periodic variability characteristic of stars with dark,
non-axisymmetrically distributed spots.  A handful were found to be
eclipsing binaries, and another handful (all relatively high mass
stars) are pulsators.   However,  twenty-three of the stars showed
periodic photometric variability which did not fit into any of those
categories.   It is those stars that are the subject of this paper.  
All of these stars are M dwarfs, with most having spectral types
between M4 and M6.   All have short periods -- more than half less
than one day, and all but one less than two days.

In \S 2, we describe the sources of data for this paper.  Our process
for identifying unusual light curve classes, and the working
definitions we have developed for our three classes is provided in \S
3.   The empirical characteristics of the variability for each of the
three classes is laid out in \S 4-6.

\section{Observational Data }

\subsection{{\em K2} Light Curves}

For the first three {\em K2} campaigns, the {\em K2} Science Center did not
produce detrended light curves.  Instead, what was provided was the
raw CCD pixel data for each star at each epoch.   We (Cody \etal\
2017a) therefore created our own detrended light curves.  We produced
two primary versions -- one where the aperture centers were held fixed
over the course of the {\em K2} campaign, and another where a new center was
derived at each epoch.  For each flavor of reduction, photometry was
produced with 2 and 3 pixel radius apertures.  For each of these four
light curve versions, we then applied a version of the detrending
routine described in Aigrain \etal\ (2016) to remove the short-term,
non-astrophysical trends in the data.

After we had begun writing this paper, two other light curve
reductions became publicly available for Campaign 2, the Vanderburg
reduction (Vanderburg \& Johnson 2014) and the EVEREST reduction
(Luger \etal\ 2016).   We downloaded light curves for each of the
stars that we considered as possibly relevant to this paper.

For each star addressed in this paper, we examined all of the light
curve variants that we had available, and chose the one which appeared
to have the least non-astrophysical contribution (when several variants
were about the same, we simply adopted our own detrended light curve).   
In all cases, the
unusual features which caused us to include the star in our paper were
present in multiple light curve variants.   In a few cases, one of the
light curve variants did not have the same signature.  We believe 
this indicates that the detrending algorithm employed in that version
removed some real astrophysical signature, or that the aperture size
used in that version or the centering algorithm in that version
resulted in the light curve being dominated by light from a different
star.  We made special effort in these cases to closely examine all
the available data (finding charts, literature, all light curve
variants and different aperture sizes) in order to make sure that the
light curve we present here is the
best available for that star.

\subsection{Ancillary Data}

The Ecliptic Plane Input Catalog (EPIC; Huber \etal\ 2016) catalog
available at MAST provides accurate coordinates for all of the stars
in our sample.   Using those coordinates, we downloaded all available
near and mid-IR photometry for our stars from the 2MASS (Skrutskie
\etal\ 2006), WISE (Wright \etal\ 2010), Spitzer  (Werner \etal\ 2004)
SEIP\footnote{http://irsa.ipac.caltech.edu/data/SPITZER/Enhanced/SEIP/overview.html}
and FEPS (Meyer \etal\ 2006),  AKARI (Murakami \etal\ 2007), and SDSS
(e.g., Ahn \etal\ 2014) archives as well as $B$ and $V$ band
photometry from APASS (Henden \& Munari 2014).   We also downloaded
$grizy$ photometry for most of our stars from the PAN-STARRS database
(Flewelling \etal\ 2016). Spectral types were available in the
literature  (Preibisch \etal\ 2002, Erickson \etal\ 2011, Lodieu
\etal\ 2011, Luhman \& Mamajek 2012, Rizzuto \etal\ 2015) for 80\% of
our stars; for most of the stars, those same references also provided
H$\alpha$ and lithium (6708\AA) equivalent widths. We have used our
own Keck HIRES spectra to provide a small amount of new spectral
information, as described in the Appendix.

One goal for the ancillary data was to provide a color-magnitude diagram which
would include all the stars in our sample, and for that we decided to
attempt to determine estimated $V-K_s$ colors for every star.  Because
only a small fraction of the Upper Sco members have published $V$ band
photometry, this required transforming photometry or colors obtained at other
optical bands.  As the
first step in that process, we downloaded $G$ band photometry for all
of our stars from the newly released Gaia DR1 archive (Gaia
Collaboration 2016).  We then derived a formula to convert $G-K_s$
colors to $V-K_s$ colors based on a large sample of low mass stars
with published $V-K_s$ colors in Praesepe (Rebull \etal\ 2017,
submitted).  For the stars lacking Gaia
photometry, we obtained $I$ magnitudes from the Deep Near-Infrared
Survey of the Southern Sky (DENIS; Epchtein \etal\ 1997)  and derived
a conversion formula in a similar way between $I_{\rm (DENIS)}-K_s$ and
$V-K_s$.

We used all of the available photometry and spectral types to produce
spectral energy distributions (SEDs) for each star (the individual
SEDs are shown in the Appendix). From those SEDs, we identified stars
with possible IR excesses.   For each of the sources thought to have
an IR excess, we looked at the images, where available, to assess the
quality of the photometry.  The longest wavelength data we have is
either WISE 22 \mum\ or, for just four stars Spitzer 24 \mum. In
several cases, the 22 \mum\ point from WISE, in particular, was not a
secure detection, and the apparent IR excess therefore vanished.  
Only one star in our final sample has a relatively secure, though
small  IR excess (only at 24 \mum), and that is EPIC 205024957. 
Carpenter \etal\ (2009) also determined that EPIC 205024957 has a 24
\mum\ excess and identified it as a debris disk.    For the stars
where our longest wavelength detection is at 12 \mum, the lack of an
excess at that wavelength precludes them from having primordial
disks.  Because of the paucity of good data longward of 12~$\mu$m for
most of our stars, a larger fraction could harbor debris disks.

\section{Sample Selection and Light Curve Taxonomy}

\subsection{Sample Selection}

The first step in our sample selection was to eliminate stars with
active accretion (CTTs) from the Upper Sco and $\rho$\ Oph {\em K2} sample,
since the accretion process in CTTs and interactions with the
primordial circumstellar disk can produce unusual light curves of many
types, and addressing those types of variability was beyond the scope
we wanted here (but see Cody \etal\ 2017a).  None of the stars we
retained for this paper have SEDs consistent with Class I or II
primordial disks (see the SED plots in the Appendix, and compare, for
example, to the SEDs of Class I and II sources in Upper Sco shown in
Cody \etal\ 2017a or similar sources in Taurus as shown in Rebull
\etal\ 2010).   Two of the stars are considered as $\rho$\ Oph
members in the literature (EPIC 203962559 and EPIC 203927435); all of
the others are considered as members of Upper Sco. 

Once we had identified the full set of Upper Sco weak-lined
T~Tauri (WTTS) members with {\em K2} data, three of the authors (TJD,
LMR, AMC) independently examined the data for each star, including
each of the light curve variants described above.  They also
independently searched for periodicities in those light curves and
examined the phased light curves. Those efforts were aimed primarily
at identifying possible planet transits, or eclipsing binary
candidates, or variability related to accretion or circumstellar disk
structure, or in compiling stellar rotation periods. During the course
of those efforts, a few dozen stars were identified as having light
curves that were not easily explained by any of those mechanisms.  In
those cases, further analysis was done and other members of the team
were brought in to perform more manual detrending of the light
curves.  This more detailed analysis resulted in some of these light
curves being ascribed to artifacts in the data or to allowing them to
fit into one of the established light curve classes.  However,
23 of these stars survived.  These stars are listed in Table~1.

\subsection{Definition of Light Curve Classes and Their Morphologies}

We have sorted our unusual light curves into three 
classes in order to facilitate our discussion of their properties.  We
acknowledge that this sorting is probably imperfect, and that the boundaries
between classes are indistinct.   An alternative view is that all of our
stars belong to a single group, and their differing light curve morphologies
arise due to a range in our view angle to their rotation axes or to a
range in their rotation rates (e.g., because that affects the location of the
Keplerian co-rotation radius relative to the sublimation radius) or to
some other hidden parameter.   After assembling the data we had available,
we nevertheless felt it was helpful to sort the stars into the three groups
in order to emphasize the shared properties within each group.

\begin{figure}[ht]
\epsscale{0.9}
\plotone{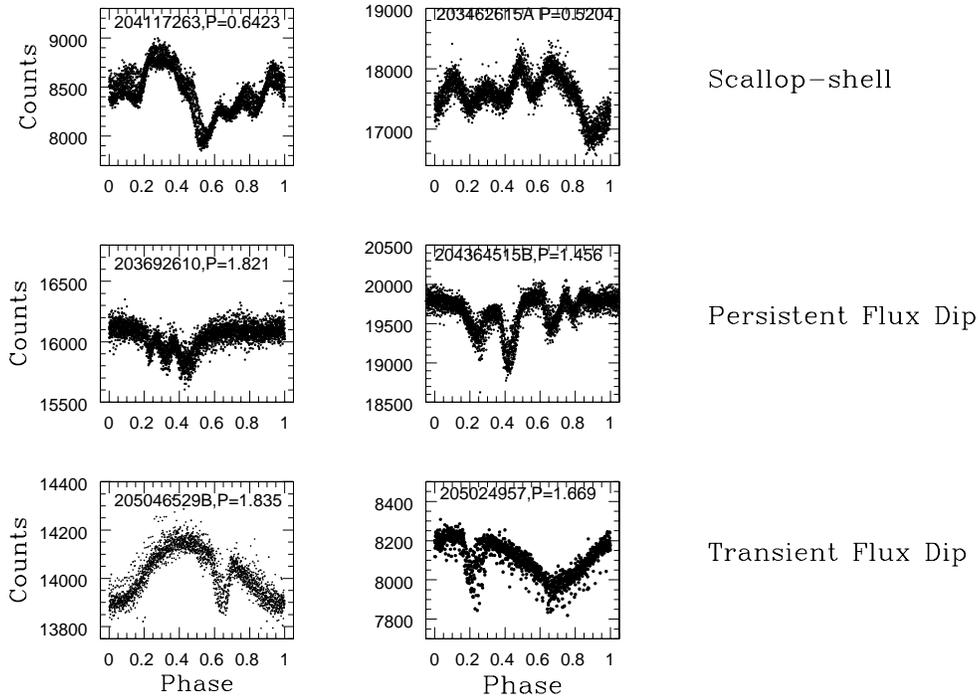}
\caption{Phased light curves  for exemplary members of the ``scallop-shell"
light curve class (top row), the persistent, short-duration flux dip
class (middle row), and the transient, short-duration flux dip class (bottom
row).
\label{fig:prototype_light_curves}}
\end{figure}

Table 1 provides a list of all 23 stars we have identified as
belonging to these three classes, along with some of their overall
defining properties. Spectral types are known for 20 of these stars;
all are M dwarfs with 17 of the 20 having spectral types between M4.0
and M5.5.  Most of the Upper Sco {\em K2} sample are M dwarfs, but the stars
in Table 1 are predominantly weighted towards later M sub-types (this
is also reflected in their CMD locations -- see Figure
\ref{fig:UScoCMD.Pvmk}).  Eighteen of the stars in Table 1 have known
H$\alpha$ equivalent widths, all but one of which have equivalent
widths appropriate for WTTS ($-$3.0 \AA\ $<$ EqW $<$ $-$16 \AA).
Thirteen of the stars in Table 1 have measured Li 6708\AA\ equivalent
widths, all of which are $\sim$\ 0.5 \AA, also as appropriate for WTTS
mid-to-late M dwarfs (the spectral types for two stars and the lithium
data for six stars are from our own HIRES spectra, as described in the
Appendix).   

Eleven of 23 (48\%) of the stars in Table~1 have two Lomb-Scargle
periods (i.e. are apparent binaries), which seems high.  From the full
sample of {\em K2} M dwarf WTTs with 2 $<$ P $<$ 8 days ($\sim$280 stars), only
16\% are apparent binaries by the same measure.  However, for the $\sim$140
{\em K2} M dwarfs with P $<$ 1 day, 38\% are apparent binaries.   Thus, the 
stars in Table~1 may have a high binary fraction simply because the
rapidly rotating low mass stars in Upper Sco are often binaries.  This
is consistent with the conclusion from Stauffer \etal\ (2016) that M 
dwarfs in the Pleiades with two Lomb-Scargle peaks were both more rapidly
rotating and the two periods were closer together than would have been
true if their periods were drawn at random from the single star population.

Figure \ref{fig:prototype_light_curves} shows phased light
curves for two examples of  each of the three classes; phased light
curves for all members of the scallop-shell and persistent flux dip
classes are provided in the Appendix, along with a description of the 
processing we have done to the original light curves.    The class
definitions are as follows:
\begin{itemize}
\item {\bf scallop-shells} -- phased light curves that look like the
  rim of a scallop shell.  They have many wave-shaped undulations in
  their phased light curves.  Usually the shape remains nearly
  constant for the entire {\em K2} campaign, but in some cases it changes
  over time.  Peak to trough amplitudes from a few percent to more
  than 10\%; periods always less than 0.65 days.  Eleven of our Upper
  Sco stars fall in this category.
\item {\bf persistent flux dip class} -- discrete, triangularly shaped 
  flux dips, generally covering less than 0.15 in phase (for each of
  up to  4 dips). The flux dips are often more or less stable in depth
  over the {\em K2}  campaign, with dip depths generally a few percent. 
  Outside of the dips, the phased light curves are either nearly flat
  or consistent with sinusoidal variation due to cool starspots. 
  Periods range from 0.5 to 1.8 days. A half dozen rapidly rotating,
  late M dwarf members of the Pleiades with this light curve signature
  were identified from {\em K2} Campaign 4 data (Rebull \etal\ 2016, \S 3.6;
  Collier Cameron \etal\ 2017).  Eight of our Upper Sco stars fall
  in this category.
\item {\bf transient, narrow dip class} -- more or less triangularly
   shaped flux dips, narrow in phase, but always variable in depth, on
   both short and long timescales,  over the 78 day {\em K2} campaign.  
   Only one prominent dip present in the phased light curve.   Dip
   depths generally a few percent, but up to 20\% for EPIC 205483258
   (RIK-210).  All four stars also show a spotted-star waveform, whose
   period is identical to the dip period within our uncertainties. 
   Periods ranging from 1 to 5.5 days.
\end{itemize}

In the next sections, we discuss specific examples of these categories.

\floattable
\begin{deluxetable*}{lccccccccccccc}
\tabletypesize{\footnotesize}
\tablecolumns{13}
\tablewidth{0pt}
\tablecaption{List of Upper Sco Stars with Unusual Periodic Light Curves\label{tab:basicdata}}
\tablehead{
\colhead{EPIC ID} & 
\colhead{Name\tablenotemark{a}} &
\colhead{RA } & 
\colhead{Dec} &
\colhead{$K_s$} &
\colhead{$V-K_s$ \tablenotemark{b}} &
\colhead{$[W1]-$[W3]} & 
\colhead{$\lambda_{\rm max}$\tablenotemark{c}} &
\colhead{SpT\tablenotemark{d}} & 
\colhead{H$\alpha$ EqW} & 
\colhead{Li EqW} & 
\colhead{P1\tablenotemark{e}} & 
\colhead{P2\tablenotemark{e}} & 
\colhead{Class\tablenotemark{f}}  \\
& & \colhead{(deg)}& \colhead{(deg)}
& \colhead{(mag)}& \colhead{(mag)}
& \colhead{(mag)}& \colhead{(\mum)}
& &\colhead{(\AA)} &\colhead{(\AA)}
& \colhead{(days)}& \colhead{(days)} & }
\startdata
 204918279 & RIK-21  & 239.1046 & -20.2711 &  9.86 & 6.90d & 0.38 & 12 & M5.0  & -10.2 & 0.61  &  0.4594 &  0.4665* & 1 \\ 
 204066898 & UScoCTIO 80A & 239.6509 & -23.8005 & 10.19 & 5.59  & 0.27 & 12 & M3    & -9.9 & \nodata   &  0.3956* &  0.5386 & 1 \\
 203462615 & RIK-42 & 239.9086 & -26.0565 & 10.25 & 6.13  & 0.47 & 12 & M5.5  & -16.2 & 0.81  &  0.5201* &  0.4421 & 1 \\
 204897050 & UScoCTIO 56 & 240.4208 & -20.3689 & 10.86 & 7.16d & 0.42 & 12 & M5 & \nodata & \nodata   &  0.2639 &  \nodata & 1 \\
 202724025 & RIK-90 &  242.2373 & -28.5993 &  9.63 & 5.78  & 0.33 & 12 &  M4.5  & -10.2 & 0.21  &  0.2595* &  0.2795 & 1 \\
 204117263 & LHJ-65 & 242.7589 & -23.5973 & 10.95 & 6.42  & 0.35 & 12 & M5    & -11.6 & \nodata  &  0.6423 &  \nodata & 1 \\
 204367193 & LHJ-77 & 242.9766 & -22.6137 & 13.30 & 8.33  & \nodata & 4.6 & M6.3  & -10.3 & \nodata  &  0.4835 &  \nodata & 1 \\
 203534383 & & 243.5125 & -25.8148 & 11.71 & 7.73  & 0.73 & 12 & M4:   &  -9.5  & \nodata &  0.2784* &  0.3234 & 1 \\
 205110559 & & 243.8322 & -19.3520 & 10.46 & 7.74  & 0.61 & 12 & \nodata & \nodata & \nodata   &  0.4031 &  \nodata & 1 \\
 203050730 & RIK-246 & 247.7741 & -27.4295 & 10.53 & 5.87  & 0.53 & 12 & M4.5  & -9.1 & 0.38  &  0.4865* &  0.7740 & 1 \\
 203185083 & RIK-253 & 248.6464 & -26.9675 & 10.48 & 6.62  & 0.18 & 12 & M4.5 & -27.1 & 0.54 & 0.4400 & \nodata & 1 \\
  \hline \\
 204364515 &      & 240.3398 & -22.6240 & 10.05 & 6.84 & 0.34 & 24 &M4   &  -8.4 & 0.55    &  3.0863 &  1.456* & 2 \\
 203849738 & RIK-100 & 242.4703 & -24.6982 & 10.93 & 6.69d & 0.37 & 12 & M5.5  & -15.1 & 0.40  &  0.6190 &  \nodata & 2 \\ 
 203692610 &  &242.6318 & -25.2671 & 11.60 & 5.92  & 0.28 & 12 & M4   &   -3.0 & 0.54   &  1.821 &  \nodata & 2 \\
 205374937 &     & 242.8255 & -17.9580 &  9.33 & 6.35  & 0.34 & 24 &M4    & -4.8 & 0.55  &  0.6345* &  0.5436 & 2 \\
 202873945 & &  245.4164 & -28.0518 & 11.24 & 8.21 & 0.75 & 12 &  \nodata &  \nodata & \nodata &  0.6258 &  \nodata & 2 \\
 204296148 & & 246.2158 & -22.8952 & 10.99 & 8.06k  & 0.10 &  12 & \nodata & \nodata & \nodata   &  0.5314* &  0.4717 & 2 \\ 
 203962559 & AOC 64 & 246.7103 & -24.2312 & 10.80 & 8.83d & \nodata & 8 & M4  & -6.0 & \nodata  &  1.5402 &  \nodata & 2 \\ 
 203927435 & WMR 2-23 &  247.1794 & -24.3812 & 10.14 & 9.11d & 0.13 & 12 & M4.5  & -6.8 & 0.76  &  0.4820* &  0.4162 & 2 \\
  \hline \\
 205024957 &      & 242.5459 & -19.7678 & 11.38 & 6.53  & 0.46 & 24$^+$ & M5    & -4.4 & 0.56  &  1.6656 &  \nodata & 3 \\
 205046529 &     & 242.6099 & -19.6642 & 10.40 & 6.52k & 0.35 & 24 & M4    & -4.4 & 0.54  &  2.5619 &  1.8358* & 3 \\
 204143627 & LHJ-120 & 243.9969 & -23.4934 & 11.31 & 6.29  & 0.13 & 12 & M5 & \nodata & \nodata   &  1.125 &  \nodata & 3 \\
 205483258 & RIK-210 & 245.8523 & -17.2909 &  9.65 & 5.04  & 0.37 & 12 & M2.5  & -5.3 & 0.59  &  5.667 &  \nodata & 3 \\
\enddata
\tablenotetext{a}{sources of the names are:
RIK: Rizzuto, Ireland and Kraus 2015;
UScoCTIO:  Ardila, Martin \& Basri 2000;
LHJ:  Lodieu, Hambly, Jameson et al. 2007;
AOC:  Alves de Oliveira \& Casali 2008;
WMR:  Wilking, Meyer, Robinson \& Greene 2005.}
\tablenotetext{b}{$V-K_s$ estimated preferentially using a conversion based on the Gaia DR1 `G' magnitudes
and 2mass K$_s$ values.  A `d' is appended if instead the $V-K_s$ estimate is based on Denis `I'
magnitudes and 2mass $K_s$.  A `k' is appended if the {\em K2} countrate
itself was used to estimate $V$.  }
\tablenotetext{c}{Longest wavelength at which there is a secure
detection, in \mum. A plus sign indicates if there is a plausible,
small IR excess}
\tablenotetext{d}{Spectral type, H$\alpha$\ equivalent width, and Li equivalent width for
EPIC 203534383 and 203692610 are from our own HIRES spectra.  The Li equivalent widths for
EPIC 203849738, 204364515, 205024957 and 205046529 are also from our own HIRES spectra.  See
discussion in the Appendix.}
\tablenotetext{e}{An asterisk is attached to the period
  (P1 or P2) to designate which star in a binary has the unusual light curve which is discussed
  in this paper.  All of the binary companions have normal, spotted-star light curve
  morphologies. }
\tablenotetext{f}{Light curve classes:   1 = scallop-shell; 2 = persistent short-duration flux dip;
  3 = transient, short-duration flux dip. }  
\end{deluxetable*}
\noindent

\section{Stars with Scallop-Shell Phased Light Curves}

As illustrated in the top row of Figure \ref{fig:prototype_light_curves},
these stars show complex phased
light curves, with many local maxima and minima.  For half of these
stars, their waveform is relatively stable for the entire 78 day {\em K2}
campaign.   However, for five of the eleven stars we put in this
category, there is distinct evolution in the waveform within the
campaign.  Because we believe the nature of this evolution provides
important clues as to the physical mechanism responsible for creating
this light curve morphology, we discuss in detail these five stars and
their light curve evolution during the campaign.  First, however, we
illustrate the class by describing the light curve of one of the six
members whose phased light curve shape does not change appreciably
over the duration of the {\em K2} campaign.

\subsection{EPIC 204066898a}

\begin{figure}[ht]
\epsscale{0.65}
\plotone{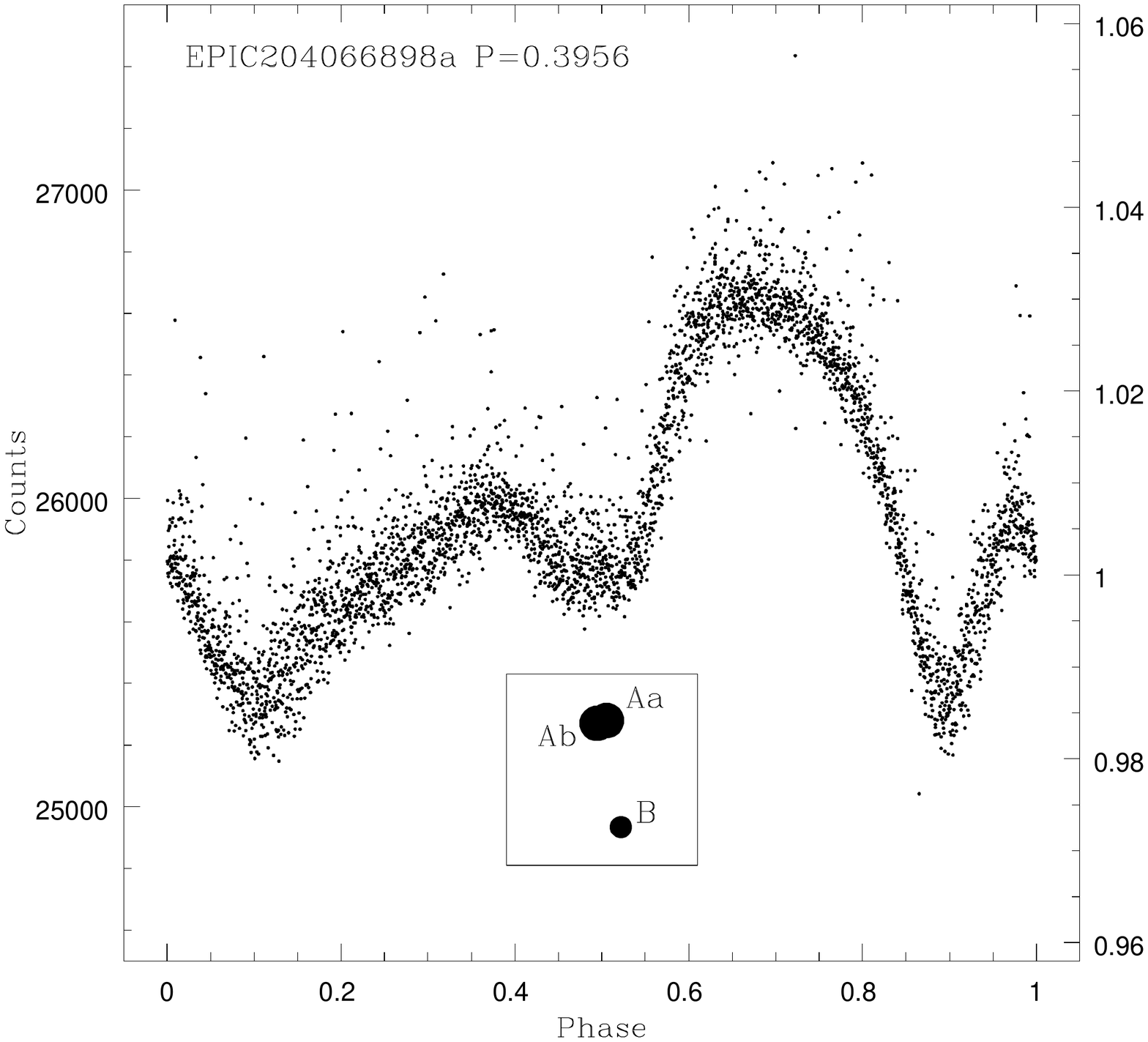}
\caption{The phased light curve  of EPIC 204066898a (UScoCTIO 80Aa). 
The left-side y-axis is in "counts"; the right-side y-axis normalizes
to the median-count rate in order to allow easy determation of
amplitudes in terms of per cent change in the light curve flux.   The
light curve shows no evidence for significant change  over the 78 days
of the {\em K2} campaign.  The phased light curve shows more structure than
can be produced by cool spots on the photosphere.  The inset is a
not-to-scale spatial cartoon of the orientation of the Upper Sco CTIO 80
triple system; the actual separation between ``Aa" and ``Ab" is
0.05$\arcsec$ and between ``Aab" and ``B" is 13$\arcsec$.  This cartoon
does not represent the {\em K2} stamp image for this system nor the aperture
we have used for the photometry.
\label{fig:epic204066898}}
\end{figure}

EPIC 204066898 (a.k.a.\ UScoCTIO 80A\footnote{KH2009 actually call
this star USco 80A, but in keeping with the SIMBAD usage, the more
appropriate name is UScoCTIO 80A.  The star designated UScoCTIO 80 in
Ardila \etal\ 2000 is UScoCTIO 80B, as explained in the note to Table
9 of KH2009.};  Kraus \& Hillenbrand
2009=KH2009; Kraus \& Hillenbrand 2012=KH2012) is an M3 member of
Upper Sco with no apparent IR excess.
It has been identified as a wide binary, with
separation  about 13$\arcsec$ and $\Delta K_s$ = 1.9 mag (Kraus \&
Hillenbrand 2009=KH2009).  KH2012 also identify UScoCTIO 80A as a
close (0.05$\arcsec$), nearly equal mass binary (see their Table 3).    
We find two  independent peaks in the Lomb-Scargle periodogram with
periods of 0.3956 and  0.5386 days.   We have experimented with one
pixel and 1.5 pixel apertures for the light curve extraction in order
to assess the contribution of UScoCTIO 80B to the EPIC 204066898 light
curve; our conclusion is that the two Lomb-Scargle peaks are both
correctly associated with UScoCTIO 80A (UScoCTIO 80B is also EPIC
204066097, for which we derive a period of 0.36d).  We thus assume the
two stars associated with the two Lomb-Scargle peaks are also the
close binary pair identified in KH2012.  The longer period system for
EPIC 204066898 (which we designate as the secondary component)  has a
more or less sinusoidal light curve with about 1\% peak-to-trough
amplitude.  After removing the signal from this system and phasing the
remaining signal to the 0.3956 day period of the primary component,
the light curve shape for this star (Figure~\ref{fig:epic204066898})
is fairly typical of the scallop-shell class, with a full amplitude of
about 5\% and multiple peaks and troughs in its phased light curve.

\subsection{EPIC 205110559}

\begin{figure}[ht]
\epsscale{0.65}
\plotone{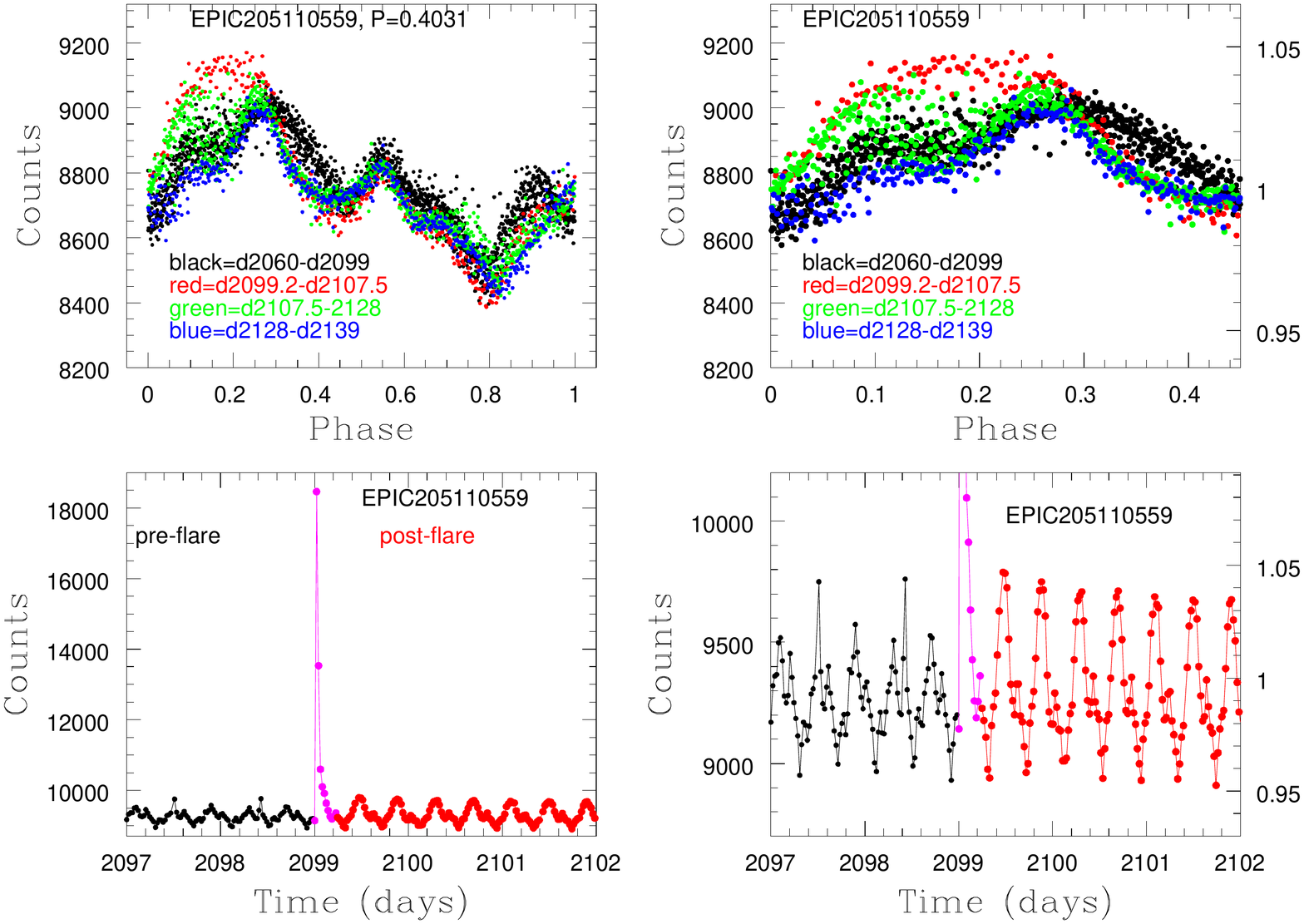}
\caption{The phased light curve and other plots to illustrate the
light curve evolution of EPIC 205110559.  Top left: full phased light
curve.  The different colors correspond to the windows in time noted
within the plot.  Top right: an expanded view of just the first half
of the phased light curve.  Bottom left: A strong flare occurred at
day 2099 (by far the strongest for this star during the {\em K2} campaign).
Bottom right: expanded view of the light curve around the time of the
flare, showing that the periodic waveform changed following the flare.
For the top-right and bottom-right plots, y-axis labels for both counts
and counts normalized to the median value are provided.
\label{fig:epic205110559}}
\end{figure}

There are no published spectra for EPIC 205110559, but its colors
suggest it is a mid to late M dwarf. Its light curve
(Figure~\ref{fig:epic205110559}) is also fairly typical of the
scallop-shell class, with a full amplitude of about 5\% and multiple
peaks and troughs in its phased light curve.  For phase 0.45 to 1.0,
the waveform is relatively stable over the duration of the campaign.  
However, as emphasized in the second panel of
Figure~\ref{fig:epic205110559}, the waveform shows considerable
evolution over the campaign for the phase range 0 to 0.45.  The
primary shift in waveform shape appears to occur almost exactly at the
boundary between the first window (day 2060-2099) and the second window
(day 2099.2-2107.5).\footnote{The times listed here and throughout the
remainder of the paper are in days since January 1, 2009, or as
JD - 2454833.  {\em Kepler} was launched on March 7, 2009.}  
The third panel of Figure \ref{fig:epic205110559}
shows an expanded view of the timeframe when the waveform changed
state, which reveals that a strong flare occurred at day 2099.   This is
by far the strongest flare for this star during the {\em K2} campaign, and
it occurs at exactly the point that the waveform made its state
transition.  This latter point is illustrated in the last panel in
Figure \ref{fig:epic205110559}.  The ``excess" flux at phase
$\sim$0.15 gradually declines, returning to approximately the
pre-flare level by day $\sim$2128.  We conclude that there is very
likely a physical link between the occurrence of the flare and the
change in morphology of the phased light curve.

\subsection{EPIC 204918279B}

The {\em K2} light curve for EPIC 204918279 has a
beating appearance and the Lomb-Scargle periodogram shows two strong
peaks at very similar periods, 0.4594 and 0.4665 days, which we
interpret as indicating it is a binary with the two components having
nearly the same period.  

\begin{figure}[ht]
\epsscale{0.65}
\plotone{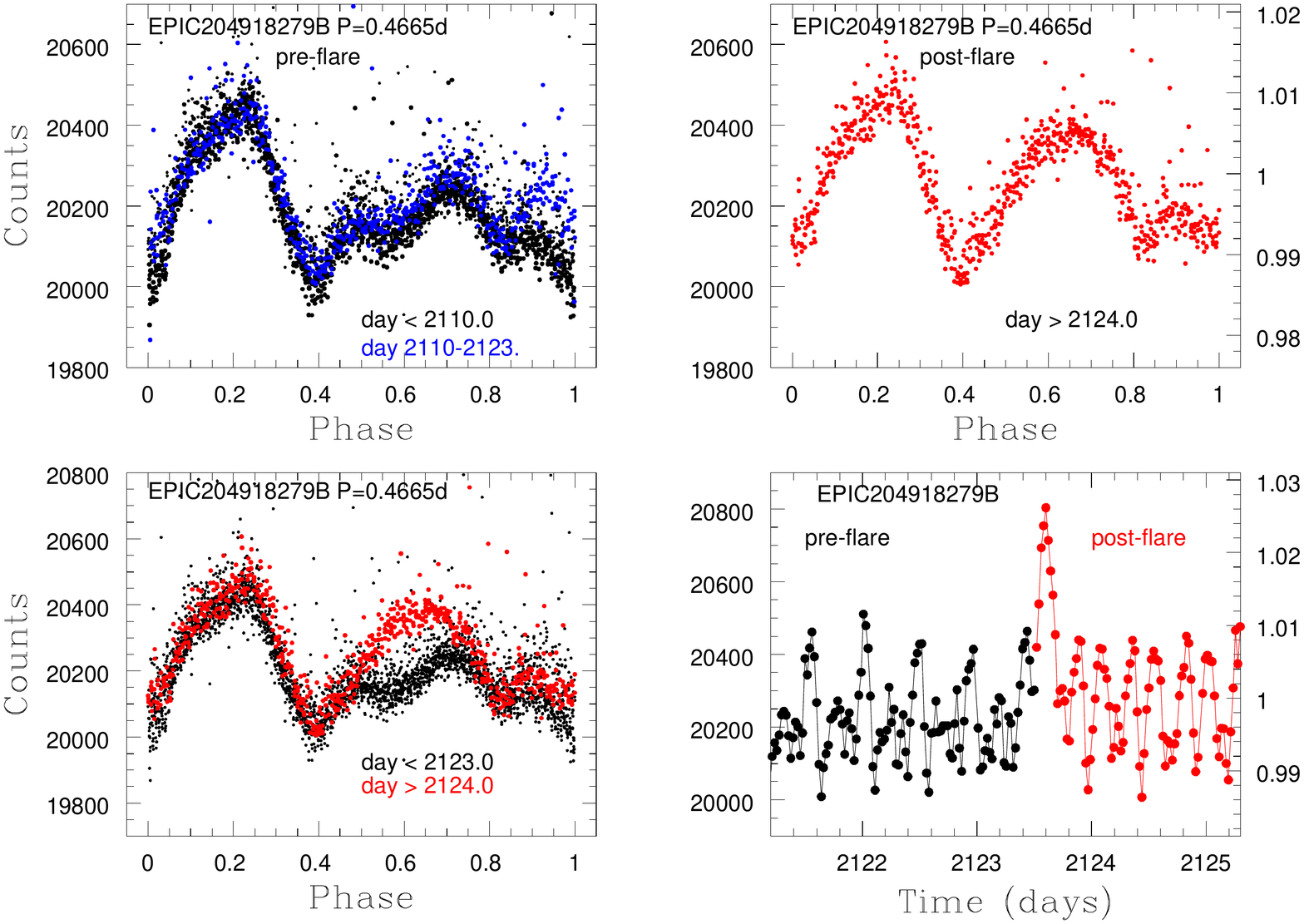}
\caption{The phased light curve and other plots to illustrate the
light curve evolution of EPIC 204918279B.  Top left: phased light curve
for d $<$ 2123. Top right: phased light curve for d $>$ 2124.  Bottom
left:  superposition of the two previous panels, showing that the
change in waveform was restricted to the phase range 0.45 to 0.7. 
Bottom right: expanded view of the light curve around the time of the
phase transition, showing the presence of a small flare when the
change occurred.
\label{fig:epic204918279}}
\end{figure}

After separating the variability at those two periods, we find that
the star with $P$=0.4665 days has a light curve shape which places it in
the scallop-shell category at least for the first $\sim$60 days of the
{\em K2} campaign, as shown in the first panel of
Figure~\ref{fig:epic204918279}.  The phased light curve changed
abruptly in shape at day $\sim$2123.5, as illustrated in the second
and third panels of Figure \ref{fig:epic204918279}. The state
transition of the light curve shape again appears to have been
triggered by a flare; see the fourth panel of Figure
\ref{fig:epic204918279}. In this case, this is a much weaker
flare than for EPIC 205110559\footnote{Because this is a binary, we
cannot know for certain which of the two stars was the site of the
flare.  However, because of the close coincidence between the flare
and the change in light curve waveform for the star with $P$=0.4665
days, we believe it is highly probable the flare originated on that
star.}.   Nevertheless, this was the strongest flare seen in the {\em K2}
light curve for EPIC 204918279, at least during the last 60 days of
the campaign.  Prior to and after the flare, the light curve shape
appears quite stable during the {\em K2} campaign.    

\subsection{EPIC 203185083}

EPIC 203185083 is an M4.5 YSO Upper Sco member  with unusually strong
H$\alpha$ emission (equivalent width = $-$27 \AA) for a WTTS of that
spectral type (Rizzuto \etal\ 2015).  However, it appears to have no
IR excess, so we consider it to be a WTTS.  The Lomb-Scargle
periodogram indicates a period of 0.4400 days.   

\begin{figure}[ht]
\epsscale{0.65}
\plotone{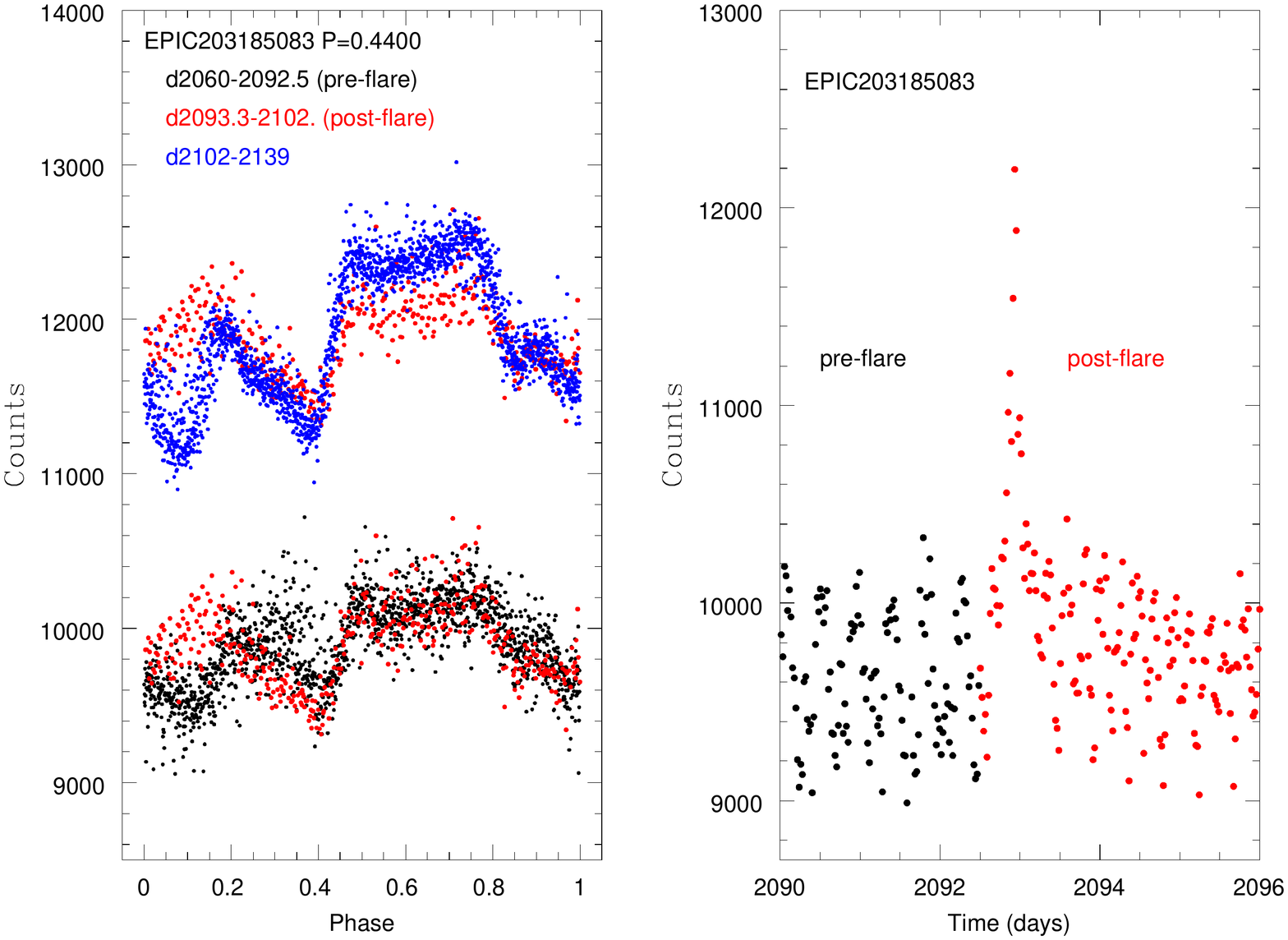}
\caption{Plots illustrating the
light curve evolution of EPIC 203185083.  Left: phased light curve for
three time intervals: blue (day 2060-2092.5), green (day 2093.3-2102),
red (day 2102-2139). There is a transition in waveform shape at day
2093, and another transition in shape at day 2102. The light curves
at the top of the plot are offset in y by 2000 counts for clarity.
Right: expanded view of the light curve near the 1st state-transition
in the phased waveform, showing the presence of a small flare when
the change occurred.
\label{fig:epic203185083}}
\end{figure}

After phasing the light curve to the 0.44 day period, we find a
waveform shape which is not plausibly explained by cool spots and
which seems to evolve over the duration of the {\em K2} campaign. After some
investigation, we find the waveform evolution basically occurs at two
rapid transitions on day 2093 and day 2102, as shown in the left panel
of Figure \ref{fig:epic203185083}.   The first state transition of the
light curve shape again appears to have been triggered by a flare
(right panel of Figure \ref{fig:epic203185083}), though again a much
weaker flare than for EPIC 205110559. Nevertheless, this was the
strongest flare seen in the {\em K2} light curve for EPIC 203185083. The
second state transition appears to happen without any obvious trigger
as seen from our vantage point.   Finally, for both EPIC 203185083 and
EPIC  204918279, the flares at the state-transition seem unusual in
that they have nearly symmetric shape (rather than a rapid rise and
slow decay as true for most flares and for the flare on EPIC
205110559).

\subsection{EPIC 204117263}

\begin{figure}[ht]
\epsscale{0.6}
\plotone{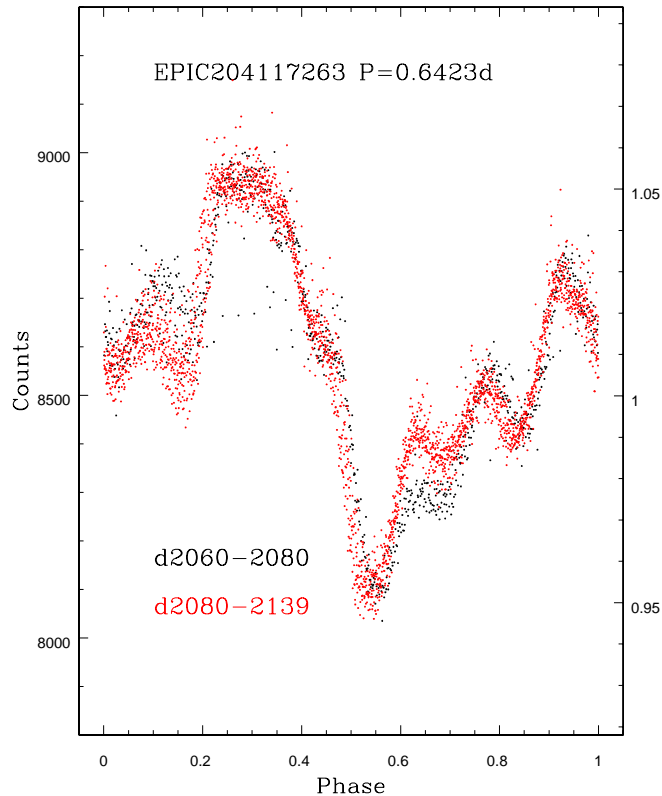}
\caption{The phased light curve of EPIC 204117263, illustrating the
abrupt change in shape at day 2080. Black points correspond to the time
window for days 2060-2080; red points correspond to data after day
2080.    The largest change in waveform shape is near phase=0.65, but
there is also a small change near phase 0.15.   Otherwise, the shape
of the phased light curve was very stable over the entire {\em K2} campaign.
\label{fig:epic204117263}}
\end{figure}

The Lomb-Scargle periodogram for EPIC 204117263 shows a single strong peak at
$P$=0.6423 days; when phased to that period, the light curve has a
scallop-shell appearance for the entire campaign.  There was, however,
a small, abrupt change in the phased light curve shape at day 2080.0, as
illustrated in Figure \ref{fig:epic204117263}.   In this case, the
state transition in the light curve shape does not appear to be
associated with a flare (at least on the side of the star that is
facing us).

\subsection{EPIC 204897050}

\begin{figure}[ht]
\epsscale{0.6}
\plotone{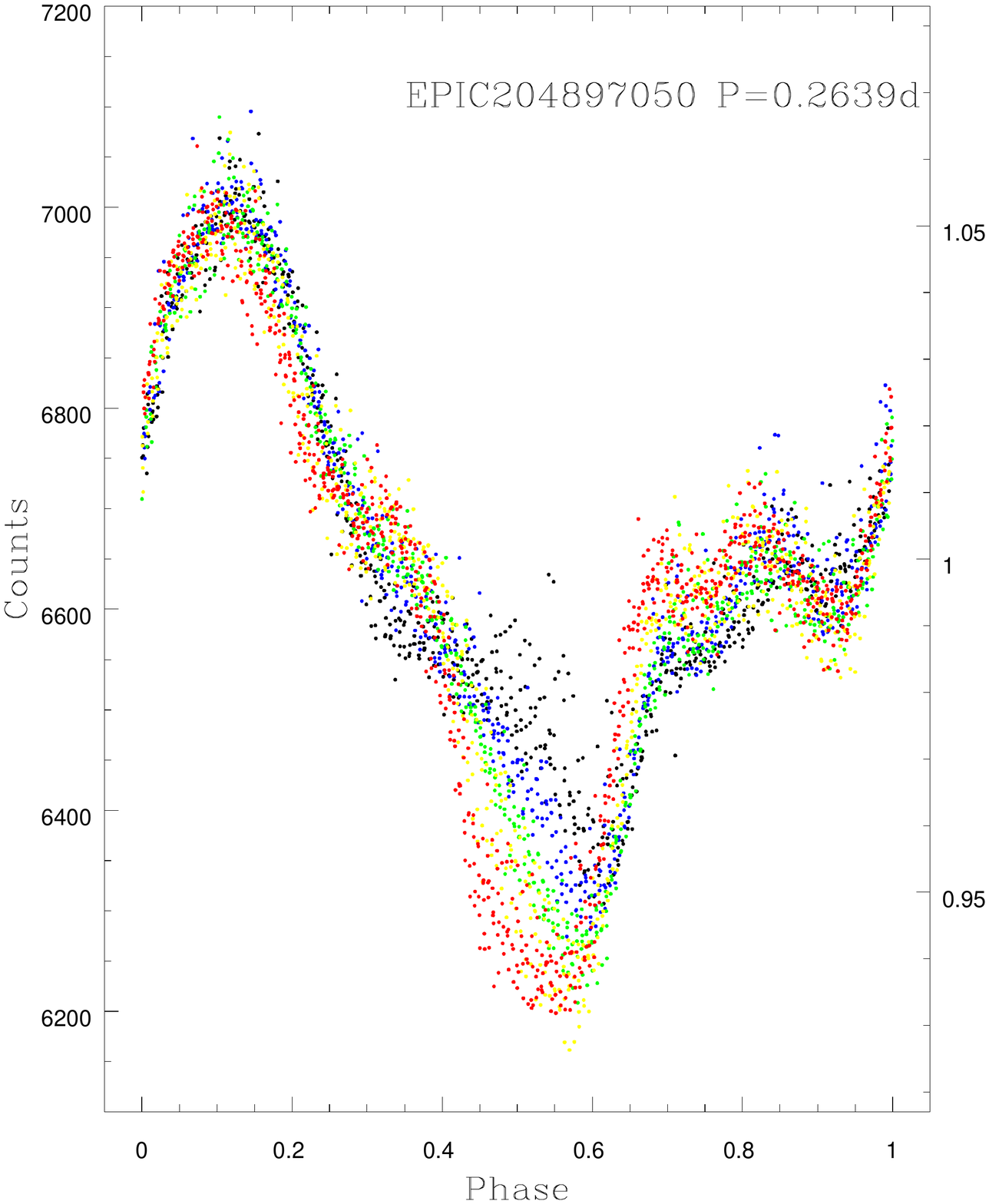}
\caption{The phased light curve of EPIC 204897050, illustrating the
gradual change in the shape of the phased light curve near phase=0.60.
Otherwise, the shape of the phased light curve was stable over the
entire {\em K2} campaign.  Colors (black, blue, green, yellow, red, in
order) correspond to approximately fifteen day long windows.  The flux
minimum near phase 0.5 became deeper and broader as the campaign
progressed.
\label{fig:epic204897050}}
\end{figure}

The Lomb-Scargle periodogram for EPIC 204897050 shows a dominant
strong peak at a very short period of  $P$=0.2639 days; when phased to
that period, the light curve has a scallop-shell appearance for the
entire campaign.  As illustrated in Figure \ref{fig:epic204897050},
there is a small portion of the phased light curve around phase=0.6
where the waveform shape gradually changes throughout the length of
the campaign.   This could be consistent with the other three stars
discussed above, if there was a trigger event just prior to the
beginning of Campaign 2; no significant flares are seen during the
time period of our light curve.


\subsection{Shared Characteristics of the Stars with Scallop-Shell Light Curves}

The stars having scallop-shell light curves have many shared characteristics,
which presumably provide clues to the physical process driving their unusual
light curve morphology.  Below we collect the most salient of those shared
characteristics:
\begin{itemize}

\item  All have periods less than 0.65 days.  We do not know that this is,
 in fact, the stellar rotation period, but we will assume that is the case.
 With that assumption, their periods place them among the most rapidly
 rotating M dwarfs in Upper Sco (see also \S 7).

\item All have mid to late M spectral types (M4 to M6.5).
 
\item They have full amplitudes for their {\em K2} light curves of 2\% to 10\%.

\item Six of eleven of them show relatively stable phased light curve shapes
   over the entire 78 day {\em K2} campaign.

\item Five of them show evolving waveforms. In all but one case, there
  is an abrupt transition from one waveform to another with the switch
  happening in generally less than a day.   In three cases, the phase
  change takes place when a strong flare occurs.  In four of the five
  cases, the change in shape only significantly affects 20 to 30\% of
  the waveform, with the remaining waveform staying essentially
  constant in shape.  For EPIC 205110559, the fractional flux change
  between states at phase=0.1 is of order 4\% (Figure
  \ref{fig:epic205110559}, second panel).  Any physical mechanism to
  explain this light curve morphology must be able to accomodate
  events of this size.  For all three events with triggering flares,
  the affected portion of the light curve brightened after the flare.

\item The number of distinct peaks (or troughs) in their phased light curves
  ranges from 3 to 6.

\item There are 83 WTTs in our sample with P $<$ 0.65 days and  $(V -
K_s) >$ 5. The fraction that have scallop-shell light curve
morphologies for that period range is therefore 13\%.

\item We have also examined this WTT control group for obvious flares.  Based
on a simple visual check of the detrended light curves of our Table~1 stars
and our WTT control group, we find no obvious difference in either the frequency
or strength of flares.

\end{itemize}

\section{Stars with Short-Duration, Persistent Flux Dips}

There are eight stars in Table~\ref{tab:basicdata} in this category.
The phased light curves of these stars are dominated by shallow,
triangularly shaped flux dips, superposed either on a flat ``continuum"
or on a waveform consistent with that due to rotational modulation of
cool spots at the same period as the flux dips.   Four of these stars
show little or no evolution in their phased light curve morphology over
the time frame of the {\em K2} campaign; we describe one of these four
stars (EPIC 205374937) in detail.  The other four members of the
class show demonstrable evolution in their light curve morphology
over the {\em K2} campaign, and we discuss each case individually in
order to illustrate the nature of this evolution.

\subsection{EPIC 205374937}

\begin{figure}[ht]
\epsscale{0.6}
\plotone{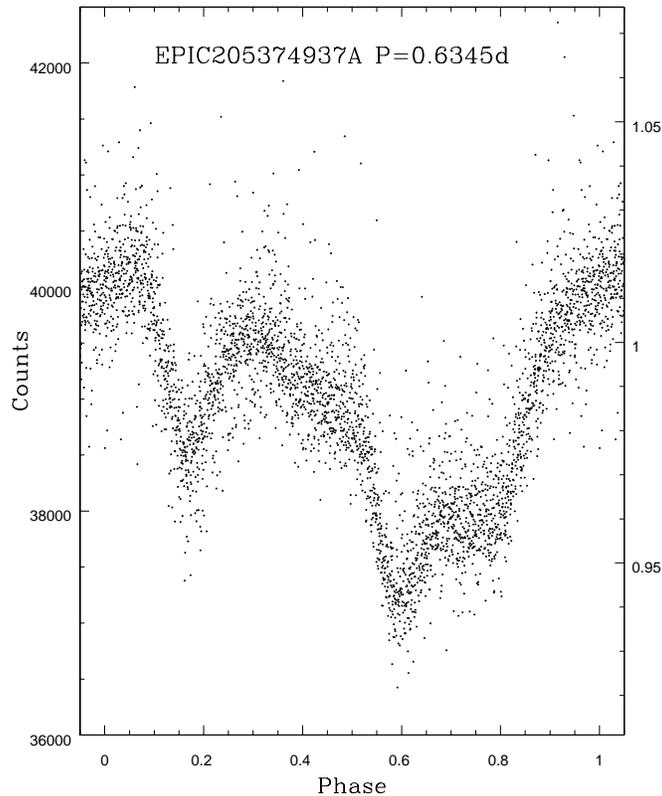}
\caption{The phased light curve of EPIC 205374937A.  The two most
prominent features are short duration, triangularly shaped flux dips
at phases of about 0.2 and 0.6.   These are superposed on a more
slowly varying waveform which is broadly consistent with the 
approximately sinusoidal shape often produced by non-axisymmetrically
distributed star spots.
\label{fig:epic205374937}}
\end{figure}

Preibisch \etal\ (2001) first identified EPIC 205374937 as a low mass
member of Upper Sco.  Carpenter \etal\ (2009) and Luhman \&
Mamajek (2012) both reported no IR excess.  Using Gemini North
adaptive optics (AO) imaging, Lafreniere \etal\ (2014) identified EPIC
205374937 as a nearly equal mass binary with a 0.1$\arcsec$
separation.   The {\em K2} light curve shows two approximately equal
strength close peaks, corresponding to periods of 0.6345 and 0.5436
days.  We assume the two nearly equal mass components seen with the AO
imaging give rise to these two periodogram peaks; we somewhat
arbitrarily assign the the larger periodogram peak as the A
component.  The phased light curve for the A component is shown in
Figure \ref{fig:epic205374937}.

We interpret the phased light curve for EPIC 205374937A as the
superposition of a more or less sinusoidal variation due to star spots
and two (or possibly three) short-duration flux dips.  The two most
prominent flux dips are centered at phases about 0.2 and 0.6; the full
width at zero intensity (FWZI) of the dips are 0.16 and 0.12 in phase
(or about 2 hours in time) and their depths are about 3\% of the
continuum.  There is no suggestion of the dips varying in depth or
width over the course of the campaign (that is, the noise level in the
phased light curve appears the same inside and outside the dips). 
The  period that phases the dips best also produces the least scatter
in the rest of the light curve, indicating that the dip period is
essentially the same as the spot period.   Two of the other stars in
this group (EPIC 204296148 and 203849738) also have both spotted star
waveforms and persistent flux dips where the periods are the same (see
Figure \ref{fig:batwing_medians}).  

\subsection{EPIC 203692610}

\begin{figure}[ht]
\epsscale{0.65}
\plotone{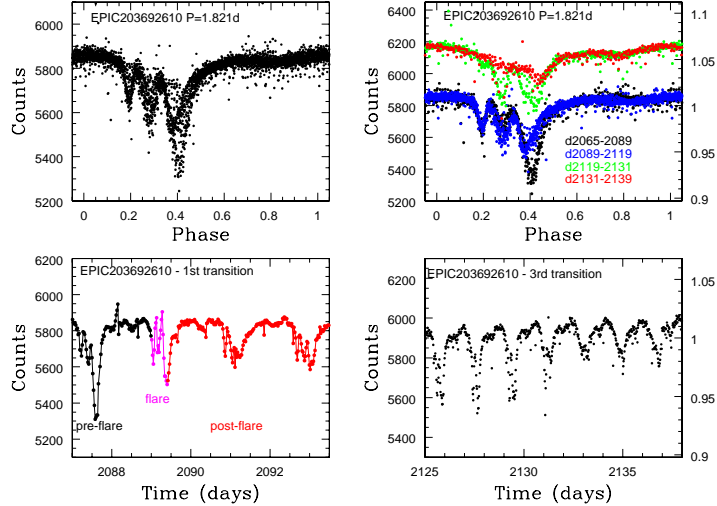}
\caption{(top left) The phased light curve of EPIC 203692610 showing data
for the full {\em K2} campaign, illustrating the three short-duration flux
dips that are present.  (top right) Phased light curve for EPIC 203692610,
with the data now broken up into four time windows: black (d2065-2089),
blue (d2089-2119), green (d2119-2131), and red (d2131-2139). The green and
red curves are shifted vertically by 300 counts.  The three
flux dips appear to evolve in shape with different timescales, with the
main changes appearing to happen in sudden events. (bottom left) Expanded
view of the light curve at the time of the first state-transition in the
flux dips, showing two possible flares at day 2089 and the much reduced 
amplitude of the flux dips after that day.  (bottom right) Expanded view
of the time period associated with the 3rd state-transition of the flux
dip depths (day 2131); no obvious flare or other trigger is present at
that time.  For the top-right panel, the normalized count rate values
(RHS of the plot) apply to the blue and black points.
\label{fig:epic203692610}}
\end{figure}

EPIC 203692610 is an essentially anonymous, red star with proper
motion consistent with Upper Sco membership.  Our own Keck HIRES
spectra for this star shows it to have a spectral type of about M4,
with weak H$\alpha$ emission (equivalent width = $-$3.0 \AA), and
strong lithium absorption (equivalent width = 0.54 \AA), typical of
Upper Sco M dwarfs.  The Lomb-Scargle periodogram for its {\em K2} light
curve shows a single strong peak at $P$ = 1.821 days\footnote{Because
there is a brighter star located about 13$\arcsec$ away, we have
analysed a light curve derived using a 1.5 pixel aperture for EPIC
203692610, rather than the 2 or 3 pixel radius used throughout the
rest of the paper.}.   When the light curve is phased to that period,
it shows a relatively flat continuum with three sharp, triangular
shaped flux dips (Figure \ref{fig:epic203692610}).   The dip depths
are about 3\% to 7\%, and the FWZI for each dip
is about 0.1 in phase.  Close examination of the light curve shows
that the dip depths vary in time, with the time dependence being
somewhat different for each dip, as illustrated in the top-right panel
of Figure \ref{fig:epic203692610}. The left-most dip is more or less
stable in its depth until day 2119, when it essentially disappears.  
The central dip is more or less stable in depth until day 2131, when
it too disappears.  The strongest (right-most) dip is quite deep until
day 2089; from 2089 to 2131 it is about half as deep (and somewhat
shifted in phase), and at day 2131 it becomes significantly weaker
again.  There are no strong flares in the {\em K2} light curve for this
star.  However, the strongest flare-like event (or events) occurs at
the time of the first state-transition of the flux dips at day 2089,
as illustrated in the bottom-left panel of  Figure
\ref{fig:epic203692610}.  There are no obvious flares of any sort at
the other two state-transitions.  Finally, we note that the dip
properties and the general appearance of the phased light curve for
EPIC 203692610 are quite similar to the {\em K2} light curve for HHJ 135,
one of the six late M dwarfs with persistent, short-duration flux dips
identified in the Pleiades by Rebull \etal\ (2016).

\subsection{EPIC 203962559}

\begin{figure}[ht]
\epsscale{0.6}
\plotone{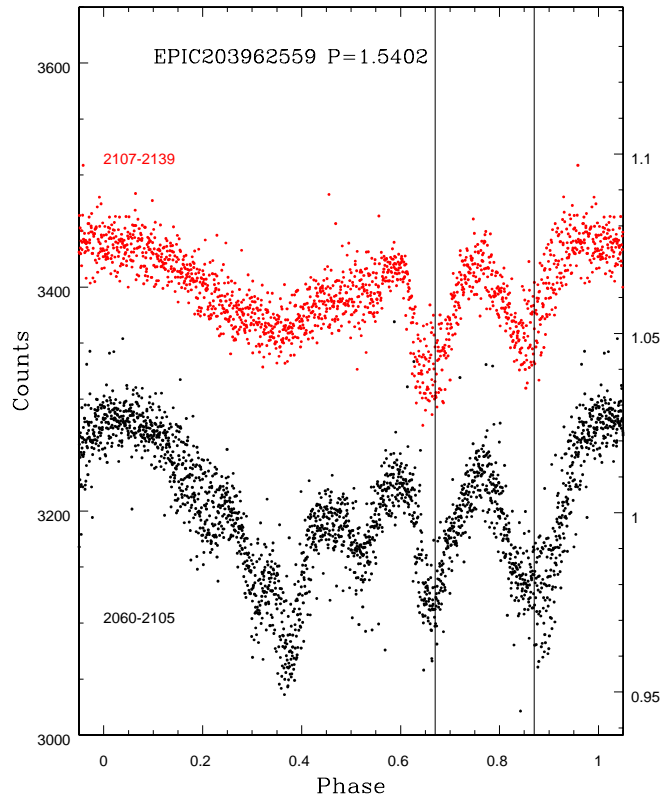}
\caption{The phased light curve of EPIC 203962559, illustrating the
abrupt change in shape at day 2106.  The largest change in waveform
shape is near phase=0.3, but the dip at phase $\sim$0.5 essentially
disappears and the dip at phase $\sim$0.85 weakens and possibly
shifts somewhat in phase.  The second time interval data have been offset
by 200 counts in the $y$-axis in order to more clearly show the
differing light curve shapes.   The RHS y-axis normalized count labels
are derived for the first time interval (the black points).
\label{fig:epic203962559}}
\end{figure}

While EPIC 203962559 has been identified as a $\rho$
Oph member (Wilking \etal\ 2005); it appears to have no IR excess and
has relatively weak H$\alpha$  emission, from which we conclude it is
a WTTS.  The Lomb-Scargle periodogram shows one strong, real peak at P
= 1.5402d.   When phased to that period and examined closely, the light
curve appears to undergo a major change in shape at about day 2106.  
This is illustrated in Figure \ref{fig:epic203962559}. Prior to that
day, there were four well-defined flux dips in the light curve; after
that day, only two flux dips remain, and one of those is significantly
narrower and less deep.   We are unable to determine if there was a
flare at day$\sim$2106 because that date also is when Mars transited the
{\em K2} FOV  (near the position of EPIC 203962559), corrupting its light
curve for $\sim$10 hours.  We can think of no possible way for the
Mars transit to induce systematic flux errors  in a specific phase range
in the extracted light curve for this star, so we assume this was
a bizarre coincidence.

\subsection{EPIC 203927435A}

\begin{figure}[ht]
\epsscale{0.65}
\plotone{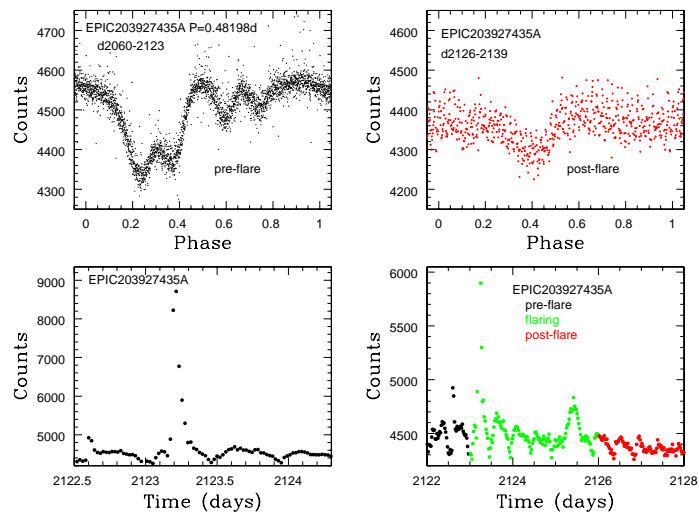}
\caption{The phased light curve of EPIC 203927435A, illustrating the
abrupt change in shape at day 2123, coincident in time with a strong
flare.  
\label{fig:epic203927435}}
\end{figure}

EPIC 203927435 has also been identified as a $\rho$ Oph member
(Wilking \etal 2005); it too appears to have no IR excess and
relatively weak H$\alpha$  emission (equivalent width $-$6.8 \AA).  
The Lomb-Scargle periodogram shows two resolved peaks, with periods of
0.4817 and 0.4162 days.   After removing the signal from the star with
$P$ = 0.4162 days (which we refer to as the B component), we find that
the phased light curve for the A component shows a waveform which is
not explainable by cold spots, but that also shows some evolution
during the {\em K2} campaign. The first panel of Figure
\ref{fig:epic203927435} shows the phased light curve for the first 63
days of the campaign; the light curve for the last 13 days of the
campaign is shown in the second panel of Figure
\ref{fig:epic203927435}.  A very strong flare occurred at the
transition point -- third panel of Figure \ref{fig:epic203927435} --
on day 2123.3; this was by far the strongest flare for this star
during the {\em K2} campaign.   Prior to that day, there were four
well-defined flux dips in the light curve; after the flare, only one
(weaker) dip remains.   In this case, the transition between states
took place over about 3 days.   During the middle of that transition
period, a second weaker flare-like event occurred; see the last panel
of Figure \ref{fig:epic203927435}. That event, however, has a
symmetrical shape (as was the case for the triggering events for EPIC
204918279 and EPIC 203185083).

\subsection{EPIC 204364515B}

\begin{figure}[ht]
\epsscale{0.65}
\plotone{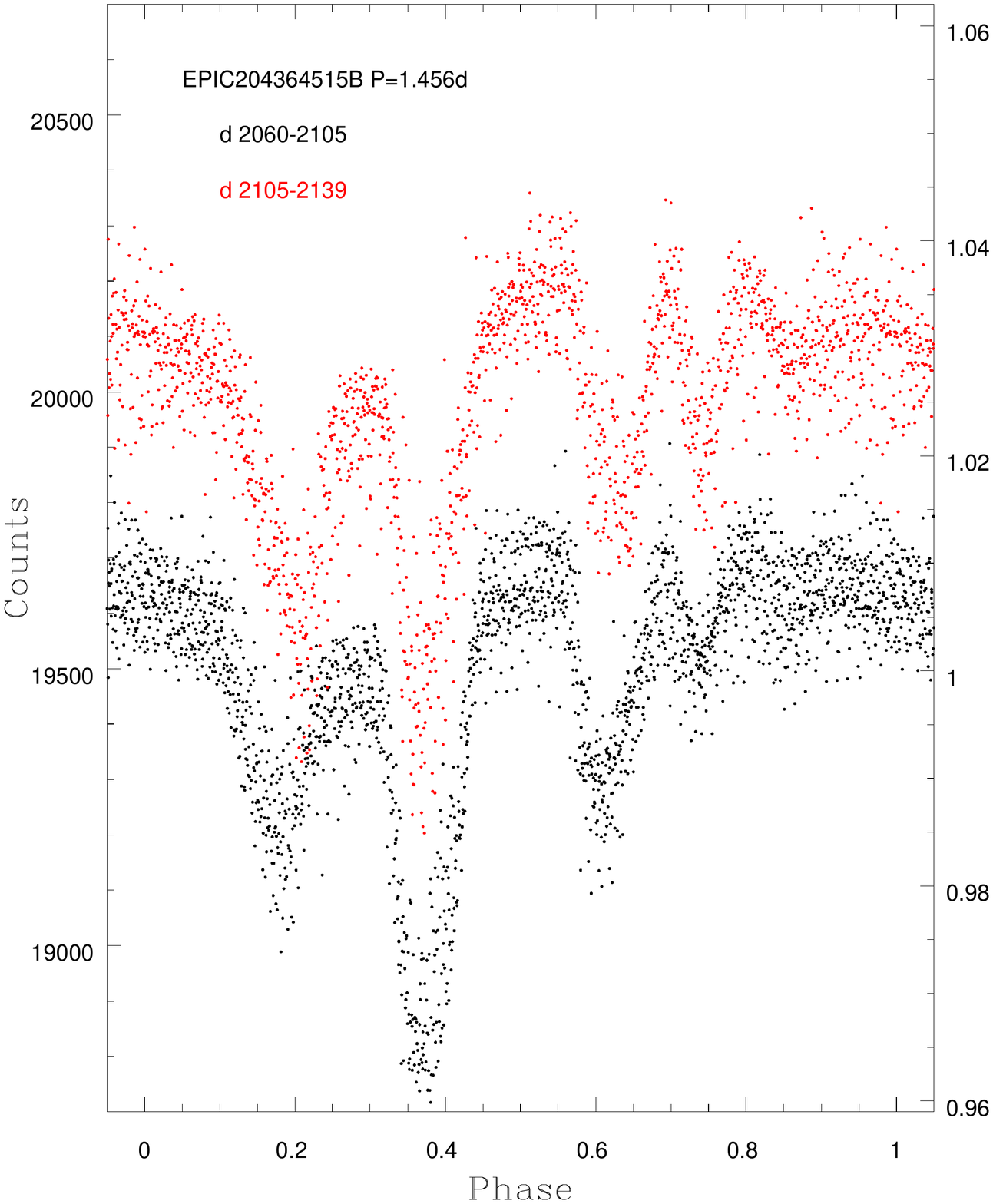}
\caption{The phased light curve of EPIC 204364515B.   There is little
change in the shape or depth of the central two dips over the {\em K2}
campaign.  The left-most dip appears to move to longer phase as
time progressed, and the right-most dip became stronger in the latter
half of the campaign. The RHS y-axis normalized count labels
are derived for the first time interval (the black points).
\label{fig:epic204364515B}}
\end{figure}

Bouy \etal\ (2006) showed that EPIC 204364515  is a possible resolved
triple system, consisting of a close, equal brightness pair with a
0.15$\arcsec$ separation and a fainter possible companion at a
separation of about 1$\arcsec$ (with its $\sim$4$\arcsec$ pixels, the
{\em K2} light curve includes light from all three stars).  The Lomb-Scargle
periodogram for EPIC 204364515 shows two dominant periods of 3.0863
and 1.456 days.   After removal of the light from the star with
$P$=3.086 days, the phased light curve for the star with $P$=1.456 days
shows four well-defined flux dips.  Figure \ref{fig:epic204364515B}
shows the phased light curve, where we have split the data into two
windows of time (black for day 2060-2105; red for day 2105-2139). The
two strongest flux dips appear to be approximately stable in position,
shape and depth.   The flux dip at phase $\sim$0.2, however, appears
to shift in phase by about 0.03 in phase (10 degrees in azimuth)
between the two windows.   The fourth flux dip at phase $\sim$0.75
appears to become considerably deeper in the second time window.

\subsection{Shared Characteristics of the Stars with Persistent Flux Dips}

We have measured widths and depths for all the prominent flux dips in
these eight stars; see Table~\ref{tab:persistentdata}.  The widths we
provide are full-width, zero-intensity values, FWZI, in units of
fraction of the period -- meaning that they are intended  to indicate
the entire duration of the event (rather than, for example, a FWHM).  
Most of the dips have FWZI of 0.1 to 0.15 in phase; for a typical star
with $P$ = 0.6 days, a FWZI of 0.12 corresponds to about 1.7 hours.  
The depths are the difference in counts between a nearby ``continuum"
level and the bottom of the dip relative to the continuum level,
expressed as a per cent. Most of the dips have depths of 1\% to
3.5\%, with the strongest being 7\% (the deepest dip for EPIC 203692610).   

\clearpage

\begin{deluxetable}{lcccc}
\tablecaption{Quantitative Measures for Persistent Flux Dips\label{tab:persistentdata}}
\tablehead{\colhead{EPIC ID} & \colhead{Dip Number}& \colhead{Phase}&
\colhead{FWZI\tablenotemark{a}} & \colhead{Dip Depth\tablenotemark{b} }}
\startdata
204296148A & 1  & 0.32  & 0.16  &  3.0  \\
204296148A & 2  & 0.62  & 0.11  &  1.4  \\
204364515B & 1  & 0.19  & 0.12  &  1.5  \\
204364515B & 2  & 0.38  & 0.11  &  3.1  \\
204364515B & 3  & 0.62  & 0.11  &  1.9  \\
204364515B & 4  & 0.74  & 0.07  &  0.9  \\
202873945  & 1  & 0.10  & 0.09  &  0.8  \\
202873945  & 2  & 0.42  & 0.10  &  2.3  \\
202873945  & 3  & 0.63  & 0.09  &  1.2  \\
203962559W1 & 1  & 0.37  & 0.08  &  3.2  \\
203962559W1 & 2  & 0.52  & 0.06  &  0.9  \\
203962559W1 & 3  & 0.67  & 0.11  &  3.1  \\
203962559W1 & 4  & 0.87  & 0.15  &  3.5  \\
203962559W2 & 1  & 0.65  & 0.12  &  3.2  \\
203962559W2 & 2  & 0.85  & 0.15  &  2.2  \\
203692610   &  1  & 0.19  & 0.06  & 2.5  \\
203692610   &  2  & 0.29  & 0.06  & 2.5  \\
203692610   &  3  & 0.41  & 0.13  & 7.5  \\
203849738   &  1  & 0.13  &  0.19  & 3.9  \\
203849738   &  2  & 0.79  &  0.07  & 1.2  \\
205374937A  &  1  & 0.17  &  0.16  &  3.2  \\
205374937A  &  2  & 0.60  &  0.12  &  3.0  \\
203927435A  &  1  & 0.26  & 0.2  &  4.5  \\
203927435A  &  2  & 0.39  & 0.2  &  3.6  \\
203927435A  &  3  & 0.63  & 0.14  &  1.4  \\
203927435A  &  4  & 0.78  & 0.14  &  1.3  \\
\enddata
\tablenotetext{a}{Full-width, zero-intensity values.}
\tablenotetext{b}{Note that for the stars which we interpret as likely physical
binaries because of the presence of two significant peaks in the Lomb-Scargle
periodogram (EPIC 204296148, 204364515, and 203927435), we do not know the relative
contribution to the total counts for the two stars.  We can therefore derive the
amplitude in counts for the flux dip, but when stated as a percent in depth this
is just a lower limit.  If the two stars are similar in brightness, then the
reported depths would be half of the true value.  For EPIC 203692610, the measured
dip properties are for the time window when the dips were strongest (day 2065 to 2089).}
\end{deluxetable}

Additional shared characteristics of these stars include:
\begin{itemize}

\item All eight of these stars have $P <$ 2 days.  Five of them have $P
  <$ 0.65 days. In three cases, the period derived for the flux dips
  agrees with the period inferred from a spotted-star waveform,
  confirming that the dip period is also the stellar rotation period. 
  For the other five, we believe it is likely that the dip period is
  the rotation period, but that is not proven.

\item All have mid to late M dwarf spectral types (M4 to M5.5).

\item The deepest flux dips for each star have depths from 2\% to 7\%; their
   widths (FWZI) are generally 0.1 to 0.15 in phase, corresponding to about 1 to 5 hours
   duration in time.

\item They have two to four discrete flux dips in their phased light
  curves (at the sensitivity level of the {\em K2} data).

\item For four of these stars, the dip depths appear fairly stable over the
  duration of the {\em K2} campaign.  In three of the other stars, one or more of the
  flux dips disappears abruptly (or becomes markedly weaker) during the course
  of the campaign.  At least two of the abrupt changes in flux dip depth occurs
  immediately following a flare (or flare-like event).  

\end{itemize}

\section{Stars with Short-Duration, Transient Flux Dips}

There are only four members of this class; we discuss each of them here.

\subsection{EPIC 205046529B}

\begin{figure}[ht]
\epsscale{0.6}
\plotone{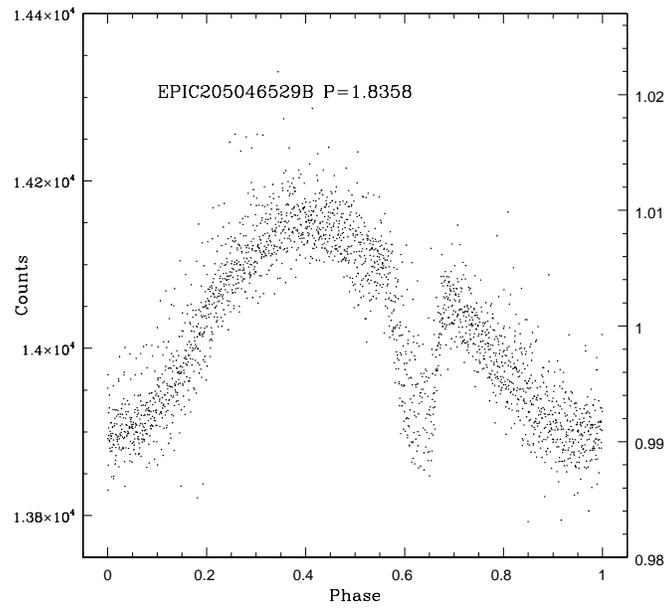}
\caption{The phased light curve of EPIC 205046529B.
The edges of the dip are quite sharp, but the interior of the dip
appears filled-in.   This is primarily because the dip depths vary
significantly with time -- as illustrated in the next figure.
\label{fig:epic205046529B}}
\end{figure}

\begin{figure}[ht]
\epsscale{0.65}
\plotone{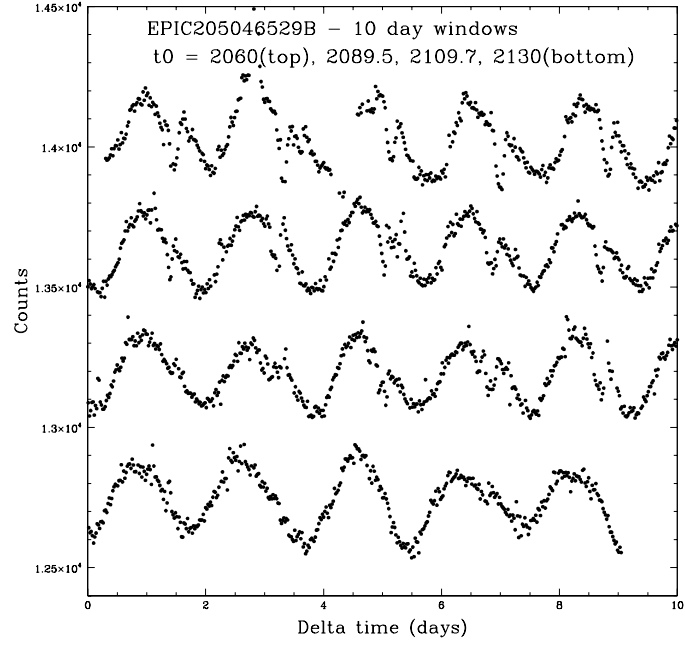}
\caption{Four 10 day segments of the Light curve of EPIC 205046529B.
The four segments (top to bottom) are d2060.0-2070.0; d2089.5-2099.5;
2109.7-2119.7; and d2130-2139; they are aligned in phase in x and 
displaced arbitrarily in y
in order to allow easy comparison of the flux dips.  
The dips are present but variable
in strength in the first three time windows, and completely absent
from the last time window.
\label{fig:epic205046529B.4win}}
\end{figure}

\begin{figure}[ht]
\epsscale{0.4}
\plotone{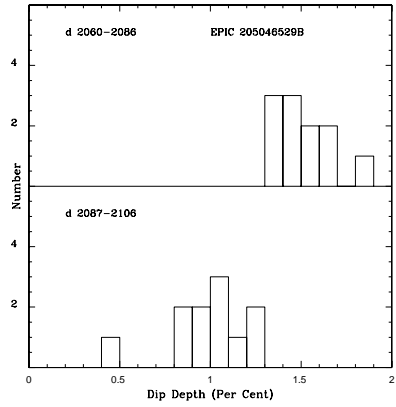}
\caption{Histograms of the dip depths for EPIC 205046529b for
two date ranges -- top: day 2060-2086; bottom: day 2087-2106.
After day 2106, the dip was either not detectably present or
was relatively weak.
\label{fig:epic205046529B.histo}}
\end{figure}

\begin{figure}[ht]
\epsscale{0.6}
\plotone{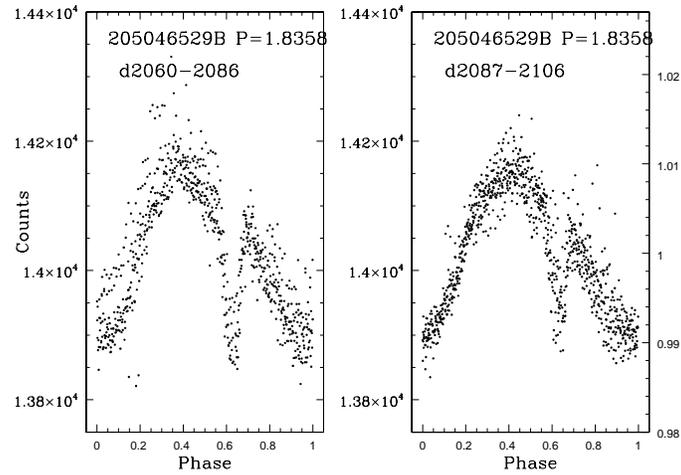}
\caption{Phased light curve for EPIC 205046529b, for the same
two time windows as used in Figure\ref{fig:epic205046529B.histo} --
left: day 2060-2086; right: day 2087-2106.
\label{fig:epic205046529B.phasedgrps}}
\end{figure}

The  Lomb-Scargle periodogram for EPIC 205046529 shows two strong peaks 
indicating it is a binary, with $P_1$
= 2.5619d and $P_2$ = 1.837d.   This star appeared as a close pair in
the Keck/HIRES guider camera, and we assume the two spatially resolved
stars seen with Keck correspond to the two periodogram peaks.
After removing the signal from the
longer period star and then phasing to 1.837 d, we noted the presence
of an apparent flux dip in the light curve for the second star.  After
further processing, we determined a best-fit period to highlight the
flux dip of $P$ =1.8358 days.  Figure \ref{fig:epic205046529B}  shows the
phased light curve;  the dominant feature is a spotted star waveform,
superposed on which is a well-defined flux dip centered at phase
$\sim$0.6.   The flux dip, however, appears somewhat noisy either due
to jitter in the individual dip depths or their timing or both.  We
believe this fuzziness is primarily due to variations in the depth of
the dip both on short and long timescales. 
Figure~\ref{fig:epic205046529B.4win} shows four 10 day windows in the
light curve for EPIC 205046529B.  Close examination of those traces
shows that the dip depth seems to decrease slowly with time over the
{\em K2} campaign, but also varies in depth from cycle to cycle.   In order
to determine better the nature of the variations in the dip times and
depths, we measured the depths and first-moment times of occurrence of
all the dips where we could identify a dip was present.   Those times
and depths are provided in Table~\ref{tab:205046529Bdepths}.
Histograms of the dip depths in two time windows (day 2060-2086 and
day 2087-2106) are shown in Figure \ref{fig:epic205046529B.histo},
demonstrating the decreasing dip depth with time.   After day 2107,
the flux dip was no longer detected with certainty. Figure
\ref{fig:epic205046529B.phasedgrps} shows the phased light curves for
those same two time windows, also illustrating the decrease in dip
strength as the {\em K2} campaign progressed.  For both time windows, the
dip profile appears to show a shallower ingress slope compared to the
egress slope.   For the stronger, better defined dips, we measure a
mean width (FWZI) of 0.195 days, or as a fraction of the period,
(dip duration)/period = 0.106.

\begin{deluxetable*}{lccc}
\tabletypesize{\scriptsize}
\tablecolumns{5}
\tablewidth{0pt}
\tablecaption{Times and Depths for Flux Dips in EPIC 205046529B\label{tab:205046529Bdepths}}
\tablehead{
\colhead{Date} & 
\colhead{Dip Depth (percent)} & 
\colhead{Date} &
\colhead{Dip Depth (percent)} 
}
\startdata
 2061.45 & 1.5  & 2087.18 & 1.2 \\
 2063.31 & 1.6  & 2089.05 & 0.4 \\
 2065.14 & 1.3  & 2090.89 & 1.0 \\
 2066.99 & 1.4  & 2092.69 & 0.8 \\
 2068.81 & 1.3  & 2094.52 & 1.0 \\
 2070.66 & 1.4  & 2096.34 & 0.8 \\
 2072.48 & 1.8  & 2098.19 & 0.9 \\
 2074.31 & 1.3  & 2100.04 & 0.9 \\
 2076.15 & 1.5  & 2101.87 & 1.0 \\
 2083.50 & 1.4  & 2103.70 & 1.2 \\
 2085.34 & 1.6  & 2105.52 & 1.1 \\ 
\enddata
\end{deluxetable*}

\clearpage

\subsection{EPIC 205024957}

\begin{figure}[ht]
\epsscale{0.6}
\plotone{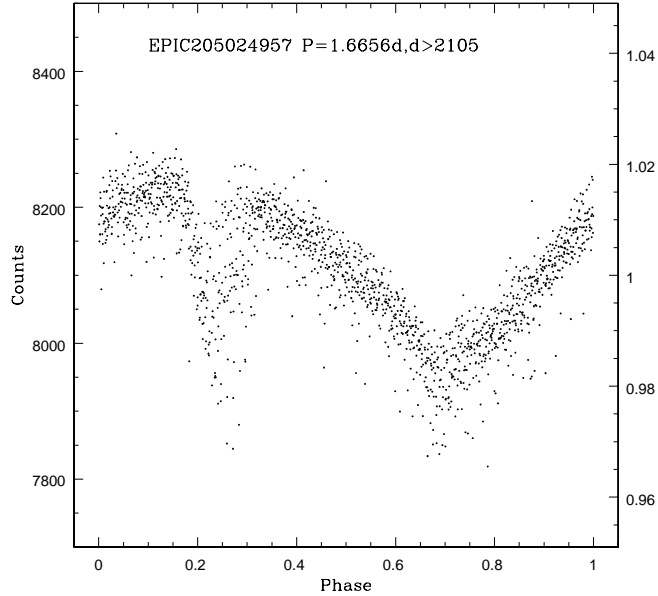}
\caption{The phased light curve of EPIC 205024957.   As for EPIC
205046529B, the edges of the dip are quite sharp but the interior of
the dip appears filled in.  This is again due primarily to rapid and
seemingly random variations in the dip depth with time, as illustrated
in the next figure.
\label{fig:epic205024957}}
\end{figure}

\begin{figure}[ht]
\epsscale{0.65}
\plotone{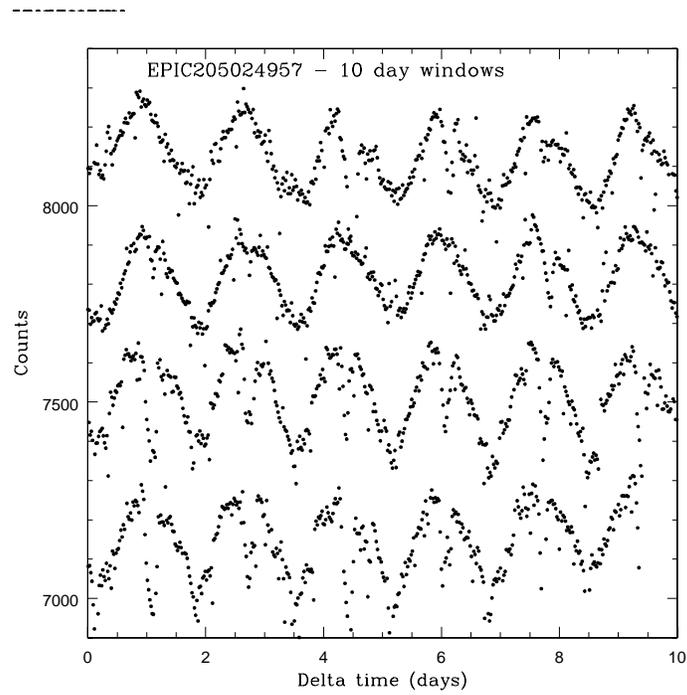}
\caption{Four 10 day segments of the light curve of EPIC 205024957.
From top to bottom, the four time windows shown are d2068.0-2078.0;
d2083.0-2093.0; d2119.7-2129.7; d2129.7-2139; the light curve segments
are aligned in phase in x and shifted arbitrariy in y in order to
highlight the evolution of the flux dips.   The dip depths vary
greatly on both short and long timescales.
\label{fig:epic205024957.4win}}
\end{figure}

\begin{figure}[ht]
\epsscale{0.4}
\plotone{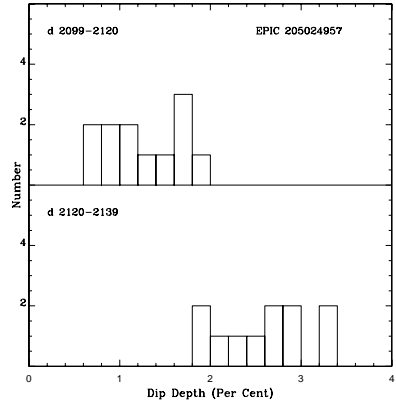}
\caption{Histograms of the dip depths for EPIC 205024957 for
two date ranges -- top: day 2099-2120; bottom: day 2120-2139.
Prior to day 2099, the dip was either not detectably present or
was relatively weak.
\label{fig:epic205024957.histo}}
\end{figure}

\begin{figure}[ht]
\epsscale{0.6}
\plotone{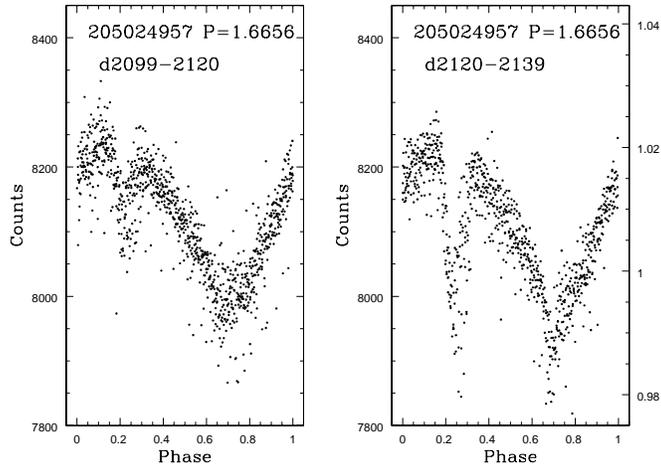}
\caption{Phased light curve for EPIC 205024957, for the same
two time windows as used in Figure~\ref{fig:epic205024957.histo} --
left: day 2099-2120; right: day 2120-2139.
\label{fig:epic205024957.phasedgrps}}
\end{figure}

EPIC 205024957 is the one star in our sample that appears to
have a significant 24 $\mu$m IR excess; however, as originally
concluded by Carpenter \etal\ (2009), the lack of an
excess at shorter IR wavelengths indicates that this is likely
a debris disk and not a primordial disk.  Its weak H$\alpha$ emission
(Table 1) also supports it being a WTTS and not a CTTS.
The Lomb-Scargle periodogram for EPIC 205024957
shows a single strong peak at $P$ = 1.6672d, largely driven
by a nearly sinusoidal spotted-star waveform; however, the phased
light curve also shows a definite but fuzzy flux dip (Figure
\ref{fig:epic205024957}).  We have analysed the light curve for this
star in the same way as for EPIC 205046529, and reach similar
conclusions.  That is, we believe the fuzzy appearance of the flux dip
is due to both short and long term variations in the dip depth, with
little contribution from variations in the occurrence times for the
dips.  Figure \ref{fig:epic205024957.4win} illustrates the short and
long term variability of the flux dip depths with four 10 day portions of
its light curve.   Based on Gaussian fits to the well-detected flux
dips, we derive a best period for the dips of $P$=1.6656d. For the
stronger, better defined dips, we measure a mean dip width (FWZI) of
0.203 days, or as a fraction of the period, (dip duration)/period =
0.122. Times and depths for the individual flux dips we could measure
are provided in Table~\ref{tab:205024957depths}.  As for EPIC 205046529,
in Figure \ref{fig:epic205024957.histo}
and Figure \ref{fig:epic205024957.phasedgrps}
we also provide histograms of the dip depths and phased light curves
for EPIC 205024957 in two time windows to illustrate the evolution of
the dip character during the {\em K2} campaign.

\clearpage

\begin{deluxetable*}{lccc}
\tabletypesize{\scriptsize}
\tablecolumns{5}
\tablewidth{0pt}
\tablecaption{Times and Depths for Flux Dips in EPIC 205024957\label{tab:205024957depths}}
\tablehead{
\colhead{DATE} & 
\colhead{Dip Depth (percent)} & 
\colhead{DATE} &
\colhead{Dip Depth (percent)} 
}
\startdata
 2072.48 & 4.3 &  2115.75 & 1.7 \\
 2074.13 & 2.5 &  2117.43 & 1.6 \\
 2080.78 & 1.0 &  2119.06 & 1.9 \\
 2084.11 & 0.9 &  2120.78 & 1.8 \\
 2090.83 & 1.6 &  2122.45 & 1.9 \\
 2095.9  & 1.0 &  2124.08 & 2.3 \\
 2099.08 & 0.8 &  2125.77 & 2.1 \\
 2100.75 & 0.7 &  2127.44 & 2.8 \\
 2102.43 & 1.0 &  2129.08 & 3.2 \\
 2104.1  & 0.7 &  2130.78 & 3.2 \\
 2105.75 & 1.6 &  2132.43 & 2.5 \\
 2109.1  & 1.5 &  2134.12 & 2.6 \\
 2110.8  & 0.9 &  2135.73 & 2.6 \\
 2112.5  & 1.0 &  2137.42 & 2.9 \\
 2114.1  & 1.3 &          &     \\
\enddata
\end{deluxetable*}

\subsection{EPIC 204143627}

\begin{figure}[ht]
\epsscale{0.6}
\plotone{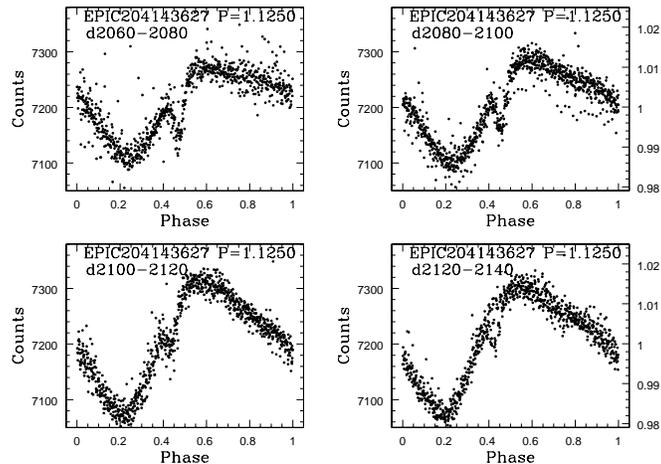}
\caption{The phased light curve of EPIC 204143627.   
The four panels show successive 20 day sections of the day,
all phased to a period of 1.125 days.  The flux dip at phase
about 0.45 became less distinct and shallower as the campaign
progressed.
\label{fig:epic204143627}}
\end{figure}

The Lomb-Scargle periodogram for EPIC 204143627 shows a single strong
peak at $P$ = 1.125 days, largely driven by a nearly sinusoidal
spotted-star waveform; however, the phased light curve also shows a
definite but weak flux dip (Figure \ref{fig:epic204143627}).  We have
analysed the light curve for this star in the same way as for EPIC
205046529, and reach similar though more limited  conclusions due to
the lower S/N for this star.  That is, we believe the fuzzy appearance
of the flux dip is due to variations in the dip depth, with little
contribution from variations in the occurrence times for the dips. 
The four panels of Figure \ref{fig:epic204143627}  show that the flux
dip was present and well-defined in the first 20 days of the {\em K2}
campaign, and slowly diminished in depth over the duration of the
campaign.  For the first time window in Figure
\ref{fig:epic204143627}, we measure a dip width (FWZI) of 0.06 days,
or relative to the period, (dip duration)/period = 0.055.

\subsection{EPIC 205483258 (RIK-210)}

\begin{figure}[ht]
\epsscale{0.65}
\plotone{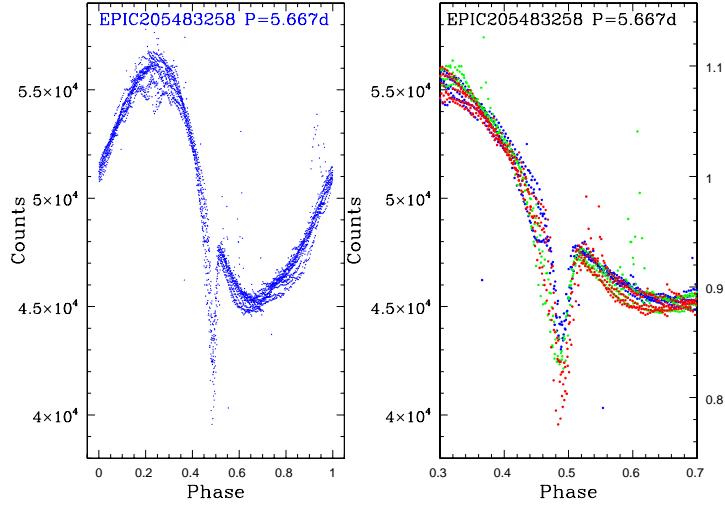}
\caption{(left) The phased light curve of EPIC 205483258 (aka RIK-210).   (right) An
expanded view of the light curve, highlighting the narrow flux dip. 
Different colors correspond to different time windows -- illustrating
that the shape of the flux dip varies significantly on short
timescales.
\label{fig:epic205483258}}
\end{figure}

EPIC 205483258 is a relatively anonymous (only two references 
in SIMBAD as of this date) Upper Sco member.  Its literature name is RIK-210
(Rizzuto \etal\ 2015).  It is both the earliest type star in Table 1
and has the longest period.  Its phased light curve is dominated by a
well-defined, stable sinusoidal variation presumably due to a highly
non-axisymmetric spot distribution.   Superposed on that waveform is a
single strong, narrow flux dip -- as shown in
Figure~\ref{fig:epic205483258}.   The dip width (FWZI) varies from
epoch to epoch, but is about 0.4 to 0.5 days, or relative to the
period, (dip duration)/period $\sim$ 0.08.  The dips are generally
asymmetric, with a shallower and more variable ingress profile.

We have much more followup data for RIK-210
than for any of our other stars.  In David \etal\ (2016c),
we provide a detailed examination of the {\em K2} light curve for RIK-210
and an analysis of synoptic spectra for it which we obtained with
Keck HIRES.  Radial velocites at 15 epochs exclude any EB scenario
to explain the narrow flux dips.   The variability in the shape, depth
and duration of its flux dip can only be explained by some type
of cloud of material orbiting the star at the Keplerian co-rotation
radius and with physical dimension similar in size to the stellar
radius.  We return to the implications of these data in \S 8.

\subsection{Shared Characteristics of the Stars with Transient Flux Dips}

All of the stars in this class share these characteristics:
\begin{itemize}

\item All four stars have $P >$ 1 day.  In all four cases, there is
  a well defined spotted star waveform in addition to the flux dip, and both
  share the same period to within the uncertainties.  Therefore, for these
  stars, we are certain that the flux dip period is the same as the rotation
  period.

\item  All have M spectral types (M2.5 to M5).

\item  In all cases, there is only one major flux dip per star.  The maximum depths
  are 1.5\%, 2.5\%, 3\%, and 20\% (the latter for RIK-210).  The widths (full
  width zero intensity) are of order 0.1 in phase or slightly less, corresponding
  to about 10 hours for RIK-210 and to a few hours for the other three stars.

\item The dip depths vary significantly on both short (cycle to cycle) and
  long (campaign duration) timescales.  From one cycle to the next, the dip
  can go from present and near maximum depth to absent.

\item In one case for sure, and probably in two of the other cases, the
  dip shape is asymmetric, with the slope of the ingress being steeper
  or shallower than the slope of the egress.

\item Of all 23 stars discussed in this paper, only one has a relatively
secure IR excess and that is EPIC 205024957.

\end{itemize}

\section{Further Characterization}

\begin{figure}[ht]
\epsscale{0.8}
\plotone{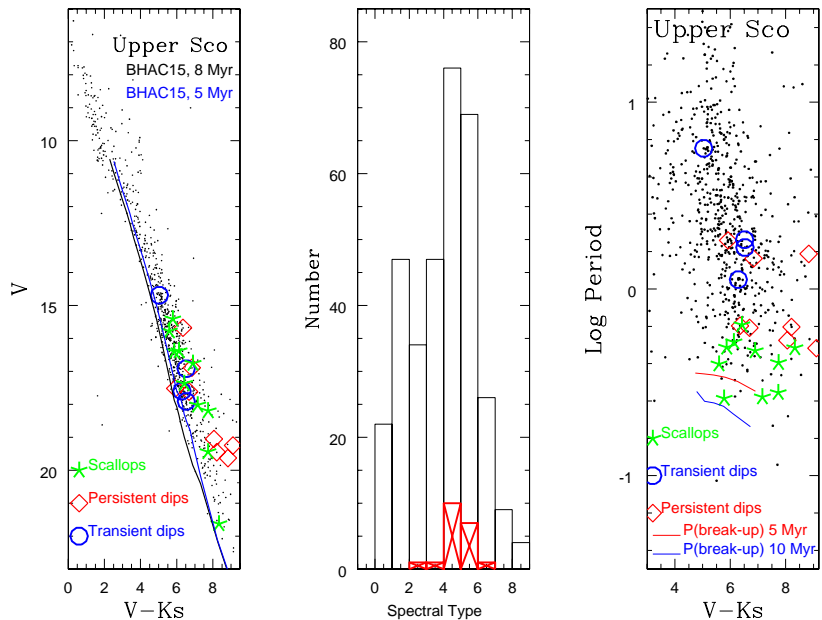}
\caption{(left) $V$ versus $(V-K_s)$ color-magnitude diagram for the
Upper Sco WTTs with {\em K2} light curves.  All of the stars from this paper
are in the lower half of the diagram (i.e., mid to late M dwarfs), but
we see no striking differences among our three groups.  
(middle) Spectral type histogram for the Upper Sco members with known
spectral types (not shown are FGK stars).   The x-axis are M subtypes
(e.g. ``0" is M0; "8" is M8).
The stars with unusual light curves from this paper are
shown in red; they preferentially occur among the stars with spectral
types M4 and later.  (right) Log Period versus $V-K_s$ diagram for the
same set of stars.  The most obvious correlation here is that the
stars with scallop-shell light curve morphology have among the
shortest periods in Upper Sco; they also have systematically shorter
periods than our other two classes. Some of their periods are very
near or at the theoretical breakup period for an age of 5-10 Myr.
\label{fig:UScoCMD.Pvmk}}
\end{figure}

How do the stars we have discussed in the previous three sections
compare to other low mass Upper Sco WTTs in terms of their basic
physical properties?  Figure \ref{fig:UScoCMD.Pvmk} provides three
such comparisons -- a color-magnitude diagram, a spectral type
histogram, and a period-color diagram.   The color-magnitude diagram
shows that despite their unusual light curves, our stars do not appear
to have unusual locations in the $V$ vs.\ $V-K_s$ CMD, other than
their being concentrated to colors corresponding to mid to late M
spectral types. There is also no obvious  displacement of one of our
groups relative to the other two groups in this diagram  (as might
have been true if they sampled different age or mass ranges). 
Spectral types are available for about a third of the 1100 WTTS with
{\em K2} light curves, and for $>$80\%\ of the stars in Table 1.  The
spectral type histogram reinforces the conclusion from the CMD that
our stars are concentrated to mid to late M dwarfs.  A very large
fraction of the {\em K2} Upper Sco stars are M dwarfs, so the fact that our
stars are M dwarfs is not unexpected.   However, 40\%\ of the full
sample have spectral types earlier than M4, whereas that is true for
$<$ 10\%\ of  the stars from Table 1 (i.e., if drawn at random, we
would have expected eight of our 20 stars with spectral type to have
to be early M, vs.\ only two that we observe).

Conversely, the period-color diagram does show striking correlations.
Most importantly we believe, this diagram emphasizes that the stars
with scallop-shell light curves segregate to the most rapidly rotating
stars in Upper Sco.  Several of them have periods quite close to
break-up at ages appropriate for Upper Sco.  The persistent,
short-duration flux dip class are also relatively rapidly rotating on
average, but less so than for the scallop-shells. The stars with
transient short-duration flux dips have periods that are essentially
typical of Upper Sco members for their $V-K_s$ color.

\begin{figure}[ht]
\epsscale{0.65}
\plotone{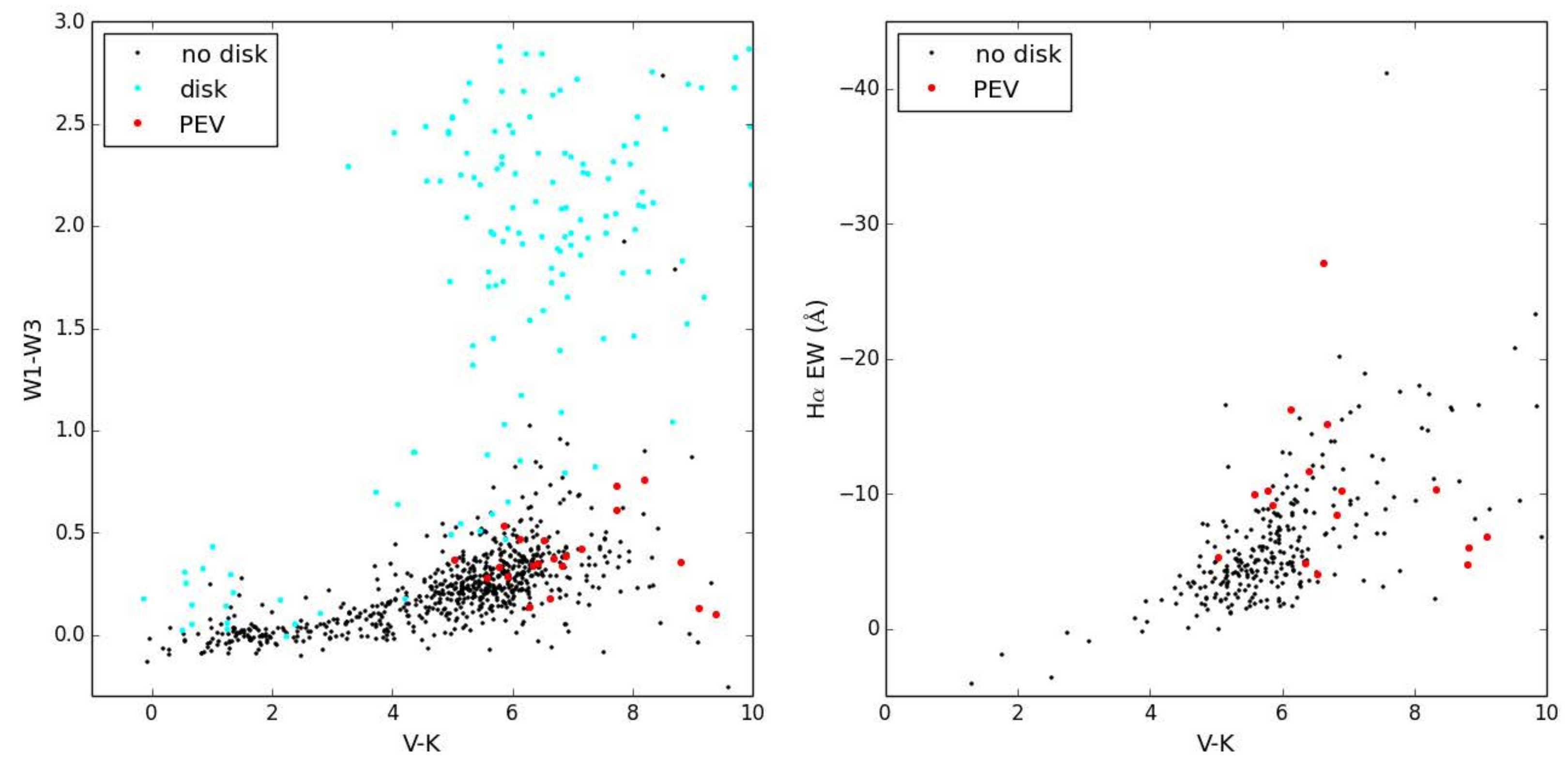}
\caption{(left) WISE $W1-W3$ color vs. $V-K_s$ for Upper Sco members
with and without primordial disks (Cody \etal\ 2017b) and the stars from
Table~1.  The stars without IR excesses are shown as black dots; the
stars with strong IR excesses are shown as cyan dots, and the Table~1
stars are shown as red dots.  The figure legend refers to the Table~1 stars
as PEVs, or Periodic Exospheric Variables.  The PEV stars appear
to track well the locus of the other WTTs.  (right)  Plot of H$\alpha$
equivalent width vs. $V-K_s$ for the stars without IR excesses and the
PEV stars.  The H$\alpha$ data comes from Preibisch \etal\ (2001),
Slesnick \etal\ (2008), Lodieu \etal\ (2011), Rizzuto \etal\ (2015),
and from Table~1.   The PEV stars are again normal for their $V-K_s$ color,
with one exception (EPIC 203185083). 
\label{fig:UScoIRexc.Halpha}}
\end{figure}

Figure \ref{fig:UScoIRexc.Halpha} provides two additional plots, both
of which relate to whether the M dwarfs with unusual phased light curves
in Upper Sco have significant signposts of circumstellar disks or active
accretion from a disk.  The left panel of Figure \ref{fig:UScoIRexc.Halpha}
compares the $[W1]-[W3]$ (3.5 \mum\ $-$ 12.0 \mum) color of our stars versus
the other Upper Sco stars with and without primordial disks.  We label
our stars as ``PEV" in this diagram, for Periodic Exospheric Variables; for
the moment, this is simply a conveniently short alias.  The PEV stars
appear to be quite normal for their $V-K_s$ color, and very well separated
from the stars of the same mass with primordial disks.  The right panel
of Figure \ref{fig:UScoIRexc.Halpha} compares the H$\alpha$ equivalent
widths for the PEV stars to the other stars without obvious IR excesses.
With one exception, the PEV stars are again unremarkable.  The one
exception is EPIC 203185083 ($V-K_s$ = 6.6, H$\alpha$ = $-$27).  Our 
conclusion from this is that the types of variability associated with
active accretion from a close-in disk (accretion hot spots, flux dips
due to an inner warped disk or dust entrained in an accretion stream) are
unlikely to be relevant for our stars.

\begin{figure}[ht]
\epsscale{0.65}
\plotone{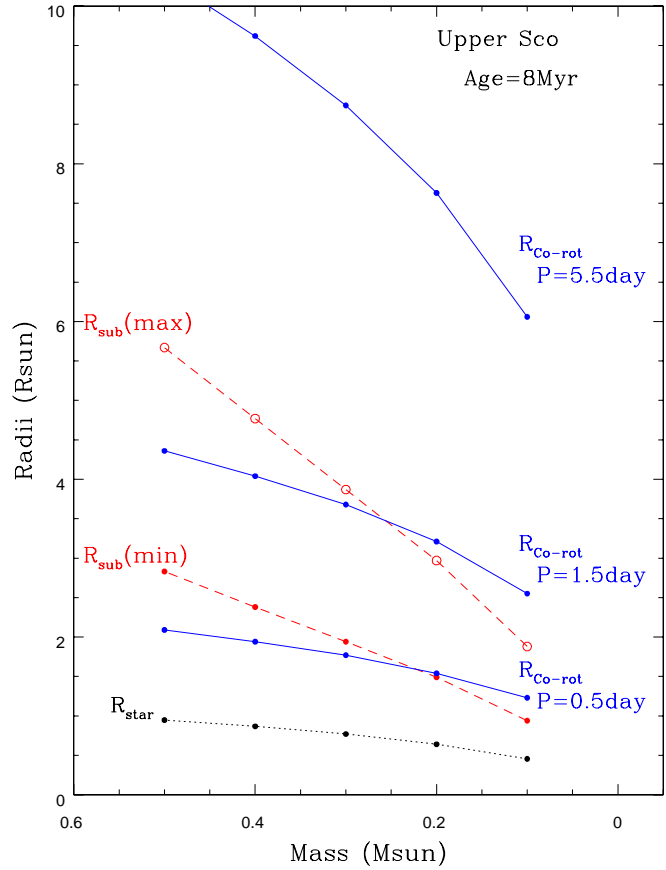}
\caption{Stellar radius, Keplerian co-rotation radius and 
estimated sublimation radii using extremes of grain properties from
Monnier \& Millan-Gabet (2002), for an assumed 8 Myr age and
using BHAC15 models.   The Keplerian co-rotation radius is provided
for three rotation periods -- 0.5, 1.5 and 5.5 days -- which
encompasses most of the period range found for the stars discussed
in this paper.   The black dotted curve provides estimated stellar radii
as a function of mass; the blue solid curves are the estimated Keplerian
corotation radii for three rotation periods that are representative of
our Upper Sco stars; and the red dashed curves illustrate approximate
sublimation radii as a function of stellar mass.
\label{fig:Mdwarf_radii}}
\end{figure}

It is also useful to compare the size of our stars to their Keplerian
co-rotation radii and to distances from the star where grains would be
expected to sublimate.   That comparison is provided in  Figure
\ref{fig:Mdwarf_radii}.   For the calculations, we have adopted the
Baraffe \etal\ (2015; hereafter BHAC15) evolutionary models to provide
stellar luminosities and radii at 8 Myr.    The radii in those models
may be underestimated because they do not include ``radius-inflation"
due to magnetic fields or spots  (MacDonald \& Mullan 2013; Jeffries
\etal\ 2016; Somers \& Stassun 2016),  but we believe the radii are
adequate for our current purposes.  We use the formula in Monnier \&
Millan-Gabet (2002) to determine sublimation radii, with the two
curves corresponding to minimum and maximum $Q_R$ values of 1.0 and
4.0, corresponding to a plausible range of sizes for silicate or
carbon grains. Based on the spectral types in Table 1, most of our
stars should have masses about 0.3 \msun.   The BHAC15 radius for such
a star is 0.8 \rsun\ at 8 Myr; the corresponding Keplerian co-rotation
radii for 0.5, 1.5 and 5.5 days are 1.8, 3.7 and 8.7 \rsun.  The
curves in  Figure \ref{fig:Mdwarf_radii} show that for the stars with
period near 0.5 days, dust is unlikely to be able to form or survive;
dust may be able to survive at the Keplerian co-rotation radius for
the stars with $P\sim$1.5 days; and dust would certainly be able to
form and survive for extended periods for RIK-210, the star with
$P\sim$5.5 days.

For the persistent and transient short-duration flux dip stars, the
duration, shape and depth of the flux dip encodes information on the size
and nature of the object that is transiting (or being eclipsed by) the star,
if that is the correct physical explanation for their variability.
For a point source at the Keplerian co-rotation radius ($R_{co}$)  whose orbit passes 
through the Earth-star line, the
duration of a transit (or eclipse) is just:
\begin{equation}
{\rm duration} = \frac{2 R_*}{(2 \pi R_{co})} P_{\rm rot} 
\end{equation}
For radii and masses in solar units and rotation periods in days, the 
durations are just:
\begin{equation}
{\rm duration} = 0.076 R_* (P_{\rm rot}/M_*)^{0.333}
\end{equation}
For an adopted age of 8 Myr, our stars have masses from the BHAC15 models
ranging from about 0.2 to 0.6 \msun, and corresponding radii of 0.64 to
1.02 \rsun.

\begin{figure}[ht]
\epsscale{0.65}
\plotone{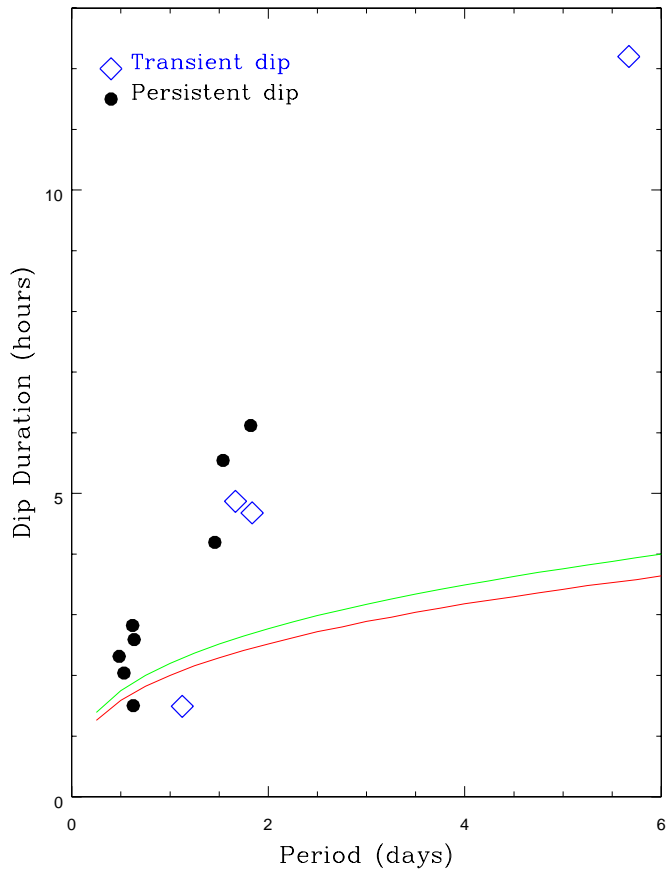}
\caption{Measured maximum flux dip duration for our Upper Sco M
dwarfs compared to the predicted (maximum) duration for a flux dip due
to a point-source object transiting its parent star (for an age of 8
Myr and mass of 0.2 and 0.6 \msun - encompassing the mass range of
our stars).   Some of the dips are consistent with
the prediction, implying a relatively small size, whereas other dips
-- particularly for the stars with longer rotation periods  -- suggest sizes
comparable  to that of the star.  RIK-210 is the outlier to the
upper-right of the diagram.
\label{fig:dip_durations}}
\end{figure}

Figure \ref{fig:dip_durations} plots the dip durations we have
measured in \S 5 and \S 6 for the short-duration flux dip stars as a
function of their periods, and compares those data to the duration
predicted for a point source for stars with mass of 0.2 and 0.6 \msun.   
For the shortest periods, most of the
stars have their maximum observed dip durations
consistent with the prediction for a point source; for periods longer
than about 1.5 days,
however, the observed durations are significantly longer than the point
source prediction,
indicating that the occulting (occulted) object has size comparable to
the star.  The relationship between dip duration and period appears 
tight based on the limited amount of data that exists, but is clearly 
much steeper than our simple Keplerian model.

To illustrate this point in another way, we have estimated stellar
radii for each of the persistent and transient flux dip stars.  For
that, we have used our spectral types and estimated $V-K_s$ color and 
Table 6 of Pecaut \& Mamajek (2013) to provide intrinsic $T_{\rm eff}$, 
$BC_J$, and $V-K_s$ color.   Those data and the Stefan-Boltzmann law yield
direct stellar radius estimates; for stars identified as binaries 
in Table~\ref{tab:basicdata}, we have divided the system luminosity by two
prior to calculating the radius.  We then derive estimates of the
Keplerian co-rotation radius, adopting masses from our estimated
luminosities and the BHAC15 8 Myr isochrone.  The maximum duration
for a point source at the Keplerian co-rotation radius then follows
from the formula above.  Table~\ref{tab:dip_durations} provides our
estimated stellar radii and compares the predicted maximum dip durations
to the longest duration flux dips for each star.  The uncertainties 
in the stellar radius estimates,
and hence in the durations, from propagating uncertainties in the
distance and age and in the stellar properties are likely 15-20\%
(see, for example, similar calculations for individual planet-hosting
stars by Mann \etal\ 2016 and David \etal\ 2016c).
Because radii two to three times as large
as we estimate would be required for the dip durations to be consistent
with observations for several of our stars (including 3 of 4 of the 
transient dip class), we conclude that radius inflation due to magnetic
effects (e.g., MacDonald \& Mullan 2017) cannot explain these discrepancies
and instead that the transiting/occulting body must have size 
comparable to the star.

\begin{deluxetable*}{lccccccc}
\tabletypesize{\scriptsize}
\tablecolumns{8}
\tablewidth{0pt}
\tablecaption{Comparison of Dip Durations to Predictions for a Point Source
   at the Keplerian Co-Rotation Radius\label{tab:dip_durations}}
\tablehead{
\colhead{EPIC\tablenotemark{a}} & 
\colhead{Spectral Type} & 
\colhead{Stellar} &
\colhead{Rotation} &
\colhead{Keplerian} &
\colhead{Pt. Src.} &
\colhead{Max Observed} &
\colhead{Class} \\
\colhead{} &
\colhead{} &
\colhead{Radius} &
\colhead{Period} &
\colhead{Co-Rotation} &
\colhead{Duration} &
\colhead{Duration} &
\colhead{} \\
\colhead{} &
\colhead{} &
\colhead{(R$_{\odot}$)} &
\colhead{(days)} &
\colhead{Radius (R$_{\odot}$)} &
\colhead{(hours)} &
\colhead{(hours)} &
\colhead{} 
}
\startdata
204364515*  & M4.0 & 1.07 & 1.46 & 4.13 & 2.90 & 4.2 & Pers  \\
203849738  & M5.5 & 0.92 & 0.62 & 1.78 & 2.45 & 2.8 & Pers \\
203692610  & M4.0 & 0.62 & 1.82 & 4.79 & 1.80 & 6.1 &   Pers \\
205374937*  & M4.0 & 1.34 & 0.63 & 2.36 & 2.72 & 2.6 &   Pers \\
203962559  & M4.0 & 1.28 & 1.54 & 4.28 & 3.52 & 5.5 &   Pers \\
203927435*  & M4.5 & 1.20 & 0.48 & 1.81 & 2.43 & 2.3 &   Pers \\
205024957  & M5.0 & 0.76 & 1.67 & 3.72 & 2.61 & 8.0 &   Trans \\
205046529*  & M4.0 & 0.88 & 1.84 & 4.82 & 2.57 & 8.6 &   Trans \\
204143627  & M5.0 & 0.75 & 1.13 & 2.87 & 2.26 & 1.6 &   Trans \\
205483258  & M2.5 & 1.30 & 5.67 & 11.24 & 5.01 & 12.2 &  Trans \\
\enddata
\tablenotetext{a}{Stars marked with an asterisk have two peaks in their periodograms,
indicating this is a likely binary, but with unknown relative brightnesses.  In order
to derive radii for the stars in these systems, we have divided their luminosity
estimates by 2, corresponding to a reduction in their estimated radii and dip
durations by 1.414.}
\end{deluxetable*}

The shape of the flux dips can also provide qualitative information on
the nature of the occulting body.   The roughly triangular flux dips for our transient
flux dip group are roughly similar in shape to the flux dip shapes found
in many eclipsing binaries (EBs).  In most EBs, the flux dips are very
stable in their depth and shape and the dip profile is symmetric.  However, 
if the primary of the EB were heavily spotted, the flux dips could be
asymmetric if the transit path crosses regions of higher or lower spot
coverage.   If, in addition, the EB orbit is precessing and the spots
are changing on timescales comparable to the orbital period, flux dips
with variable depth and shape could result.  Our four transient flux dip
stars are indeed heavily spotted. However, the spotted-star waveforms are very stable
over the entire {\em K2} campaign.  Therefore, even a precessing EB orbit would
at most produce changes in the dip shape and depth that vary smoothly
with time, which is not the case for our stars.  

The combination of dip depths of a few percent, and the above
arguments from durations and shapes therefore suggests that the
occulting (occulted) bodies are likely to have sizes comparable to
that of the star and optical depths (or surface brightnesses or
filling factors) that are small.

One can also draw inferences as to the physical mechanism(s) driving
the variability in our group of stars from the presence or absence of
similar light curve morphologies in surveys of the light curves of
other samples of M dwarfs.  First, to our knowledge none of the
$\sim$3000 M dwarfs in the original {\em Kepler} field show light curve
morphologies like those in our three groups (McQuillen, Aigrain, \&
Mazeh 2013).   This likely arises both from the relative paucity of
mid to late M dwarfs in the {\em Kepler} sample, and to a paucity of young
and/or rapidly rotating M dwarfs.  Second, {\em K2} light curves have been
obtained for several hundred mid to late M dwarfs in both the Pleiades
(age $\sim$ 125 Myr) and Praesepe (age $\sim$ 700 Myr) open
clusters.   We (Rebull \etal\ 2016) found six late M dwarfs in the
Pleiades with persistent, short-duration flux dips; we find no such stars 
in the older Praesepe cluster.   We have found no examples of
scallop-shell waveforms nor transient, short-duration flux dips in
either the Pleiades or Praesepe.\footnote{In the same mass range as
for Upper Sco, there are about 250 Pleiades stars with {\em K2} light curves
and period $<$ 0.65 d, and about 400 Pleiades stars with period $<$
2 d. For Praesepe, in that same mass range, there are about 50
stars with period $<$ 0.65 d and about 200 stars with period $<$ 2
d.}   We discuss the implications of these results in the next
section.

\section{Physical Interpretation}

Light curves from {\em CoRoT}, {\em Kepler}, and {\em K2} have identified a small but
growing population of stars with photometric variability that
often seems mysterious and often takes the form of short-duration
flux dips.   This small menagerie of objects includes pre-main sequence
stars (e.g., Cody \etal\ 2014), but also includes main sequence stars
(e.g., Rappaport \etal\ 2012, Boyajian \etal\ 2016) and even white
dwarfs (Vanderburg \etal\ 2015).  Possible physical mechanisms for
producing the short-duration flux dips range from occultations by
warped or structured circumstellar disks to dust streaming away
from a disintegrating close-in planet (as well as other even
more speculative thoughts).. 
In many cases, a consensus has not yet been reached
as to which mechanism is responsible for the observed variability.

Our goal for this paper has been to add a new set of objects to
this menagerie, and to provide detailed empirical data for these stars
in order to begin the process of categorizing their variability and
placing them in the context of the existing objects with similar
properties.   In this section, we briefly consider several broad
categories of physical models as possible mechanism to explain our
light curves.   We expect that this is just the beginning of a 
lengthy process, to which we hope many other groups will contribute.

\subsection{Could Cool Spots Be Responsible for the Variability Signatures
   We Have Identified?}

Non-axisymmetrically distributed spots on the surfaces of low mass
stars typically produce relatively simple and smooth phased light
curve shapes with one or two maxima and one or two minima per
period.  However, it is not inconceivable that some combination of
spot sizes and locations might produce light curve signatures similar
to our unusual Upper Sco stars.   In order to address this hypothesis,
we have created a spotted star modeling routine and have produced simulated
light curves for a variety of spot configurations.  We describe the
spot simulation program and some of the tests we have run in the 
Appendix.  Based on those simulations,
it {\em is} possible to get light curve shapes that are very similar
to the scallop-shell and persistent dip classes if one puts spots very
close to the upper (or lower) limb of the star as viewed from Earth.
However, the spot distributions that can mimic the light curve shapes
{\em cannot} also yield light curve amplitudes that match our observations.
For several of the scallop-shells, the full amplitude is of order
10\%, and for a number of the persistent flux dip class, the dip
depths are between 3 and 8\% with FWZI $<$0.2 in phase.
In order to yield the very structured phased light curves of the
scallop-shell class, or the very narrow (in phase) flux dips of the
persistent dip class, the spots have to be {\it{very}\rm} near the edge of the
visible hemisphere of the star.  Because of the geometric fore-shortening
and limb darkening when at the edge of the visible hemisphere, such
spots yield amplitudes or dip depths less than 1\%, even if the spots are
assumed to be totally black.   Larger spots or spots further from the
``terminator" yield light curves that are smoother and do not match
the morphologies we see.   Therefore, we believe that cool spots on
the stellar surface do not provide a suitable model to explain the
light curve morphologies for our Upper Sco stars.

\subsection{Could Variable Extinction from a Primordial Disk 
   Explain Some of our Light Curves?}

There is some similarity between the light curves of some of the
stars in Table 1 and some of the stars identified in recent years as ``dippers"
in Orion, NGC~2264 and Upper Sco (Morales-Calderon \etal\ 2011; Cody \etal\ 2014;  Stauffer
\etal\ 2015; Ansdell \etal\ 2016); these stars are also sometimes referred to as AA Tau
analogs (Bouvier \etal\ 1999; Alencar \etal\ 2010; McGinnis \etal\ 2015).  
The light curves of AA Tau stars show a wide diversity, with most of them
showing very broad and structured flux dips which are quite dissimilar from
our Upper Sco stars and thus do not serve as plausible comparison objects.
A small subset of them, however, do
have periodic, short-duration flux dips sometimes superposed on sinusoidal
variations with the same period (Stauffer \etal\ 2015), thus superficially
linking their light curve morphology with that of the stars we have discussed
in this paper.   For these stars, the periodic short-duration flux dips have
generally been attributed to extinction by localized structures in or above the star's
inner circumstellar disk, where the inner disk rim and stellar photosphere are
assumed to be locked into at least approximate co-rotation.   Could some/many of the stars
in Table 1 simply be another subset of this class, with the same basic physical
processes producing the observing photometric variability?

We believe the answer to that question is primarily no, based on one major difference
between our stars and the stars of the dipper class.   All but two of the 40 dippers in
Orion identified by Morales-Calderon \etal\ and all of the dippers identified
in NGC~2264 by Cody \etal\ and by Stauffer \etal\ have strong IR excesses and
SEDs appropriate for Class I or II YSOs.  All of the dippers in Upper Sco identified
by Ansdell \etal\ (2016) have strong 24 $\mu$m\ excesses, and all but one of them have
strong 12 $\mu$m\ excesses.   Those IR excesses arise from warm 
dust at or near the inner edge of their primordial circumstellar disks.   As
discussed in \S 2.2 and in the Appendix, with one exception our stars do not have
IR excesses and have SEDs consistent with bare photospheres.  Also, for most of 
our stars (with P $<$ 1 day), the Keplerian co-rotation radius lies interior to
the likely dust sublimation radius, and therefore any ``cloud" responsible for
the periodic light curve variability should be dust free.

The CoRoT light curves for NGC~2264 are similar in quality to the {\em K2} light curves
for Upper Sco, and CoRoT light curves were obtained for more than 300 WTT cluster
members (about twice as many WTTs as CTTs); however, none of those WTTs showed
light curve morphologies like our Upper Sco stars in Table~1.   Could that absence
somehow affect the conclusion we have just reached?  Probably not.  NGC~2264 is
five times further away than Upper Sco, and the CoRoT limiting magnitude is brighter
than for {\em K2}.  Therefore, the sample of stars with good light curves in NGC~2264
contains very few stars in the mass range of the stars in our Table~1 (Venuti
\etal\ 2016 include only two stars with $P <$ 1 day and mass $<$ 0.4 \msun\ in
their table of NGC~2264 rotation periods from CoRoT).
To the extent that the light curve signatures we see require young ages, rapid
rotation, and low masses, we would not expect to have found them in NGC~2264 because
we were not able to sample the needed mass range.

\subsection{Could Dusty Debris from Young Forming or Disintegrating Planets
     Explain the Variability for Some of our Stars?}

As discussed in \S 7, \S 8.2, and in David \etal\ (2016c), the transient short-duration
flux dips found for four of our stars cannot be produced by cool spots
on the stellar photosphere or by any other plausible phenomenon associated
with the surface of the star.  The equality of the period for the
narrow flux dip and the sinusoidal waveform (which we assume {\it{is}\rm} due to spots)
therefore places the material responsible for the narrow flux dips at
the Keplerian co-rotation radius.  That material must be distributed over a region
with size comparable to the star (Figure~\ref{fig:dip_durations} and
Table~\ref{tab:dip_durations}), but have low average optical depth given flux
dip depths generally just a few percent.   Several recently discovered stars have
light curve properties that seem reasonably analogous to our 
four transient flux dip stars.  The
closest analog is PTF 08-8695 (van Eyken \etal\ 2012; Johns-Krull \etal\ 2016), 
an M3 star in the $\sim$3 Myr old 25 Ori subgroup of
Orion.  Light curves for PTF 08-8695 show narrow flux dips with depths up to
3\%, FWZI $\sim$0.18 in phase, and a period of 0.48 days .   
The flux dip depths are variable and at some
epochs are not present at all.  The dips are superposed on a more or less
sinusoidal spotted star waveform with identical period to that for the
flux dips.  These properties are remarkably similar to our four Upper Sco transient
flux dip sources.  Van Eyken \etal\ (2012) concluded that the best interpretation
of the data was to ascribe the flux dips to transits of a recently formed
or forming planet undergoing Roche-lobe overflow mass loss.

Two main sequence field stars with {\em Kepler} or {\em K2} data have light curve
properties that make them somewhat analogous to our transient flux dip stars.
KIC 12557548 (Rappaport \etal\ 2012) is a 16th magnitude, field K dwarf with
a rotation period $P \sim$ 23 days and hence a likely age of a few Gyr.  The
{\em Kepler} data show it to have asymmetric flux dips with a period of 15.7 hours;
the flux dips vary irregularly in depth, with a maximum depth of about 1.3\%.
The FWZI of the dips is about 0.15 in phase.  {\em K2}-22b (Sanchis-Ojeda \etal\ 2015)
is a field M0 dwarf with a rotation period $P \sim$ 15 days and hence a likely
age of about a Gyr.  Its {\em K2} light curve shows asymmetric flux dips with a
period of about 9.1 hours; the flux dips are highly variable in depth, with
dip depths from zero to 1.3\%.  The FWZI of the dips is about 0.09 in phase.
The flux dips for both stars have been modeled by the above authors 
as arising from a planet that is in the process of disintegrating.
The surface of each planet would have temperatures near 2000 K.  
In the Rappaport \etal\ (2012) model, a
dusty wind from the surface generates a dust stream/tail that is responsible for
the transits, with the curved geometry of the tail explaining
the asymmetric dip profiles and their variability.

If the model proposed for KIC 1255b and {\em K2}-22b is correct, those planets are
at a kind of fiery Goldilocks distance from their star -- close enough for their
surfaces to be molten, but not so close that dust grains could not survive
in their planetary wind stream long enough to create a tail of sufficient
extent to explain the transit duration and asymmetric shape.  For our
four transient flux dip stars, using the parameters in Table~\ref{tab:dip_durations},
the surface temperature of a hypothetical planet would be in the range
1200-1500 K, and thus the specific mechanism proposed for KIC 1255b and
{\em K2}-22b is not likely applicable.  However, we believe the similarities in
light curve properties between these two stars and our four stars make a
compelling case to ascribe the flux dips for our stars to dust associated
with a planet.  In our case, the distance of the planet from the star is
not a function of the star's luminosity but is instead a function of its
rotation period and mass.  Planets at about that location could arise from
orbital migration within a circumstellar disk, and then having that migration
stop when the planet reaches the inner rim of the disk -- which is expected
to be close to the Keplerian co-rotation radius for the star (Papaloizou 2007).
What is not obvious is the mechanism for creation of the dusty cloud that
would cause the transits for our stars.   Several possibilities are explored 
in David \etal (2016c).

\subsection{Could Scattered/Emitted Light from Circumstellar Gas Clouds
   Explain Some of our Light Morphologies?}

The salient features of the persistent flux dip class are multiple short-duration
flux dips with FWZI 0.1 to 0.15 in phase and depths of a few percent. Periods
are always less than two days, usually less than 1 day.   Long term behavior
is akin to the puntuated equilibrium models in evolutionary theory -- long
periods of stability, but often with abrupt shifts where the 
depth of one or more dips changes significantly.  At some level, the scallop-shell
class is just a more extreme version of the persistent flux dip class
-- more
structure in the phased light curve, shorter periods (all $<$ 0.65 days),
the same or more frequent abrupt shifts in light curve morphology.  We therefore
hope to find a single physical mechanism that could explain the light
curve variability of both classes.

The closest analog to these stars we can find in the literature is HHJ 135,
in the Pleiades (Rebull \etal\ 2016).   HHJ 135 is a 125 Myr, dM4.5 star;
its {\em K2} data yields a period of 0.61 days and a phased light curve showing
three short-duration flux dips with a very similar appearance to our Upper
Sco persistent flux dip stars EPIC 203692610 and EPIC 204364515.  This
primarily tells us that extreme youth (low gravity) is not
required for this morpological signature.  

The next best analogy in the literature we can find for these stars is,
surprisingly, an 8 \msun\ star.  $\sigma$\ Ori E is a rapidly rotating
Be star.   Ground-based and MOST light curves for $\sigma$\ Ori E show it
to have two short-duration flux dips (widths about 0.2 in phase) superposed
on a wave-shaped undulation with a period of 1.19 days (Townsend \etal\ 2013).  
The light curve shape is apparently stable over many years.  The MOST
light curve closely ressembles that for our persistent flux dip stars
EPIC 204296148 and EPIC 205374937.  Townsend \& Owocki (2005) have proposed
that the $\sigma$\ Ori E light curve is a natural outcome for
a rapidly rotating star with a wind and a strong magnetic dipole field.  Gas from the wind
accumulates in a lumpy warped disk with inner-rim at the Keplerian co-rotation radius,
oriented so that the normal to the disk lies between the rotation and
magnetic axes.  The short-duration flux dips in the light curve arise
when the disk transits our line of sight.   Their model (see also ud-Doula
\etal\ 2006) also predicts break-out events, resulting when too much
mass accumulates onto the field lines.  These magnetic reconnections could
ressemble flares (which are otherwise not predicted for B stars).

While the Townsend \& Owacki (2005) model is for a high mass star 
with a radiative wind, it turns out that a very similar model had
been proposed to explain some properties of rapidly rotating low
mass stars.  Optical spectra of AB Doradus itself ($P$ = 0.51 days; Collier Cameron \& Robinson 1989a,b) and
Speedy Mic (HD 197890; $P$ = 0.38 days; Jeffries 1993, Dunstone \etal\ 2006) 
and other similar rapidly rotating K 
dwarfs show narrow H$\alpha$\ absorption dips which sweep through
their broad, quiescent H$\alpha$\ profiles, best explained as due to orbiting warm
gas clouds at or above the Keplerian co-rotation radius that transit the
observers line of sight.  Some Be stars show similar absorption dip
transients (Smith \& Balona 2006).  In AB Dor, often as many as six individual
features (i.e., individual clouds) are seen at a given epoch.  Many of
the H$\alpha$\ transients repeat from night to night for periods up to
the end of a week-long campaign (Dunstone \etal\ 2006).  
Similar features have also recently been detected in a rapidly-rotating 
(P=0.44 days) M dwarf, V374 Peg (Vida \etal\, 2016).  Jardine 
\& van Ballegooijen (2005) have proposed a model for these systems in which
gas from the stellar (Parker) wind accumulates in a lumpy torus at or
above the Keplerian co-rotation radius where net gravitational and magnetic
forces vanish, with density maxima at the places where the radial
component of the magnetic field
changes polarity.  These density maxima could be the ``clouds" that give
rise to the light curve structure we see in the persistent flux dip
and scallop-shell light curves.  Note that in this model, the
flux dips in the {\em K2} data arise when one of these glowing clouds is
eclipsed by the star, rather than when the cloud transits the face of
the star and produces the H$\alpha$ absorption transients.   
The scallop-shells may differ from
the persistent flux dip class because they have an additional source of
supra-photospheric gas -- gas lost through their equatorial plane because
they are rotating at breakup velocity.  

The above speculations are at most proof of concept.  The next step
is to determine if the gas in these structures have densities, ionization
fractions and temperatures consistent with that needed to generate
the light curve features we see.   We intend to address those issues
in a future paper (Collier Cameron et al. 2017, in prep).

\section{Summary and Conclusions}

By targeting the Upper Sco association, the second campaign of the {\em K2}
mission provided the largest collection of high quality light curves
for low mass pre-main-sequence stars ever obtained.  Those data made
it possible to search for rare light curve types which only occur to a
significant degree at very young ages and very low masses.  We have,
in fact, identified 23 stars with light curve morphologies which have
few or no previous counterparts in the literature.    We have attempted to
characterize these light curves as well as
possible and to identify correlations or shared characteristics among
these stars and their light curves that  might point to the physical
mechanism responsible for  creating their unusual light curve
morphologies.

The most striking shared characteristics of the 19 stars in our
two main groups are:
\begin{itemize}
\item all have mid-to-late M dwarf spectral types;
\item more than half of the stars have rotation periods within a
  factor of two of the predicted breakup period;
\item eight of these stars show rapid ``state-transitions" in their
  light curve morphology, in many cases apparently triggered by a strong
  flare;
\item the short-duration flux dips and highly structured phased light curves
   seem incompatible with star-spots as the driver of the photometric variability.
\end{itemize}
Warm gas clouds organized into a structured, toroidal shape located 
at the Keplerian co-rotation radius seems to be the most plausible
explanation for these characteristics.   A discussion of a detailed
physical model for creating and maintaining these clouds and for their
producing our observed light curve morphologies is provided in Collier Cameron
\etal\ (2017, in prep).

The four other Upper Sco stars we have identified have single
short-duration flux dips superposed on normal spotted-star light
curves, with the dip period and spot period being identical.  The flux
dips have highly variable depths and shapes, with shapes that are
often asymmetric.  These characteristics seem better matched to some
type of dusty debris, orbiting a young planet located at or
near the Keplerian co-rotation radius.   A detailed examination of the
best studied of this group (RIK-210) and a comparison of those data to
a variety of physical models is provided in David \etal\ (2016c).

While the unusual light curve morphologies we find in Upper Sco may
only be present for a small fraction of stars during a brief time
interval in their early evolution, understanding these stars may have
more far reaching implications.    If an AB Dor slingshot prominence
type model  is found to be appropriate for the scallop-shell and
persistent flux dip morphologies, then analysis of those stars may
provide important new clues to the magnetic field topologies, winds
and mass loss rates of low mass stars.  These data could also help
explain the comparatively uniform and rapid rotation found for most
field brown dwarfs.  If a young planet is present near the Keplerian
co-rotation radius for the four transient flux-dip stars, that also
could have significant implications for models of planet formation,
particularly around M dwarfs.

\begin{acknowledgements}

We thank Leslie Hebb for timely advice on debugging and testing our spot
simulation routine.  We also thank Kevin Covey, Lee Hartmann, Nuria
Calvet, Steve Saar and Jeremy Drake for helpful discussions of these
objects.

Some of the data presented in this paper were obtained from the
Mikulski Archive for Space Telescopes (MAST). Support for MAST for
non-HST data is provided by the NASA Office of Space Science via grant
NNX09AF08G and by other grants and contracts. This paper includes data
collected by the {\em Kepler} mission. Funding for the {\em Kepler} mission is
provided by the NASA Science Mission directorate. This research has
made use of the NASA/IPAC Infrared Science Archive (IRSA), which is
operated by the Jet Propulsion Laboratory, California Institute of
Technology, under contract with the National Aeronautics and Space
Administration. This research has made use of NASA's Astrophysics Data
System (ADS) Abstract Service, and of the SIMBAD database, operated at
CDS, Strasbourg, France. This research has made use of data products
from the Two Micron All-Sky Survey (2MASS), which is a joint project
of the University of Massachusetts and the Infrared Processing and
Analysis Center, funded by the National Aeronautics and Space
Administration and the National Science Foundation. The 2MASS data are
served by the NASA/IPAC Infrared Science Archive, which is operated by
the Jet Propulsion Laboratory, California Institute of Technology,
under contract with the National Aeronautics and Space Administration.
This publication makes use of data products from the Wide-field
Infrared Survey Explorer, which is a joint project of the University
of California, Los Angeles, and the Jet Propulsion
Laboratory/California Institute of Technology, funded by the National
Aeronautics and Space Administration.
\end{acknowledgements}

\facility{K2}

\clearpage

\appendix

\section{Spectral Energy Distributions}

\begin{figure}[ht]
\epsscale{0.65}
\plotone{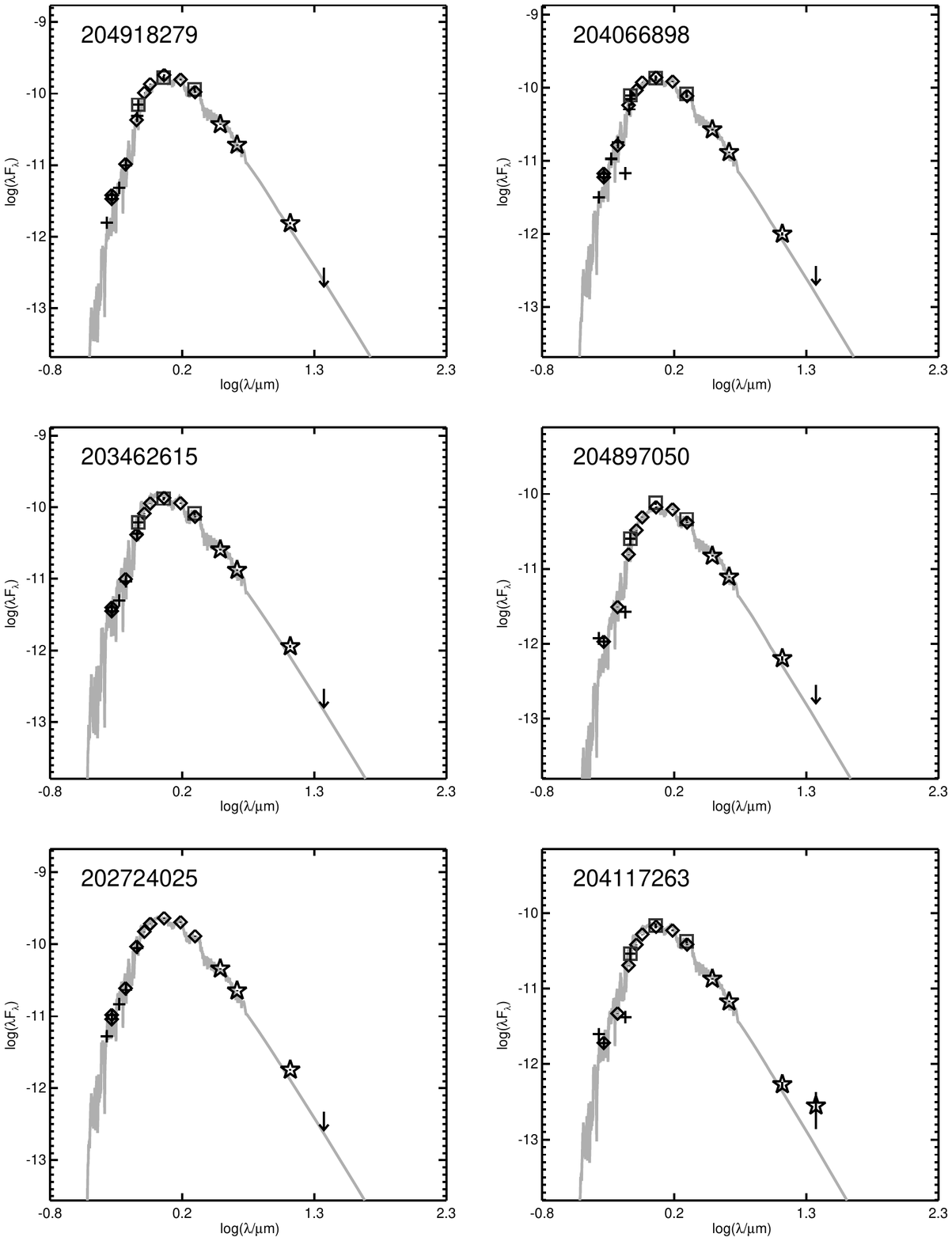}
\caption{Spectral energy distributions for the first six stars in Table 1.
See text for a description of the symbols.
\label{fig:SEDS1}}
\end{figure}

\begin{figure}[ht]
\epsscale{0.65}
\plotone{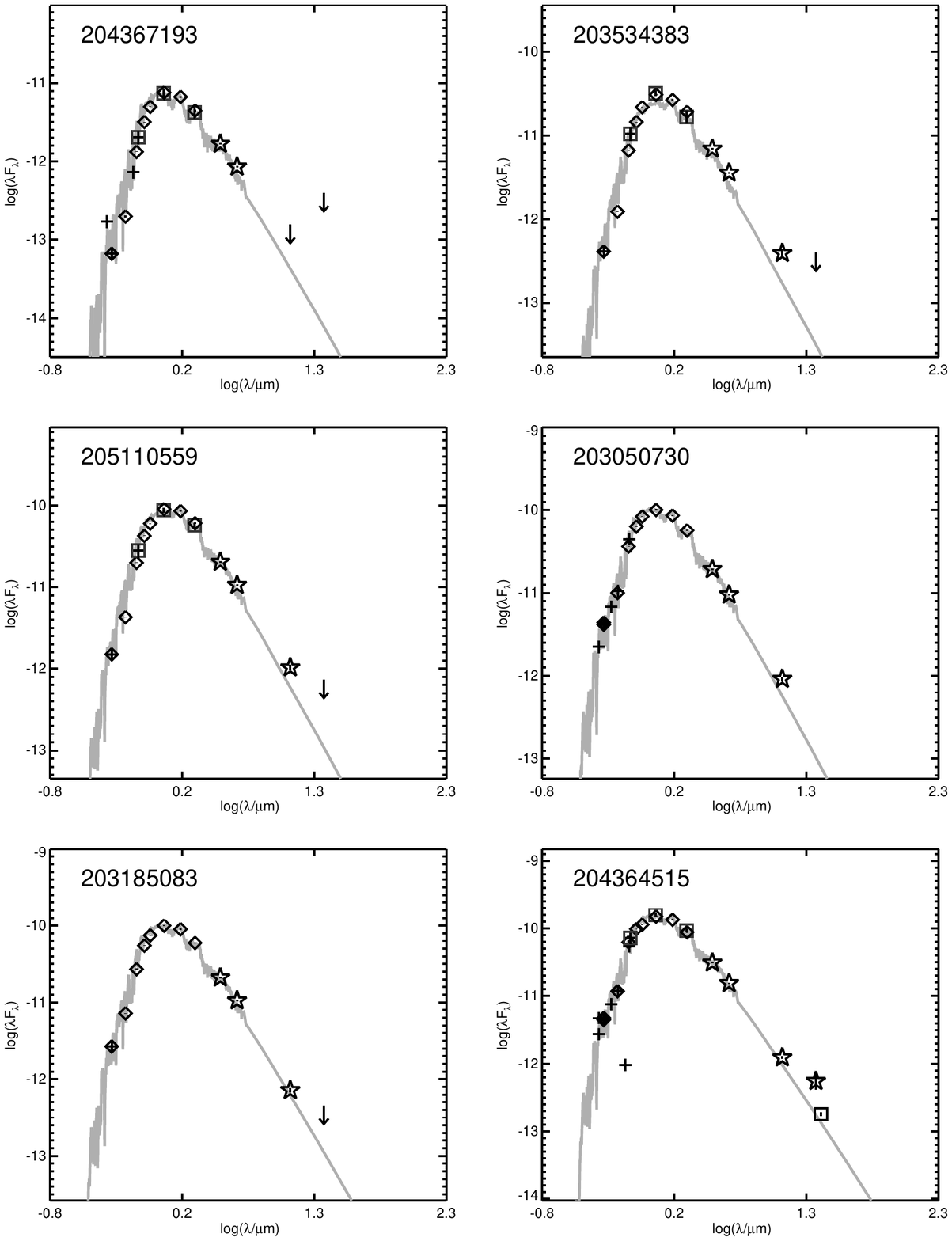}
\caption{Spectral energy distributions for the next six stars in Table 1.
\label{fig:SEDS2}}
\end{figure}

\begin{figure}[ht]
\epsscale{0.65}
\plotone{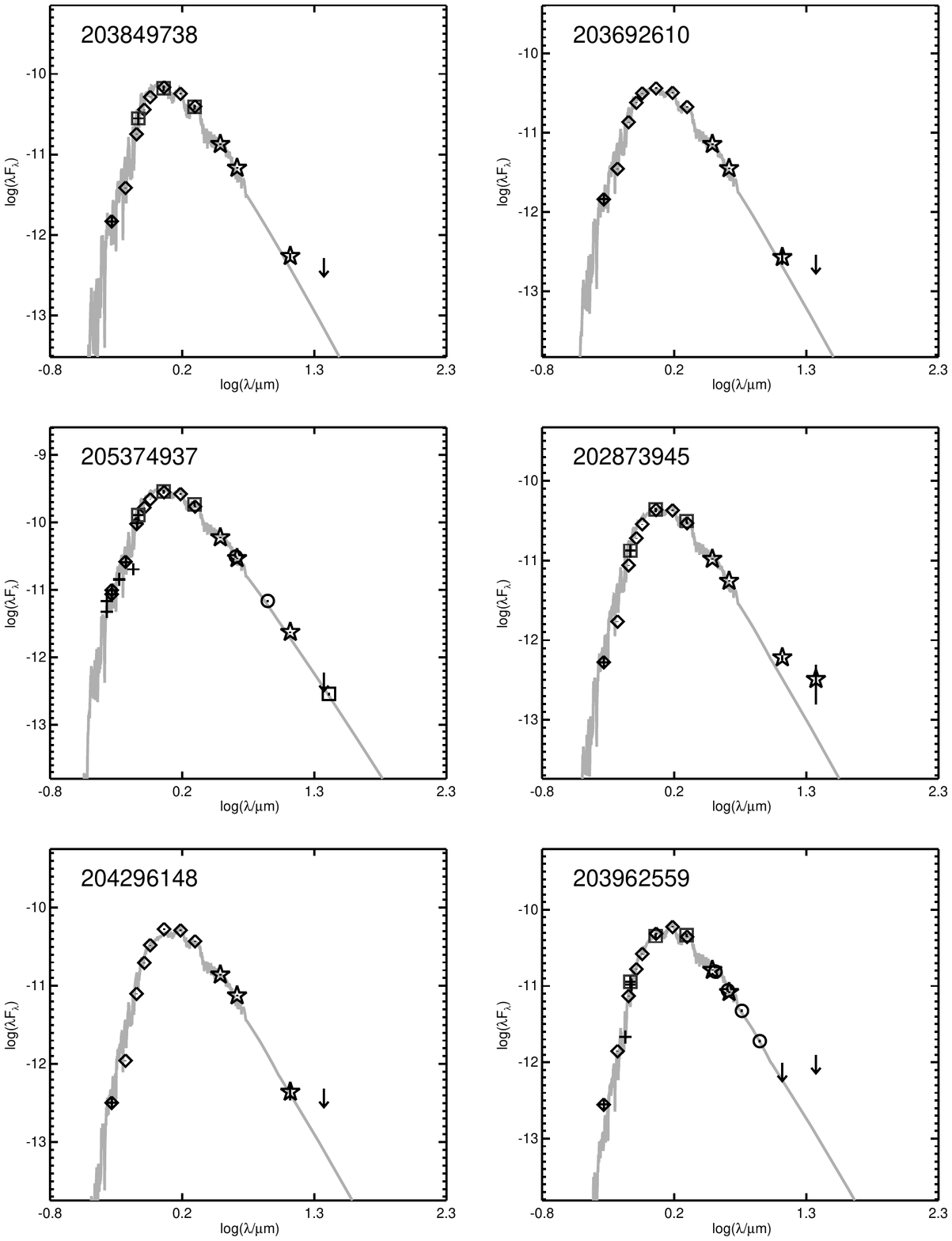}
\caption{Spectral energy distributions for the next six stars in Table 1.
\label{fig:SEDS3}}
\end{figure}

\begin{figure}[ht]
\epsscale{0.65}
\plotone{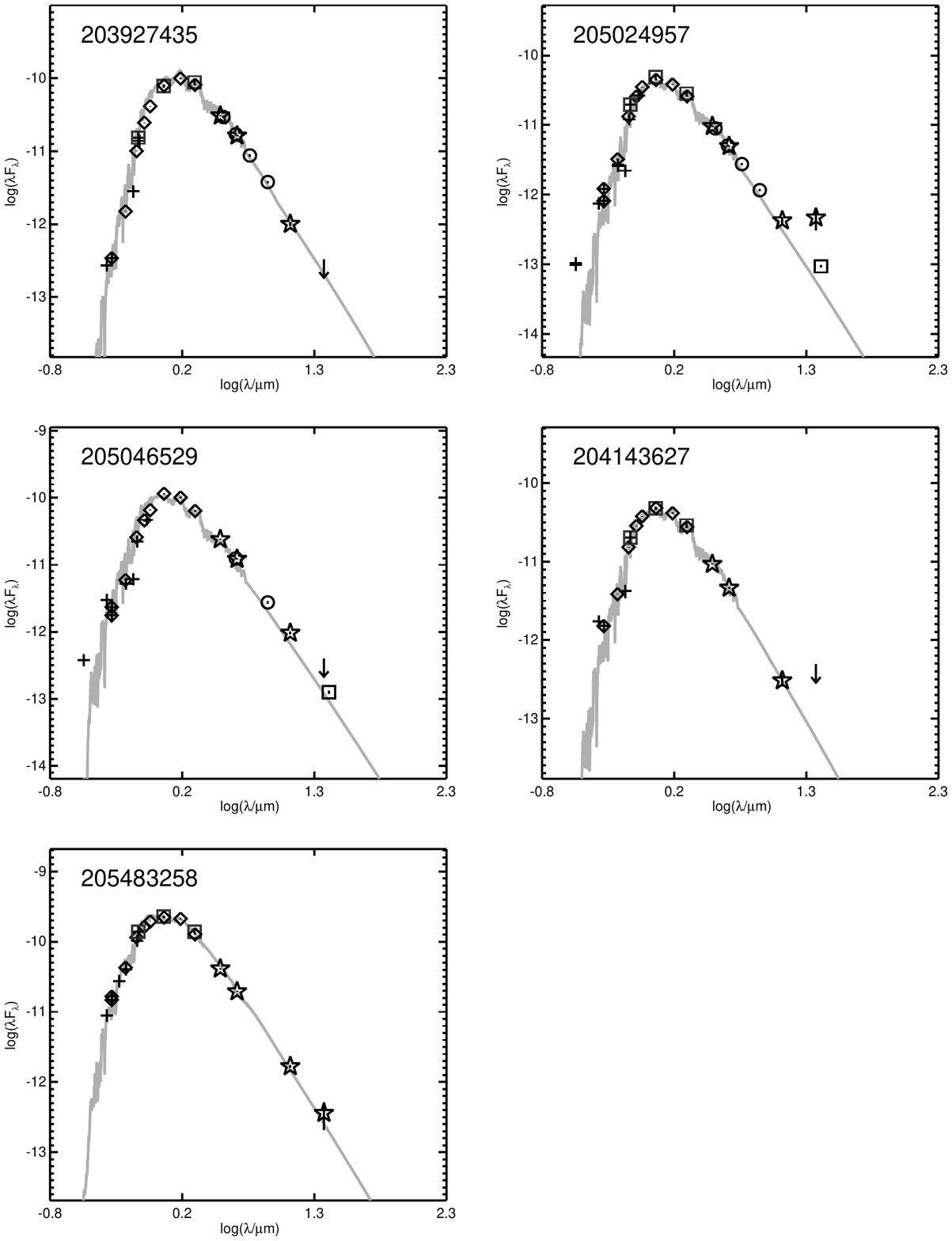}
\caption{Spectral energy distributions for the remaining stars in Table 1.
\label{fig:SEDS4}}
\end{figure}

Spectral energy distributions for all the stars from Table 1 are 
provided here as Figure \ref{fig:SEDS1} - Figure \ref{fig:SEDS4}.
Plots are log $\lambda F_{\lambda}$ in cgs units (ergs
s$^{-1}$ cm$^{2}$) against log $\lambda$ in microns. Symbols: $+$:
optical literature (SDSS, APASS); box at short bands: DENIS $IJK$;
diamond: 2MASS $JHK_s$; circle: IRAC; box at long bands: MIPS; stars:
WISE; arrows: limits; vertical bars (often smaller than the symbol)
denote uncertainties.  A Kurucz-Lejeune model for the corresponding spectral
type (taken to be M3 for those without known types) is also shown as
the grey line, normalized to the observations at $K_s$. Note that this
is not a robust fit, but just to ``guide the eye.'' Some models have
been lightly reddened to better fit the optical points ($A_J \leq 0.7$
in all cases).  Almost all the stars have SEDs consistent with  pure
photospheres; none have large IR excesses. EPICs 204117263 and
202873945 have apparent excesses, but the uncertainties are large and
we do not believe these apparent excesses are real. EPIC 203534383 has
a possible W3 detection but the PSF in the image is not circular, so
we do not believe this is a secure excess.  EPIC 203050730 has a W4
detection in the WISE catalog, but nothing is present at the source
location in the image, so we have changed it to an upper limit. EPIC
204364515 and 205024957 both have possible W4 detections that are
contradicted by the MIPS 24 $\mu$m data; since MIPS is higher spatial
resolution and more sensitive, we take the MIPS measurement as
reality, rather than the W4.  EPIC 205024957 has a small apparent
excess at 24 $\mu$m, and it is the only reasonably secure IR excess
for the stars in Table 1. Carpenter \etal\ (2009) had previously
identified this star as having a 24 $\mu$m excess, which they
interpreted as due to a debris disk (because the shorter wavelength
points are consistent with photospheric emission).

\section{New High Resolution Spectra}

We have obtained high resolution spectra for seven of the stars in Table
1  using the Keck HIRES
spectrograph (Vogt \etal\ 1994). Additional HIRES spectra of one of these star from
Table 1 -- EPIC 205483258 (RIK-210) -- are reported in David \etal\
(2016c).   The spectra cover the wavelength range roughly 4800 to 9200
\AA, at a spectral resolution of about R = 45,000. 
We measured equivalent widths for H$\alpha$ and lithium using the SPLOT
routine in IRAF.   

\begin{figure}[ht]
\epsscale{0.65}
\plotone{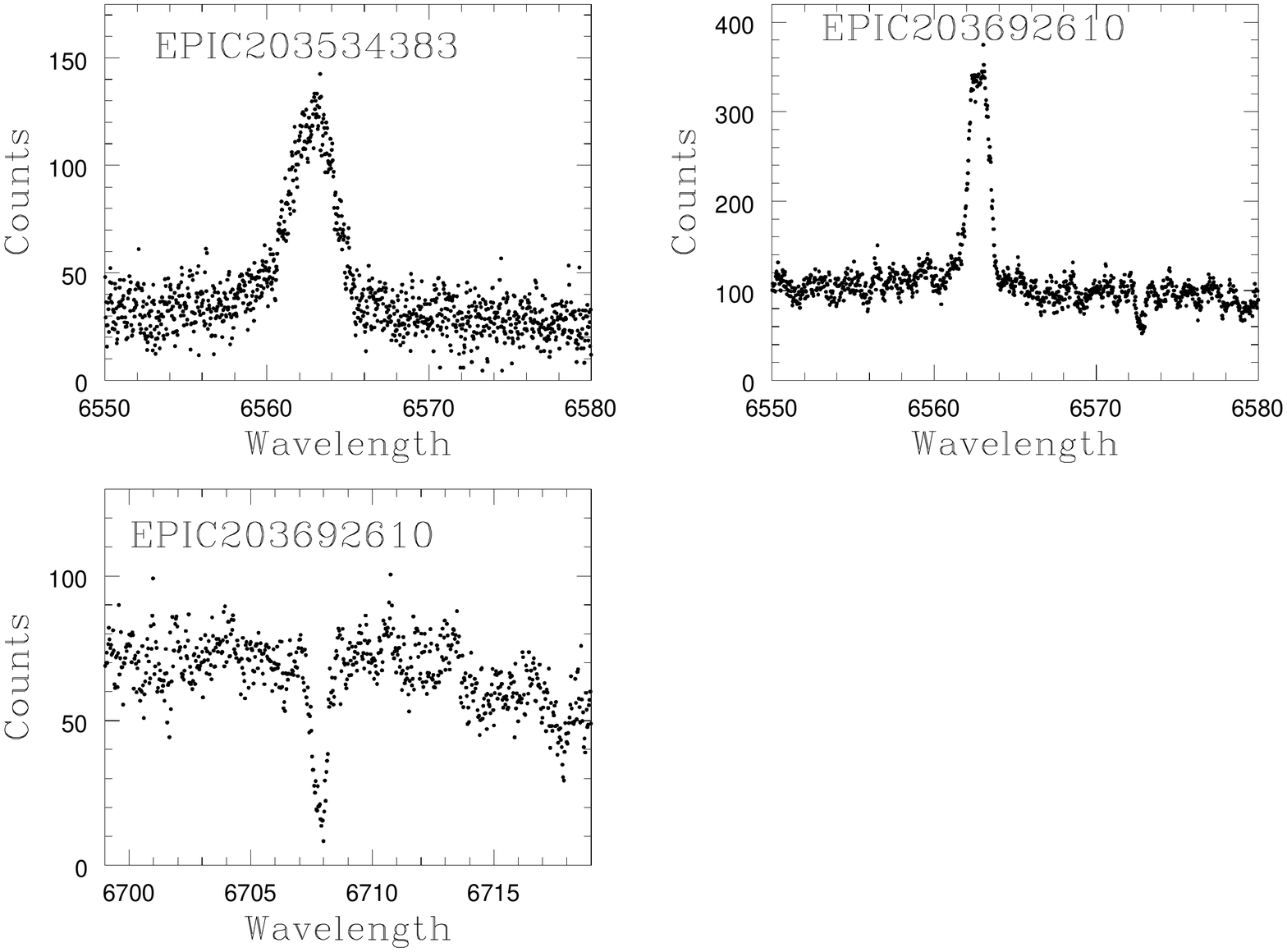}
\caption{Snippets of HIRES spectra for H$\alpha$ and for the lithium
6708 \AA\ doublet for EPIC 203692610, and for just H$\alpha$ for
EPIC 203534383.
\label{fig:hires_spectra}}
\end{figure}

For two of the stars -- EPIC 203534383 and EPIC 203692610 --
there was little spectral information in the published
literature, and we have used the HIRES spectra to determine spectral
types as well as to measure H$\alpha$\ and lithium equivalent widths.
Figure \ref{fig:hires_spectra} show the H$\alpha$ profiles for these two
stars, and the Li 6708 \AA\ feature for EPIC 203692610; the S/N at the
lithium line for EPIC 203534383 is too poor to detect lithium.
For EPIC 203692610, we determine equivalent widths of
$-$3.0 \AA\ for H$\alpha$ and 0.55 \AA\ for Li 6708.   For EPIC 203534383,
our derived H$\alpha$ equivalent width is $-$9.5 \AA.

We determined spectral types for the two stars based on the strength of the
TiO bandheads near 7100 \AA (Stauffer \etal\ 1979; Preibisch \etal\ 2001).
For both stars, we estimate a spectral type of M4, though for EPIC 203534383
this is uncertain due to the low S/N of the spectrum.

For the other four stars, EPIC 203849738, 204364515, 205024957 and 205046529 we have only used
the spectra to determine equivalent widths for the lithium 6708 \AA\ doublet.
Those lithium equivalent widths are reported in Table~1.

\section{Phased Light Curve Library}

\begin{figure}[ht]
\epsscale{0.65}
\plotone{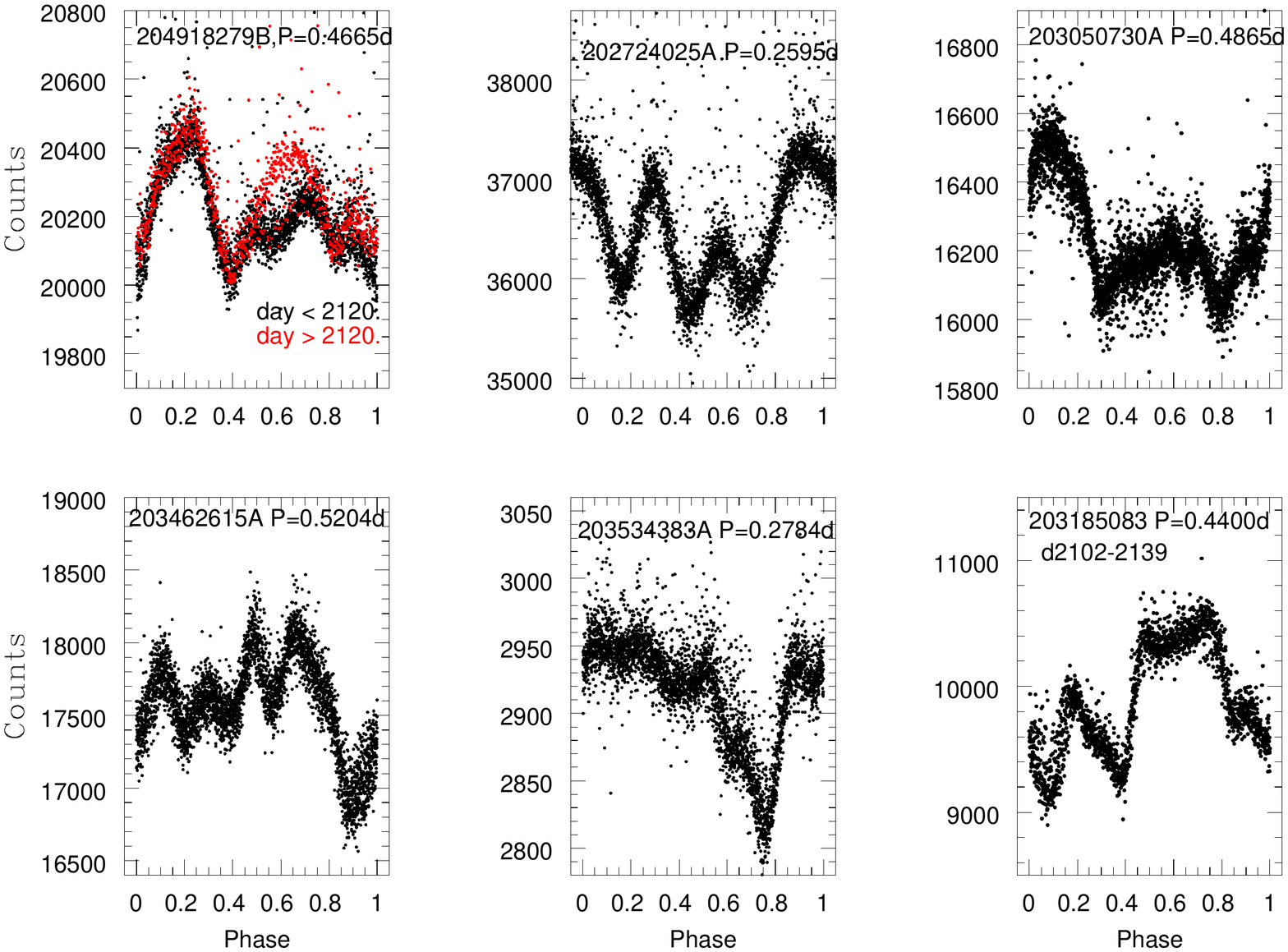}
\caption{Phased light curves for six of the scallop shell light curve class.
\label{fig:scallops1_light_curves}}
\end{figure}

\begin{figure}[ht]
\epsscale{0.65}
\plotone{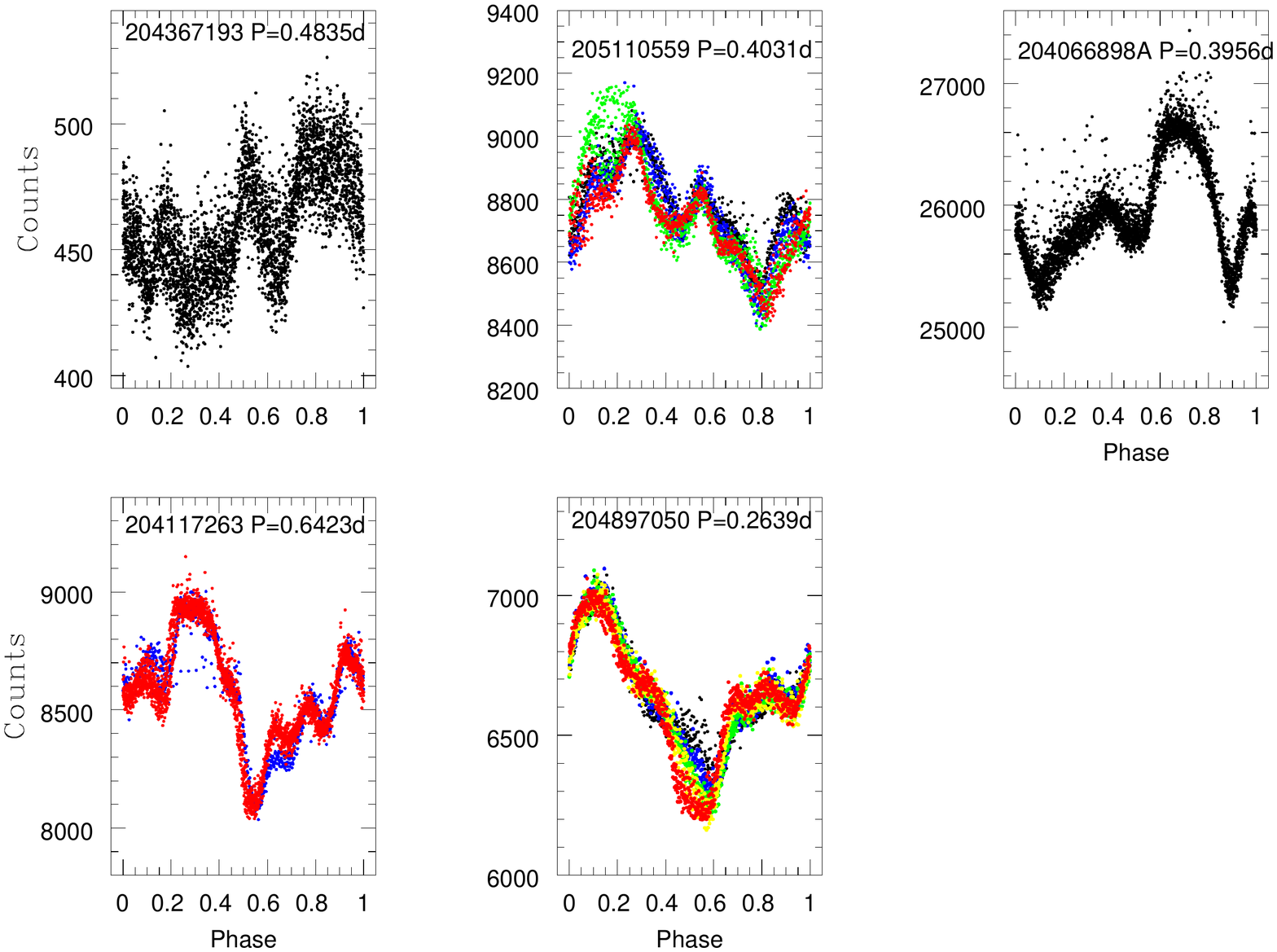}
\caption{Phased light curves for five of the scallop shell light curve class.
\label{fig:scallops2_light_curves}}
\end{figure}

\begin{figure}[ht]
\epsscale{0.65}
\plotone{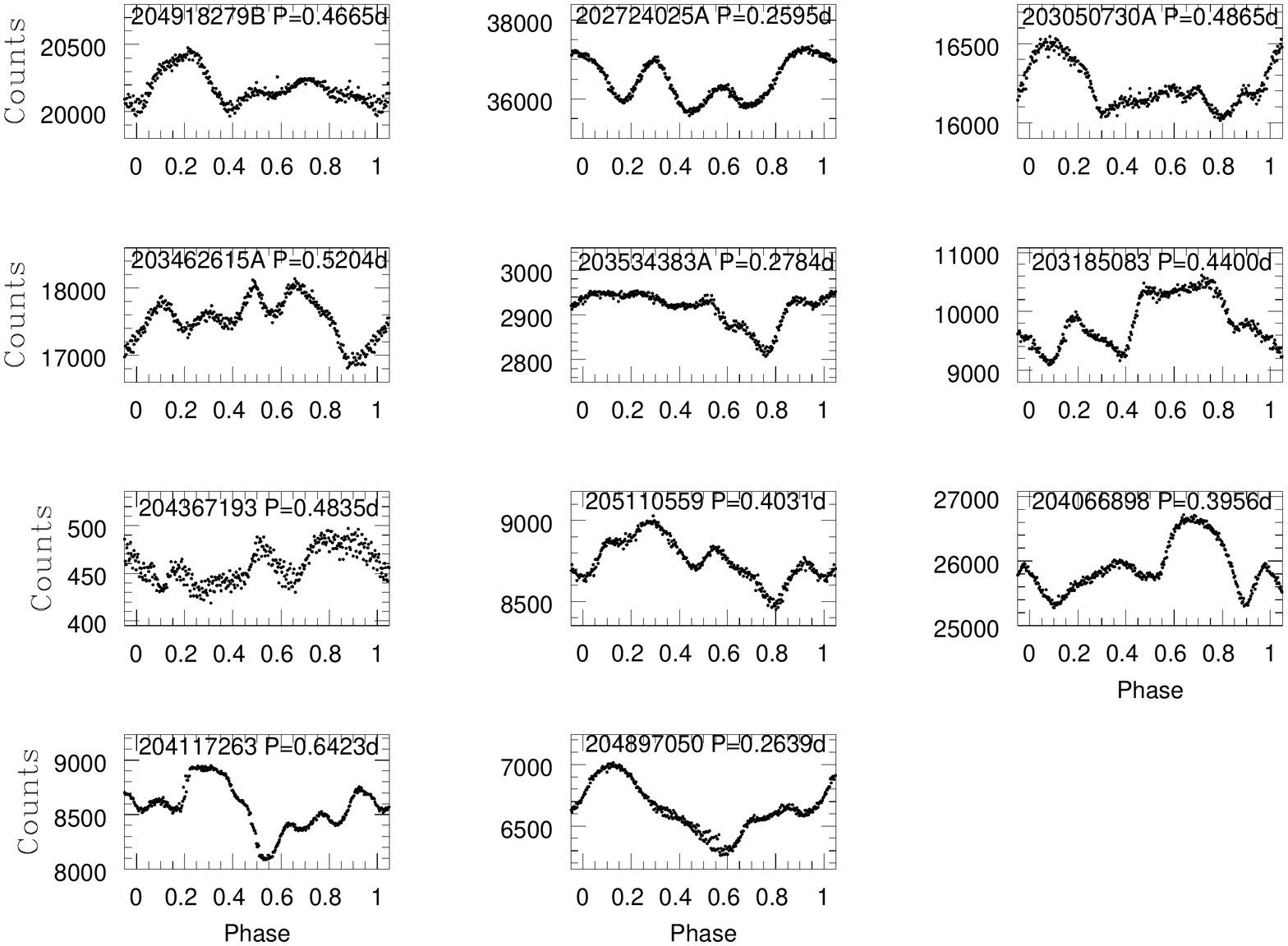}
\caption{Median-smoothed phased light curves for stars in the scallop
shell class.   Where the phased light curve changed shape significantly
during the campaign, only data from either before or after the change
was used in forming the median.
\label{fig:scallops_medians}}
\end{figure}

Phased light curves for all of the stars with scallop-shell light curves
are provided in Figure \ref{fig:scallops1_light_curves} and
Figure \ref{fig:scallops2_light_curves}.   For those stars whose phased
light morphology underwent significant evolution during the {\em K2} campaign,
we split up the 78 day {\em K2} campaign into several time windows and plot
each window with a different color point.   Note that phase = 0 is
arbitrary for these light curves.  Figure \ref{fig:scallops_medians}
shows median smoothed versions of these light curves.  For the stars
whose phased morphology evolved during the {\em K2} campaign, we derived the
median shape from only the first time window.

\begin{figure}[ht]
\epsscale{0.65}
\plotone{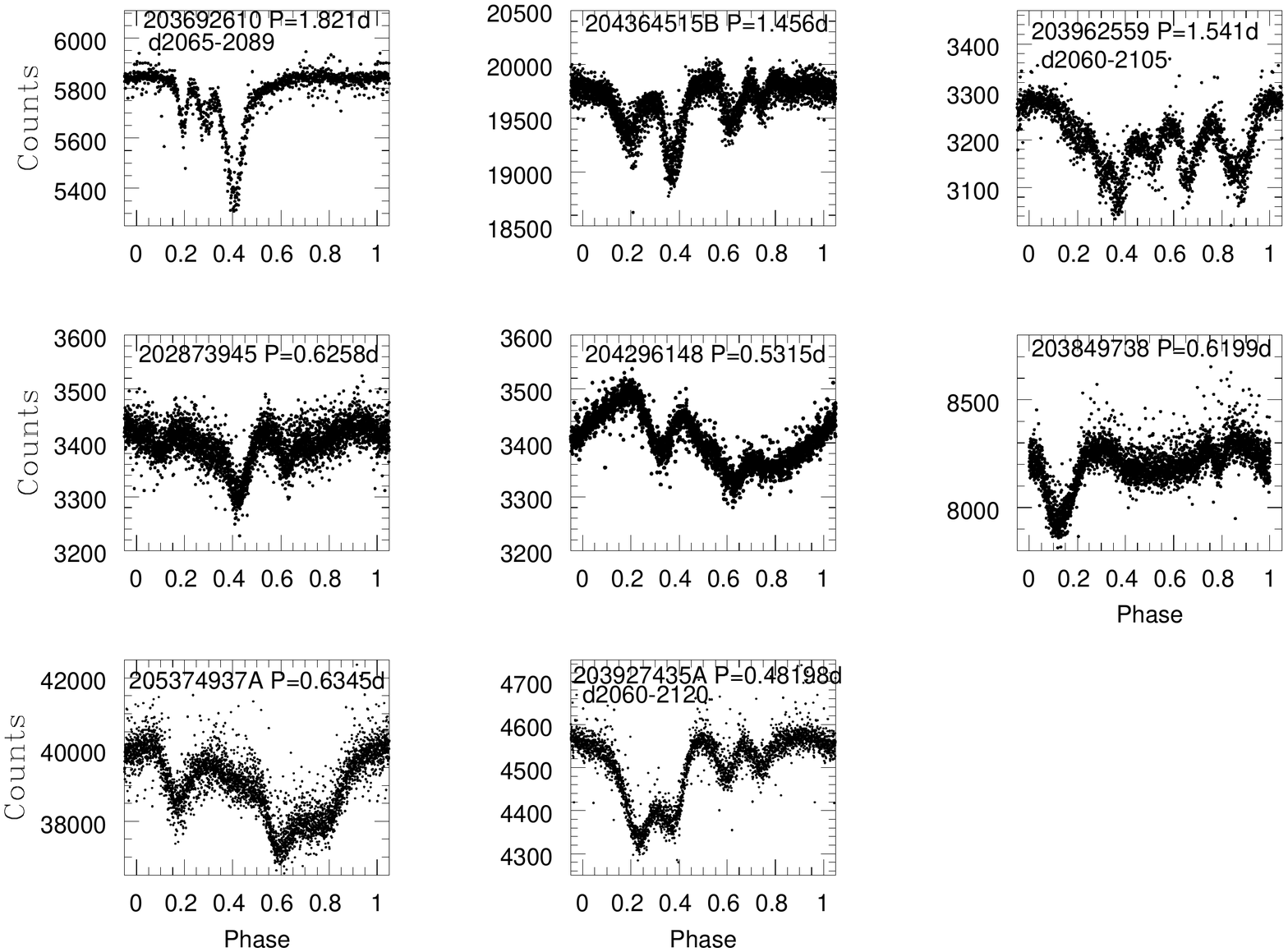}
\caption{Phased light curves for stars in persistent flux-dip class.
\label{fig:batwing_light_curves}}
\end{figure}

\begin{figure}[ht]
\epsscale{0.65}
\plotone{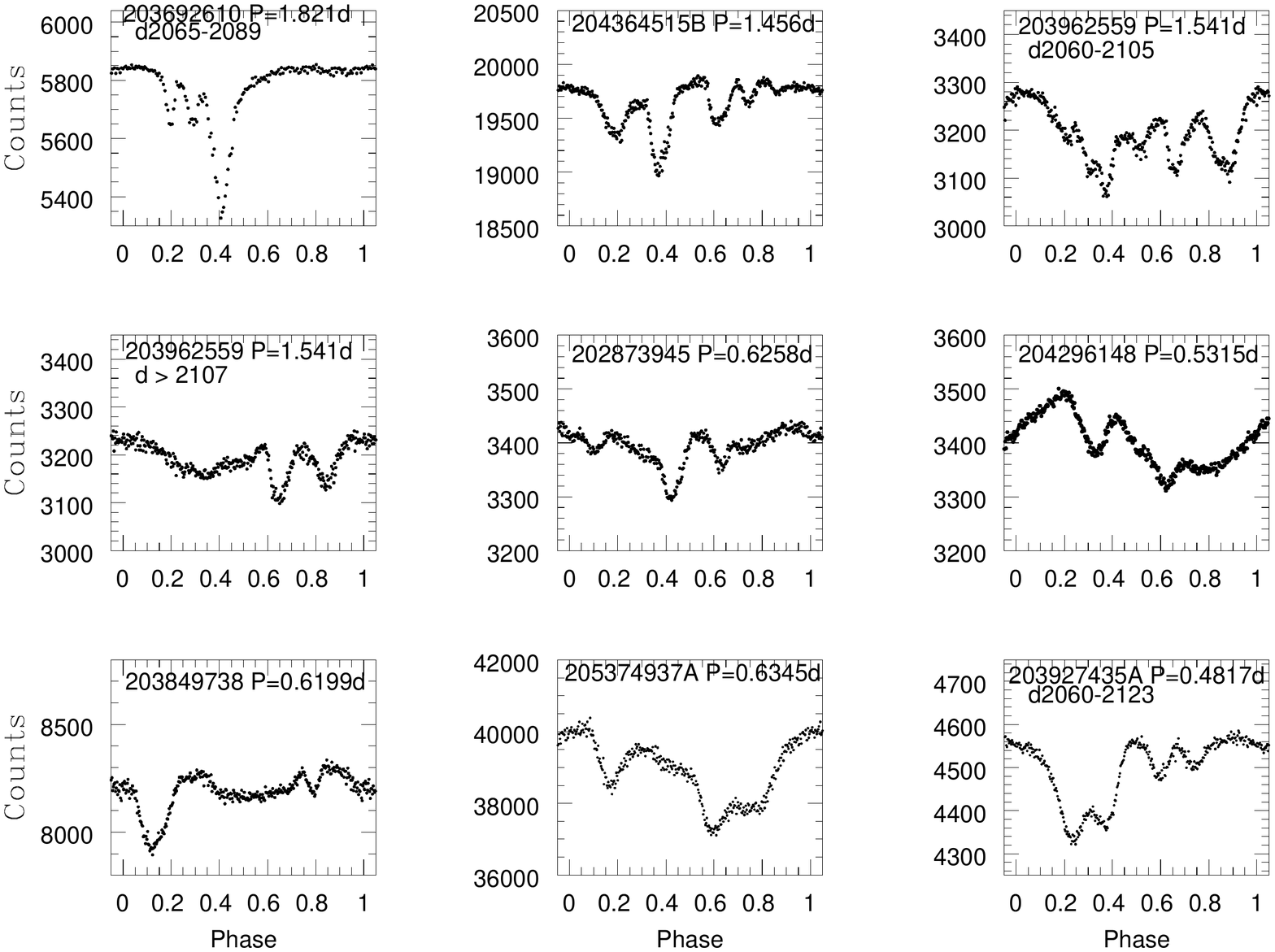}
\caption{Median-smoothed phased light curves for stars in the persistent 
flux-dip class.  EPIC203962559 is shown twice, in order to illustrate
the change in its waveform after the flare on day 2106.
\label{fig:batwing_medians}}
\end{figure}

Figure \ref{fig:batwing_light_curves} and Figure \ref{fig:batwing_medians}
show similar figures for all stars in the persistent flux dip class.

\section{Light Curve Detrending}

Our standard processing (see \S 3) produced light curves from which
short-term non-astrophysical trends had been removed and identified
the best Lomb-Scargle period (or periods) for each star.   Plots were
created showing the original light curve and the light curve phased to
each period identified for the star.   Those latter plots were the
basis for selecting stars that might be included in this paper.  

In order to produce higher quality versions of the phased light curves
and to validate stars for inclusion in the paper, we conducted
additional processing. For stars with long-term trends in their light
curve, we fit low order polynomials to the data and then removed the
long-term trend (assuming that it was likely to be non-astrophysical
-- but that in any case the presence of the long term trend would
degrade the ability to phase-fold the light curve and examine its
detailed structure).   In some cases, even after this step, the light
curve retained  bumps and wiggles on timescales much longer than the
rotation period but shorter than the {\em K2} campaign length.   In those
cases, we determined the median count-rate at each period and then
normalized the data in each period to the same rate (for stars with
very short periods, we instead adopted an interval of 4 periods for
the calculation).   This again could remove real flux variations from
the light curve, but removing these trends allowed us
to produce phased light curves with significantly less noise.

More than half of the scallop-shell and persistent narrow flux dip
stars  had two significant, narrow-peaked, independent (non-harmonic)
Lomb-Scargle periods,  indicating that they are binaries (or at least
that light from two stars is contributing to the {\em K2} light curve).   In
all cases, the plots from our standard processing showed one of the
two stars to be relevant for this paper with the other star appearing
to have a normal, spotted star light curve.   For these systems, we
adopted an iterative approach to the final processing.  First, we
phased the initial light curve to the period of the star that appears
to have the larger light curve amplitude.  We then derived a median
smoothed version of that light curve, and then subtracted that
median-smoothed light curve from the initial light curve -- resulting
in a residual light curve that should largely be the signal from the
second star.   We then phased that residual light curve to the second
Lomg-Scargle period, derived a median-smoothed version of that, and
then subtracted this median-smoothed light curve from the initial
light curve - resulting in a new residual light curve that should
largely be the signal from the 1st star (now with the signal from the
second star largely removed). This process was repeated if necessary --
with the final product being a phased light curve for the star of
interest, from which most or all of the signal from the binary
companion had been removed.

\clearpage

\section{Simulated Light Curves of Spotted Stars}

Cool spots generally produce phased light curve morphologies ressembling
sinusoids or broad dips or humps whose FWZI is approximately 50\%
in phase.   That is because spots seen from most view angles will be
visible for nearly half the rotation period of the star (or more), and
therefore the signatures they impose on the light curve will span
half or more in extent of the phased light curve.   However, for a view
angle which places a spot very near the top or bottom of the visible 
hemisphere of the star, the spot could be visible for significantly less than half
the rotation period.   Therefore, it is conceivable that in such
geometries, single spots could produce narrow flux dips (as observed
for our persistent or transient flux dip class) or that several spots
at widely spaced longitudes could produce the highly structured, wavy
phased light curves we find for the scallop-shell class.

To test this hypothesis, we have created a program to produce simulated
light curves for stars with up to four cool spots placed at arbitrary
latitudes and longitudes, and viewed from an arbitary view angle.  
The spots are required to be circular and to
have a uniform surface brightness that is a factor of between 0.0 and 1.0
of the surface brightness of the pristine photosphere of the star.
The star is assumed to be spherical.  We adopt a linear limb darkening
law, and a limb darkening coefficent (normally designated as "u") of
0.8, following the tables of Claret \etal\ (2012) for a mid-M dwarf
and for the {\em Kepler} camera.   We then allow the star to rotate, and 
measure its brightness after every 1 degree in rotation, following the
star for a complete rotation period.   For the present purposes, we
set all the spots to be "black" -- that is, non-emitting -- because that
will produce the largest possible amplitude for a given spot size.

\begin{figure}[ht]
\epsscale{0.65}
\plotone{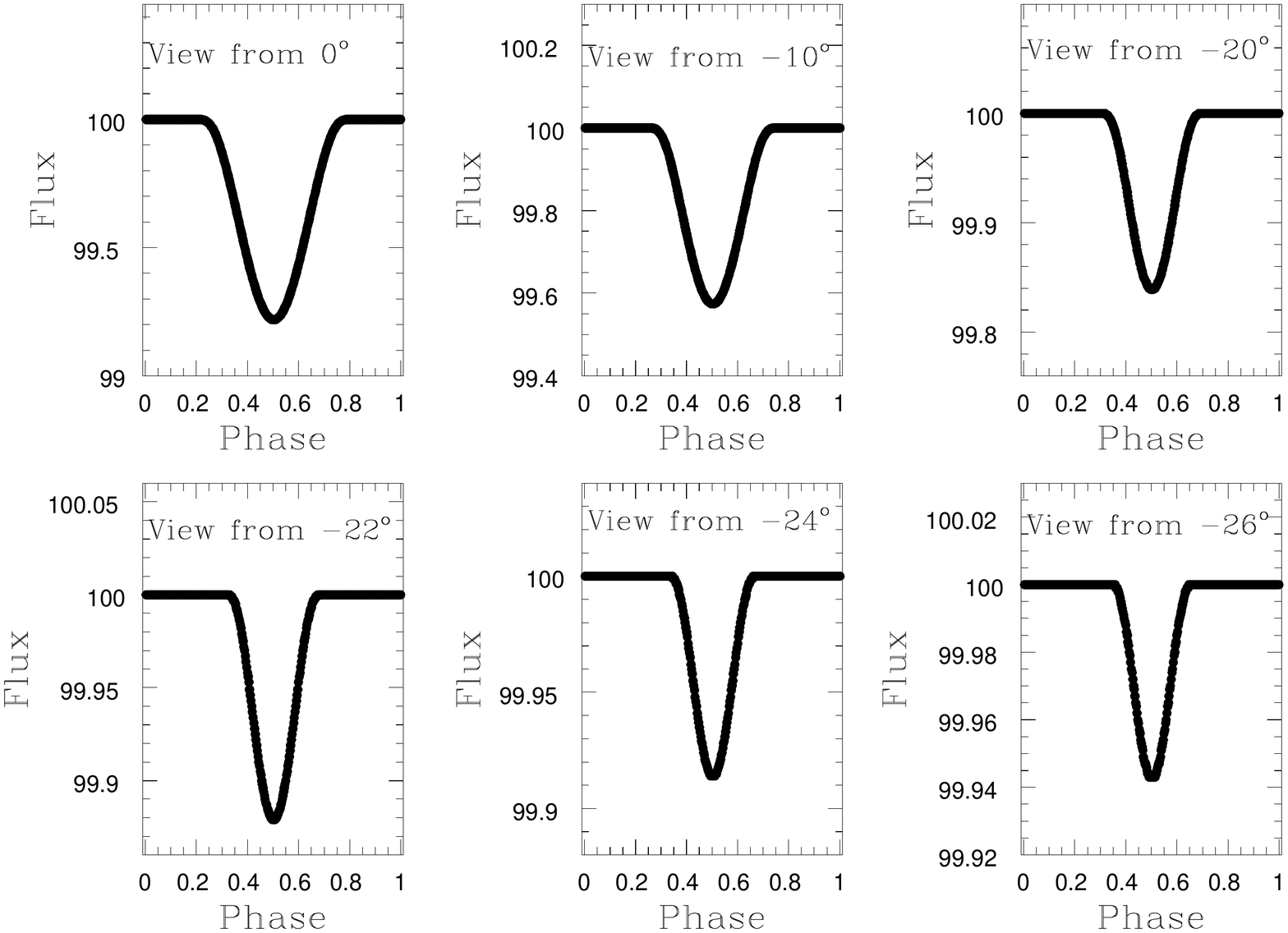}
\caption{Model phased light curves for single spots of radius 8$\arcdeg$.
The spots are assumed to be circular and ``black".  The spot was placed
at latitude 60$\arcdeg$, with the viewer located at latitudes ranging
from 0 to $-$26$\arcdeg$.  Only the case for an observer at $-$26$\arcdeg$ 
produces a dip width comparable to the broadest dips we observe -- but
in that case, the dip depth is tiny.
\label{fig:onespot_8d}}
\end{figure}

\begin{figure}[ht]
\epsscale{0.65}
\plotone{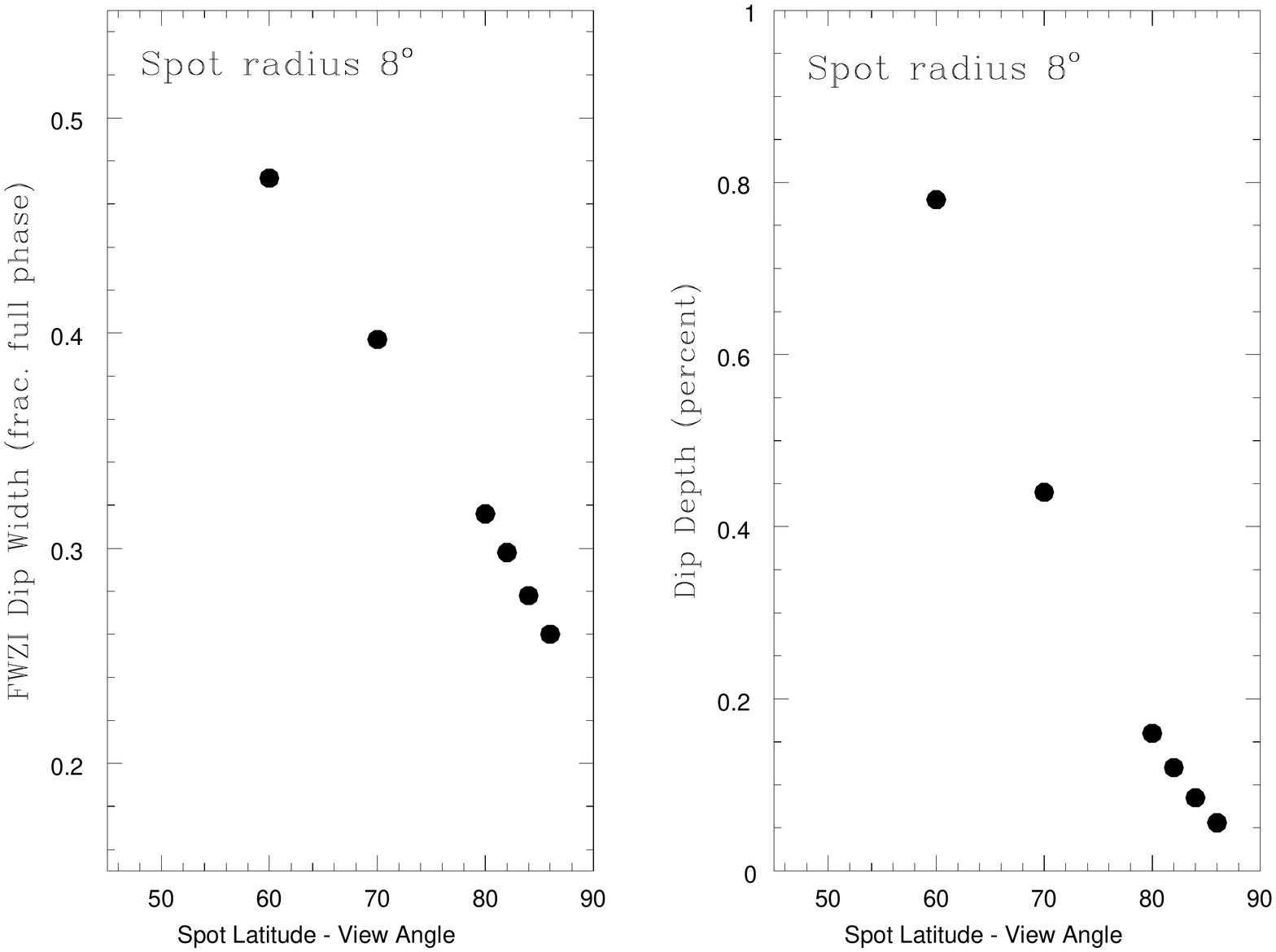}
\caption{FWZI (in fraction of the full period)  
and dip depth (in per cent) versus the angle between
the spot latitude and the observer latitude for the model stars with
a single 8 degree radius spot.  The plot shows that even for an
angle of 60$\arcdeg$, the dip will have a width covering almost half the
phase.   Only for angles $\sim$85$\arcdeg$ do the widths approach what we
observe in Upper Sco, but then the depths become $<$ 0.1\%.
\label{fig:spotwidth_depth_8d}}
\end{figure}

We have first used the program to attempt to produce narrow flux dips
that might be similar to those we observe.  The persistent flux dip
stars have FWZI for their dips ranging from 0.06 to 0.2 (in fractional
phase) and depths ranging from 1\% to 7.5\%, with an average of about 2.5\%.
The transient flux dip stars have FWZI for their dips of about 0.06,
0.08, 0.20 and 0.20 and maximum depths of 1.5\%, 2.5\%, 3\% and 20\%.
To determine if those properties can be matched by cool spots, we first
ran models with single spots of 8 degree radius, viewed
from a range of locations up to a view angle which places the spot
essentially at the upper limb of the visible hemisphere when the spot
is facing the observer.  Figure \ref{fig:onespot_8d} 
shows the flux dips resulting from those spots for the selected viewing 
angles.   Figure \ref{fig:spotwidth_depth_8d} shows how the width and
depth of these flux dips vary with view angle.

\begin{figure}[ht]
\epsscale{0.65}
\plotone{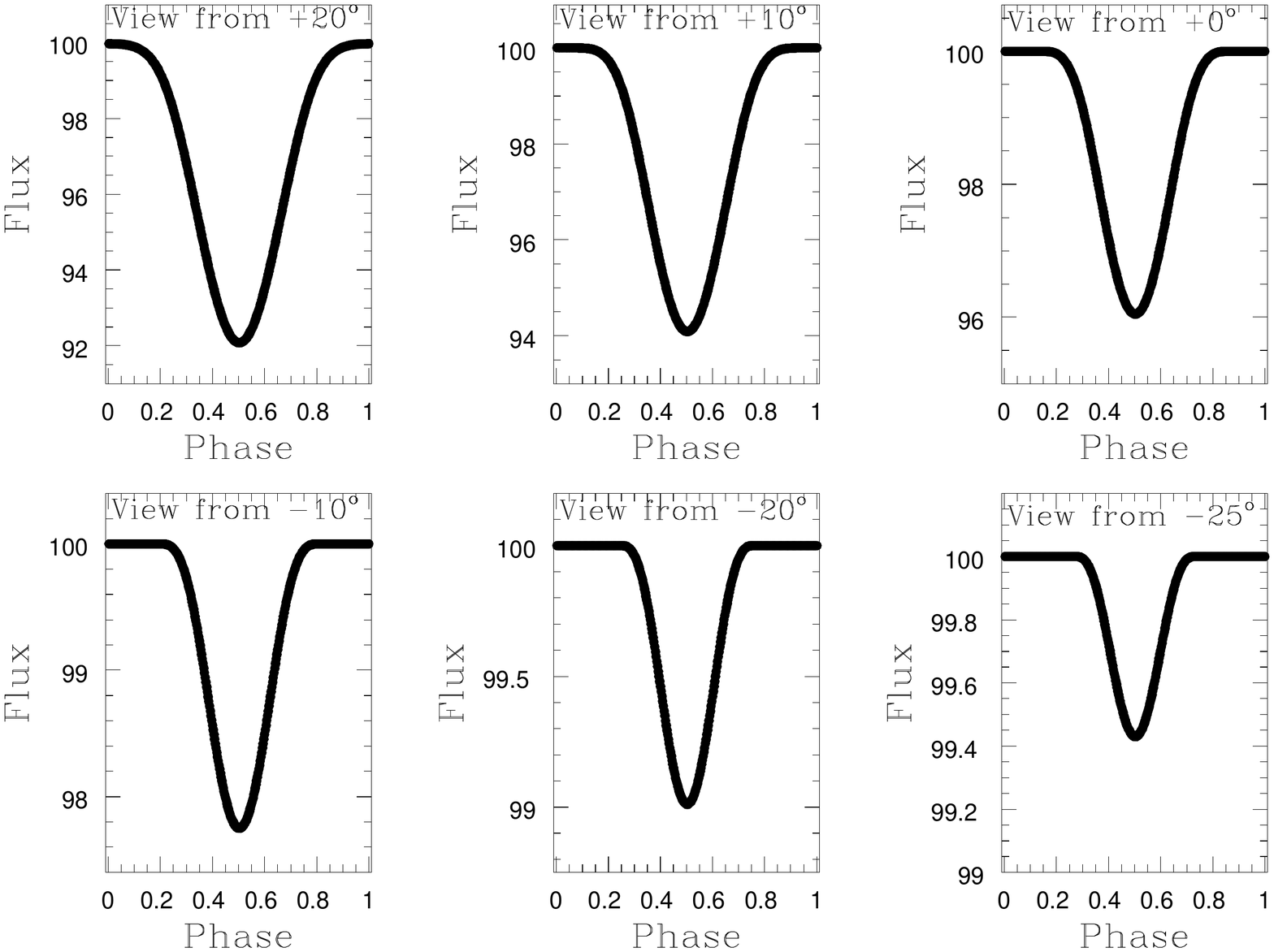}
\caption{Model phased light curves for single spots of radius 18$\arcdeg$.
The spots are assumed to be circular and ``black".  The spot was placed
at latitude 60$\arcdeg$, with the viewer located at latitudes ranging
from +20 to $-$25$\arcdeg$.  
\label{fig:onespot_18d}}
\end{figure}

\begin{figure}[ht]
\epsscale{0.65}
\plotone{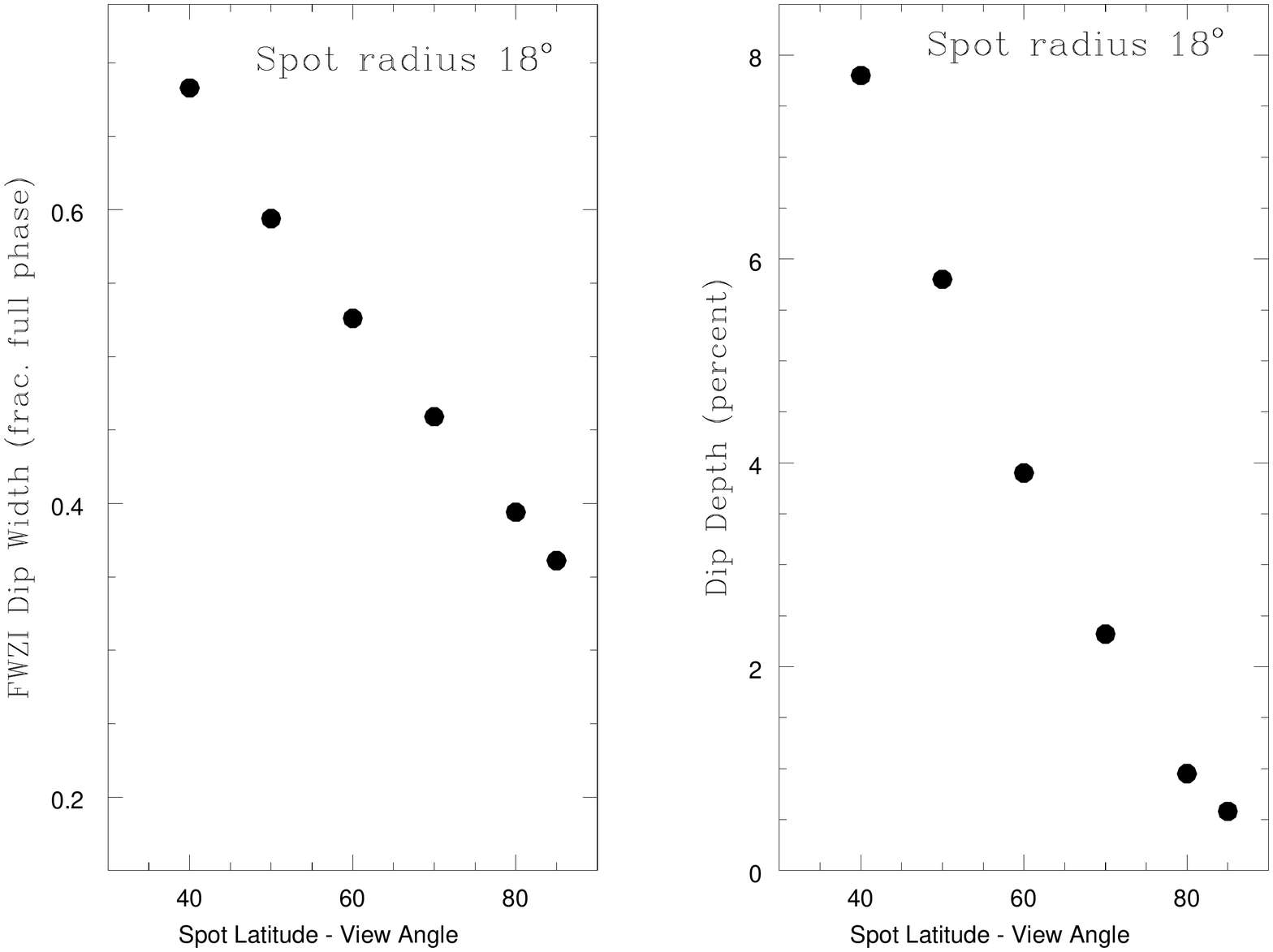}
\caption{FW5\% (in fraction of the full period)  
and dip depth (in per cent) versus the angle between
the spot latitude and the observer latitude for the model stars with
a single 18$\arcdeg$ radius spot.  
\label{fig:spotwidth_depth_18d}}
\end{figure}

Based on the two preceding figures, 
spots with radii of 8$\arcdeg$ or less cannot be responsible for our
short-duration flux dips because they are incapable of producing dip
depths of 1\% or more.  We therefore decided to try a larger spot;
specifically a spot of radius 18$\arcdeg$.  Figure \ref{fig:onespot_18d} 
and Figure \ref{fig:spotwidth_depth_18d} show the flux dip shapes
and their depth and width as a function of the observer's view to the
star.  A spot this size can produce flux dip depths of several percent;
however, the dips are relatively broad.  The plots demonstrate that
no view angle will produce a dip with depth $>$ 1\%\ and FWZI $\leq$
0.2.  We have tried other combinations of parameters, but the answers
are always the same.   For any cold spot that can produce a FWZI
of 0.2 or less, the spot depth will always be MUCH less than 1\%\footnote{
We actually measure the full-width at 5\% depth for the models, because they
are noiseless.  This more closely approximates the FWZI we measure for
the observed light curves}.
Therefore, cold spots CANNOT be the explanation for the short-duration
flux dips for either our persistent flux dip or transient flux dip
light curves.

\begin{figure}[ht]
\epsscale{0.65}
\plotone{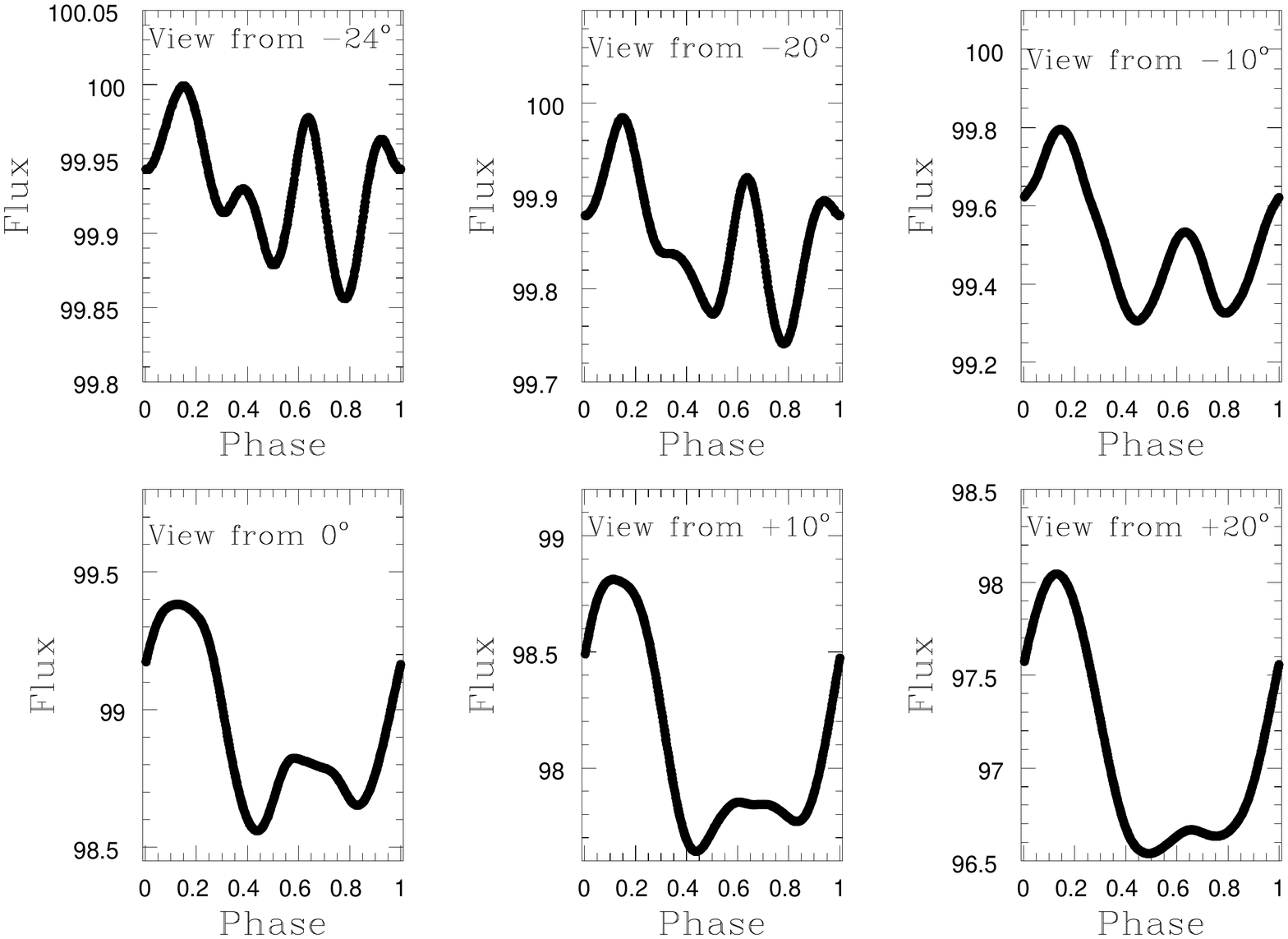}
\caption{Model light curves for a star for four cold spots near 60$\arcdeg$
latitude and widely spaced in longitude; the spots are assumed to have
radii of 8 to 10$\arcdeg$ and are completely black.  When placed very
near the upper limb of the star as seen by an observer, such an arrangement
can produce a phased light curve whose shape is reminiscent of our scallop
shell class.  However, the amplitude is tiny.  At smaller view angles, the
light curve amplitude grows but the shape becomes progressively smoother.
\label{fig:model_scallop}}
\end{figure}

As a test to determine if appropriately arranged spots could produce
phased light curves ressembling the scallop-shell class, we placed
four spots of 8 to 10$\arcdeg$ radius at latitudes between 60 and 62$\arcdeg$
and longitudes of 0, 80, 180 and 250$\arcdeg$.  We first put the observer at a
latitude of $-$24$\arcdeg$ in order for those spots to be near the upper limb of
the star as seen by the observer.   As shown in the first panel of
Figure \ref{fig:model_scallop}, this does produce a phased light curve whose
shape is reasonably similar to several of the stars in our scallop shell
class.  However, the light curve amplitude is tiny.   The other panels in
Figure \ref{fig:model_scallop} show that while the amplitude does increase
as the observer to spot angle decreases, the degree of structure in the
light curve also decreases.   By the point where the light curve amplitude
is $\sim$1.0\%, the morphology of the light curve is much smoother and no longer
ressembles the scallop-shell stars.   Choosing larger spots does not
help --
there is again a trade off between degree of structure in the light curve
and amplitude, and in no case have we been able to produce shapes which
approximate our stars and had amplitudes $>$ 0.4\%.   Because many of the
scallop-shell stars have light curve amplitudes of 5-10\%, we conclude
that cold spots also are unable to explain the scallop-shell light curve
morphologies.

\end{document}